\pdfoutput=1

\documentclass[onecolumn]{emulateapj}

\usepackage{amsmath}
\usepackage{graphicx}
\usepackage{color}
\usepackage[colorlinks]{hyperref}
\hypersetup{
    colorlinks,	
    citecolor=blue,
}

\newcommand{\be}{\begin{eqnarray}}
\newcommand{\ee}{\end{eqnarray}}

\def\jcap{JCAP}

\shorttitle{Modeling bias in SMBH spin measurements}
\shortauthors{Riaz et al.}

\begin{document}

\title{Modeling bias in supermassive black hole spin measurements}

\author{Shafqat~Riaz\altaffilmark{1}, Dimitry~Ayzenberg\altaffilmark{1}, Cosimo~Bambi\altaffilmark{1,\dag}, and Sourabh~Nampalliwar\altaffilmark{2}}

\altaffiltext{1}{Center for Field Theory and Particle Physics and Department of Physics, 
Fudan University, 200438 Shanghai, China. \email[\dag E-mail: ]{bambi@fudan.edu.cn}}
\altaffiltext{2}{Theoretical Astrophysics, Eberhard-Karls Universit\"at T\"ubingen, 72076 T\"ubingen, Germany}

\begin{abstract}
X-ray reflection spectroscopy (or iron line method) is a powerful tool to probe the strong gravity region of black holes, and currently is the only technique for measuring the spin of the supermassive ones. While all the available relativistic reflection models assume thin accretion disks, we know that several sources accrete near or above the Eddington limit and therefore must have thick accretion disks. In this work, we employ the Polish donut model for the description of thick disks. We thus estimate the systematic error on the spin measurement when a source with a thick accretion disk is fitted with a thin disk model. Our results clearly show that spin measurements can be significantly affected by the morphology of the accretion disk. Current spin measurements of sources with high mass accretion rate are therefore not reliable.
\end{abstract}

\keywords{accretion, accretion disks --- black hole physics --- gravitation}

%%%%%%%%%%%%%%%%%%%%%%%%%%%%%%%

\section{Introduction}

Most galaxies are thought to harbor a supermassive black hole at their center~\citep{bh}. Reliable mass and spin measurements are crucial to study the physics and the astrophysics of black holes. The mass is relatively easy to infer, by studying the motion of stars or gas orbiting the black hole~\citep{m1,m2}. The spin measurement is more challenging because the spin has no gravitational effects in Newtonian gravity and its measurement thus requires the study of relativistic effects near the compact object. X-ray reflection spectroscopy (or the iron line method) is a powerful technique to measure black hole spins~\citep{s1,s2,s3,s4} and even to test general relativity in the strong field regime~\citep{k1,k2,k3,k4}. Moreover, it is currently the only method to measure the spin of supermassive black holes, while spin measurements for stellar-mass black holes are also possible with the continuum-fitting method~\citep{cfm1,cfm2} and gravitational waves~\citep{gw}.

Current relativistic reflection models assume that the accretion disk is geometrically thin and that the inner edge is at the innermost stable circular orbit (ISCO) or, if the disk is truncated, at a larger radius. Numerical and observational studies show that accretion disks are geometrically thin with the inner edge at the ISCO when the source is in the thermal state accreting between a few percent to up to 30\% of its Eddington limit~\citep{nt1,nt2}. However, especially in the case of supermassive black holes, we know that a source can easily exceed this 30\% bound. The supermassive black holes in narrow-line Seyfert~1 galaxies, which are the most common target for X-ray reflection spectroscopy spin measurements, are thought to accrete near or even above the Eddington limit~\citep{e1,e2,e3,e4}. In such a case, the radiation pressure should make the accretion disk geometrically thick and the inner edge of the disk is likely inside the ISCO radius~\citep{polish}. The question then is whether we can fit the spectra of these sources with a thin disk reflection model for measuring the black hole spin, and how large is the resulting systematic error on the spin measurement.

It is quite surprising that the impact of the disk structure on X-ray reflection spectroscopy measurements has been overlooked for a long time and that thin disk reflection models are commonly used to infer the properties of all sources, regardless of their mass accretion rate. There are only a few studies in literature that investigate this issue. \citet{ww07} studied the reflection spectrum of a thick disk, but not in the case of high mass accretion rate and their disk is actually truncated before the ISCO radius. \citet{RF2008} ran 3D MHD simulations of thin accretion disks in a pseudo-Newtonian potential to study the emission of radiation from the region inside the ISCO and its impact on black hole spin measurements using X-ray reflection spectroscopy. \citet{taylor} calculated the reflection spectrum of a thin disk with finite thickness to determine the systematic error on black hole spin measurements. \citet{2019ApJ...884L..21T} ran general relativistic radiation MHD simulations of a black hole accreting above the Eddington limit to study its iron line profile, paying particular attention to the role of the funnel geometry and wind acceleration.

In the present work, we employ the Polish donut model~\citep{polish}, which is the only available semi-analytic model to describe thick disks. We use a ray-tracing code to convolve synthetic reflection spectra calculated with {\sc xillver}~\citep{xill1,xill2} and predict synthetic relativistic reflection spectra from Polish donut disks around black holes. We then fit the new spectra with {\sc relxill}~\citep{rel1,rel2}, which is a relativistic reflection model assuming a geometrically thin accretion disk. Our analysis clearly shows that using a thin disk model to fit the reflection spectrum of a source with a thick disk has important effects on the measurements of the properties of the source. In particular, spin measurements become unreliable. Our finding should warn that current X-ray reflection spectroscopy spin measurements of sources accreting near or above the Eddington rate should be taken with great caution.

The content of the present paper is as follows. In Section~\ref{s-polish}, we briefly review the basic properties of Polish donut disks. In Section~\ref{s-sim}, we present our method to investigate modeling bias in supermassive black hole spin measurements and we describe the main properties of our simulations, while more details can be found in the Appendix. In Section~\ref{s-res}, we present the results of our simulations. Concluding remarks are in Section~\ref{s-con}. Throughout the paper, we employ units in which $G_{\rm N} = c = 1$, so the gravitational radius of a black hole of mass $M$ is $r_{\rm g} = M$.

\section{Thick accretion disks}\label{s-polish}

While geometrically thin and optically thick accretion disks are well described by the Novikov-Thorne model~\citep{nt1}, there is currently no good and simple model for the description of thick disks. The only available semi-analytic model to describe the accretion disk of sources accreting near or above the Eddington limit is the so-called Polish donut model~\citep{polish}. In such a model, the disk is still non-self-gravitating, but the pressure in the accretion material is taken into account. Unlike the Novikov-Thorne model in which the disk structure is completely determined by the conservation of rest-mass, energy, and angular momentum of the accreting gas, the Polish donut model is less constrained. It requires the relation between the fluid angular velocity and the fluid angular momentum per unit energy, $\Omega = \Omega (l)$, as well as the fluid angular momentum per unit energy at the inner edge of the disk, $l_{\rm in}$.

The simplest Polish donut model is the configuration with $l = {\rm constant}$, which turns out to be marginally stable with respect to axisymmetric perturbations~\citep{font}. Accretion disks in the Kerr spacetime only exist for $l_{\rm ms} < l < l_{\rm mb}$, where $l_{\rm ms}$ and $l_{\rm mb}$ are, respectively, the angular momentum per unit energy of the marginally stable equatorial circular orbit (or ISCO) and of the marginally bound equatorial circular orbit~\citep{polish,bpt,zilong}. Indeed, for $l < l_{\rm ms}$, the fluid angular momentum is too low and there is no stable accretion disk (the fluid directly plunges onto the black hole), while, for $l > l_{\rm mb}$, the fluid angular momentum is too high and there is no accretion (there is just a fluid rotating around the black hole).

In the presence of an accretion disk, when $l_{\rm ms} < l < l_{\rm mb}$, the inner edge of the disk, $R_{\rm in}$, is between the marginally bound orbit and the ISCO, i.e. $R_{\rm mb} < R_{\rm in} < R_{\rm ISCO}$~\citep{polish,bpt,zilong}; that is, for a given black hole spin, the inner edge of a Polish donut disk is always smaller than the inner edge of a Novikov-Thorne disk. Since the exact location of the inner edge of the disk is crucial in reflection spin measurements, we can already suspect that systematic uncertainties related to the use of an incorrect disk model to fit the data of a source can be important.

\section{Simulations and fits}\label{s-sim}

In order to estimate the impact of the disk structure on spin measurements with the iron line method, we use the code presented in \citet{p1} to calculate the theoretical reflection spectrum of a Polish donut disk. We use the {\sc xillver} model~\citep{xill1,xill2} to calculate the non-relativistic reflection spectrum at the emission point in the rest-frame of the gas in the disk, and we use our ray-tracing code to convolve this non-relativistic spectrum into a relativistic spectrum of a thick disk.

At this point, we need to choose the properties of our source and of our instruments to simulate the observation of a black hole with a thick accretion disk. As X-ray mission, we consider \textsl{NICER}~\citep{nicer}, which has a good energy resolution near the iron line and is thus suitable for X-ray reflection spectroscopy measurements. We assume that the source has a very simple spectrum, just a power-law component and a reflection component. In XSPEC language~\citep{xspec}:

\vspace{0.2cm}

{\sc tbabs$\times$(powerlw + reflection)} ,

\vspace{0.2cm}

\noindent where {\sc tbabs} takes the Galactic absorption into account~\citep{wilms}, {\sc powerlw} is the power-law component generated by inverse Compton scattering of thermal photons from the disk off free electrons in some hot gas close to the black hole~\citep{d2}, and {\sc reflection} is our theoretical reflection model for the Polish donut disk. Considering a photon flux of $1.4 \cdot 10^{-10}$~erg~cm$^{-2}$~s$^{-1}$ in the energy range 1-10~keV, which is an appropriate value for a bright AGN, and an exposure time of about 420~ks, we get 50~million counts in the energy range 0.2-12~keV. The simulated data are then fitted with the XSPEC model

\vspace{0.2cm}

{\sc tbabs$\times$relxill} ,

\vspace{0.2cm}

\noindent where {\sc relxill} is the most popular relativistic reflection model and assumes that the accretion disk is thin~\citep{rel1,rel2}. When the reflection fraction parameter, $R_{\rm f}$, is free in {\sc relxill}, the model describes both the power-law component from the corona and the reflection spectrum from the disk; for this reason we do not need {\sc powerlw} here. Note that we are assuming an overoptimistic observation with 50~million counts (\textsl{NICER} cannot observe a source for a long time, but we can image to combine many observations together), but this will be used later to point out that sometimes even high quality data can be fitted well with an incorrect model.

We run two groups of simulations and all the details are reported in the Appendix. The two groups differ for the intensity profile of the reflection spectrum. For the first group (simulations~A-D), the intensity profile of the thick disk is described by a simple power-law with emissivity index $q = 9$ and we fit the simulated data with the normal {\sc relxill} with the intensity profile of the disk modeled with a simple power-law. For the second group (simulations~E-H), we assume a lamppost corona and the intensity profile is calculated with the ray-tracing code described in~\citet{relxill_nk} and the simulated data are fit with {\sc relxilllp}~\citep{rel1}. The sets of simulations A-D with the power-law emissivity profile differ by the viewing angle of the observer ($i = 20^\circ$ or $35^\circ$) and the outer edge of the Polish donut disk (either $R_{\rm out} = 20$~$M$ or 40~$M$). The sets of simulations E-H with the lamppost coronal geometry differ by the viewing angle of the observer ($i = 20^\circ$ or $35^\circ$) and the height of the lamppost corona (either $h_{\rm lamppost} = 3$~$M$ or 6~$M$). For every set of simulations, we have seven input spin parameters: $a_* = 0.4$, 0.6, 0.8, 0.85, 0.9, 0.95 and 0.98.

In these simulations, the input values of the inclination angle of the disk is relatively low, $i = 20^\circ$ and $35^\circ$, because the inner part of the disk gets obscured for higher inclination angles in the Polish donut model. However, this is true even in the case of real observations of supermassive black holes, even if for somewhat different reasons: relativistic reflection features are observed only for sources with low disk inclination angles, because dusty tori (not the inner part of the accretion disk as here) present in active galactic nuclei otherwise prevent direct observation of the reflection spectrum from the inner part of the accretion disk. For simulations A-D with power-law emissivity profile, we consider two different disk configurations. As discussed in~\cite{p1}, the location of the inner edge of the disk can be conveniently parametrized by the outer edge of the disk, $R_{\rm out}$, and here we choose $R_{\rm out} = 20$~$M$ and $R_{\rm out} = 40$~$M$ (for simulations E-F, we only consider the case $R_{\rm out} = 40$~$M$). As $R_{\rm out}$ increases, the inner edge of the disk $R_{\rm in}$ approaches $R_{\rm mb}$, but we cannot increase too much the value of $R_{\rm out}$ because, otherwise, the inner part of the disk is only visible for observers with very low viewing angles. When $R_{\rm out} = 20$~$M$, the maximum height of the Polish donut disk is $h_{\rm max} \approx 10$~$M$. In the case $R_{\rm out} = 40$~$M$, we have $h_{\rm max} \approx 25$~$M$.

\section{Results}\label{s-res}

The best-fit tables and the data to best-fit model ratio plots of our simulations are reported in the Appendix. For simulations A-D (power-law emissivity profile) and G-H (lamppost corona with viewing angle $i = 35^\circ$), we always find very good fits; that is, even if we fit the reflection spectrum of a thick disk with a model that assumes a thin disk, the fit is good, in the sense that the reduced $\chi^2$ is close to 1 and the ratio plot between the simulated data and the best-fit model does not show unresolved features. Note that we are analyzing 50~million count observations, which are very optimistic for supermassive black holes. This means that the good quality of the fit itself cannot tell us that the model employed to fit the data is incorrect. However, some model parameters are clearly estimated incorrectly, in the sense that the statistical uncertainties from the fits is much smaller than the systematic uncertainties due to the incorrect model used to fit the data. For simulations E-F (lamppost corona with viewing angle $i = 20^\circ$), the quality of the fit is bad, and it is worse for higher values of the spin parameter. The ratio plots between the simulated data and the best-fit model clearly show some unresolved features. Even in this case, the fits can recover the values of most model parameters, but still there are a few parameters that are strongly affected by the disk structure.

\begin{figure*}[t]
\begin{center}
\includegraphics[type=pdf,ext=.pdf,read=.pdf,width=8cm]{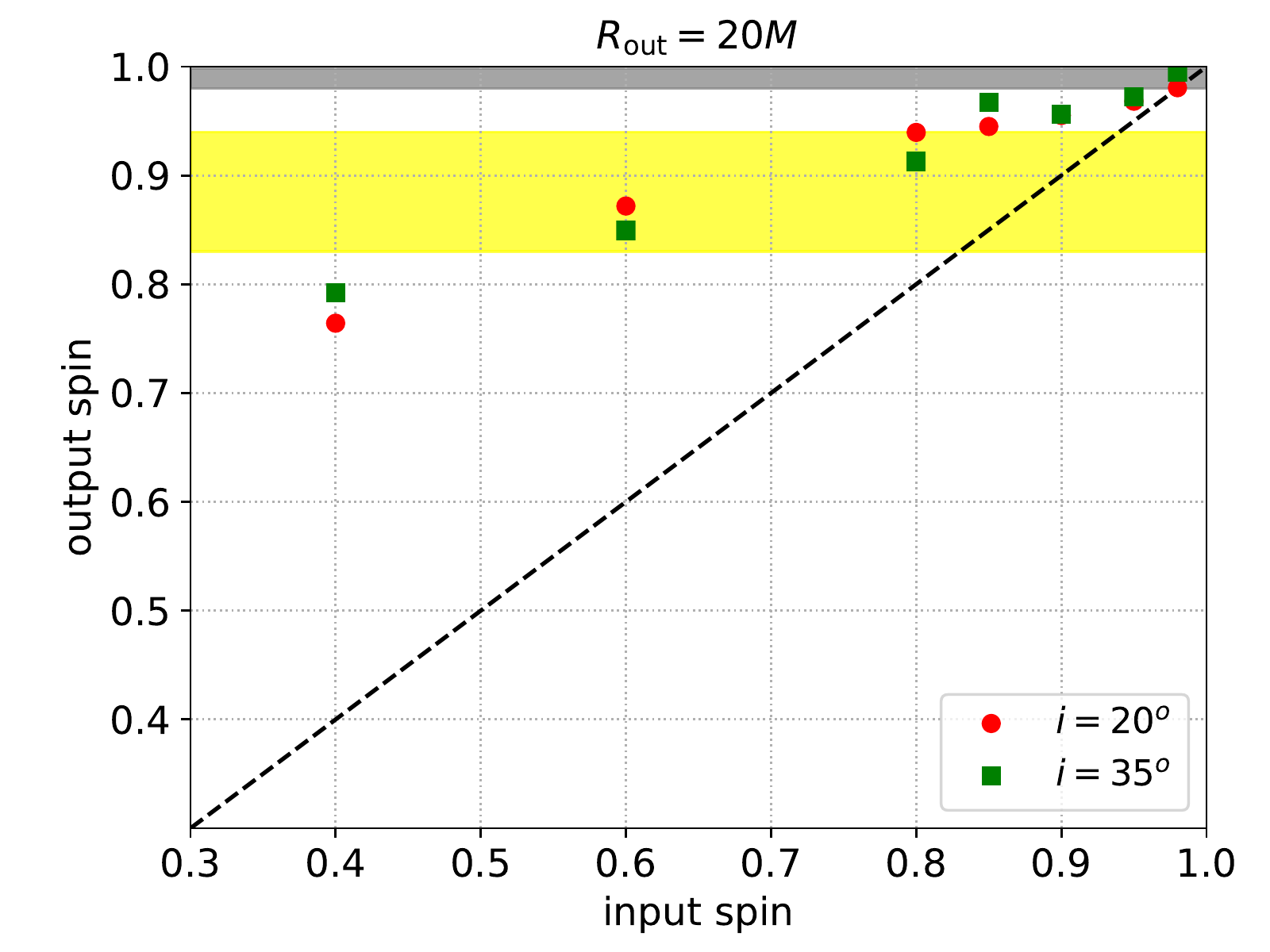}
\hspace{0.5cm}
\includegraphics[type=pdf,ext=.pdf,read=.pdf,width=8cm]{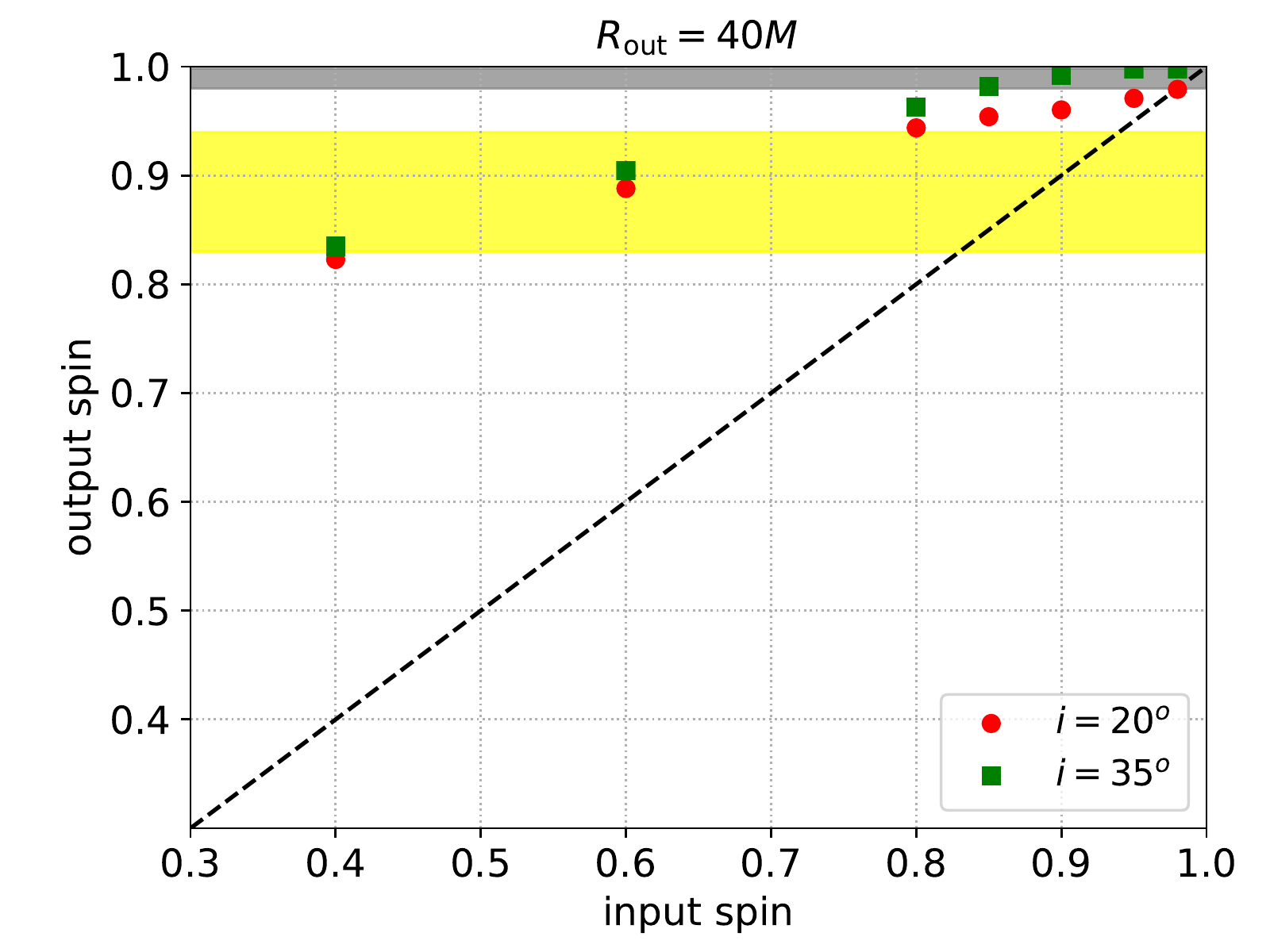}
\end{center}
\vspace{-0.4cm}
\caption{Simulations~A-D (power-law intensity profile) -- Input spin parameter of the simulations vs best-fit spin parameter obtained from {\sc relxill}. Statistical uncertainties in the spin parameter are too small at this scale. The yellow horizontal region marks the spin measurement of Ton~S180 ($a_* = 0.92_{-0.09}^{+0.02}$ from \citet{s4}) and the gray horizontal region marks the spin measurement of 1H0707--495 [$a_* > 0.98$ from \citet{s4,zog}]. \label{f-spin}}
\vspace{0.5cm}
\begin{center}
\includegraphics[type=pdf,ext=.pdf,read=.pdf,width=8cm]{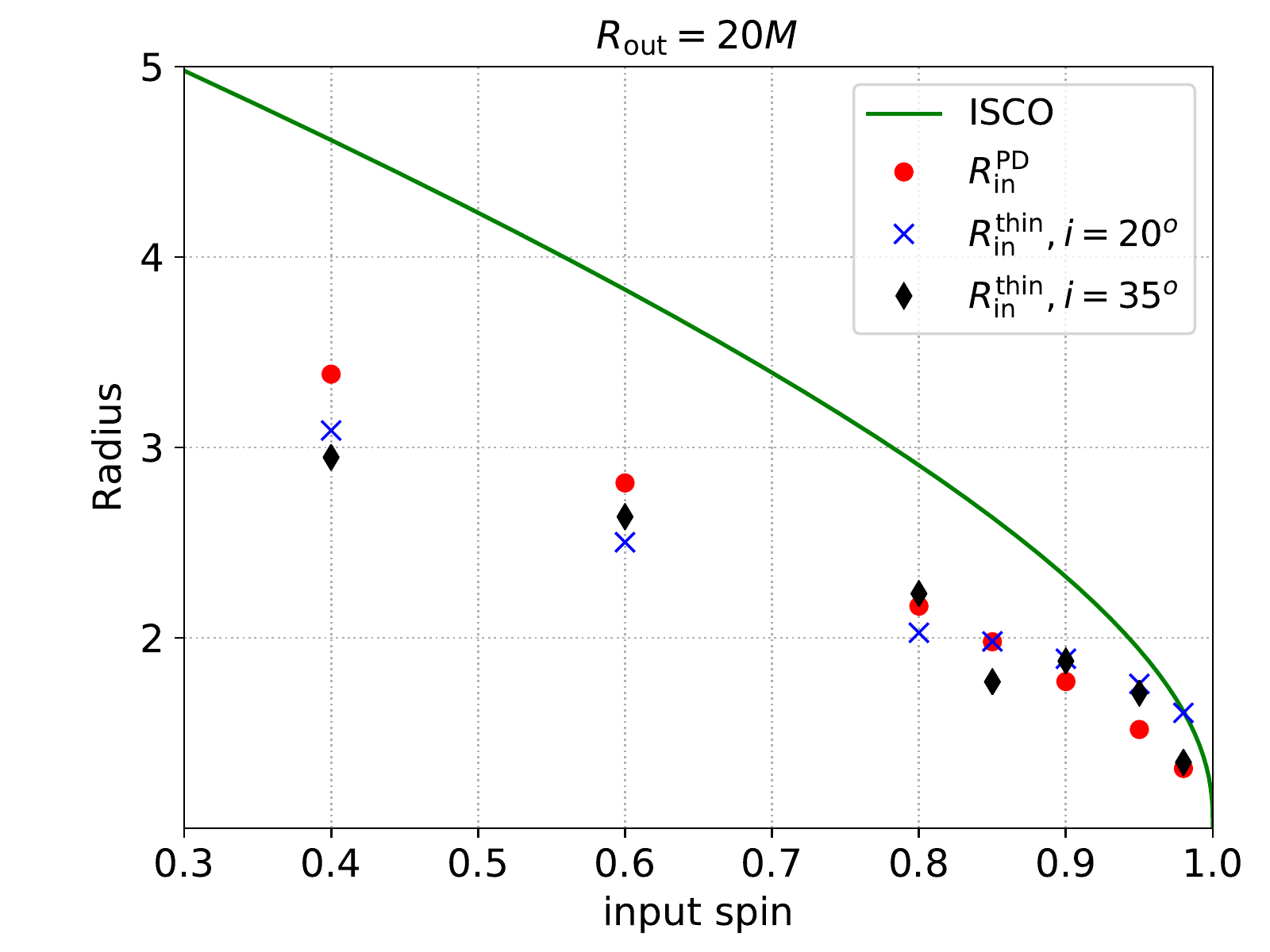}
\hspace{0.5cm}
\includegraphics[type=pdf,ext=.pdf,read=.pdf,width=8cm]{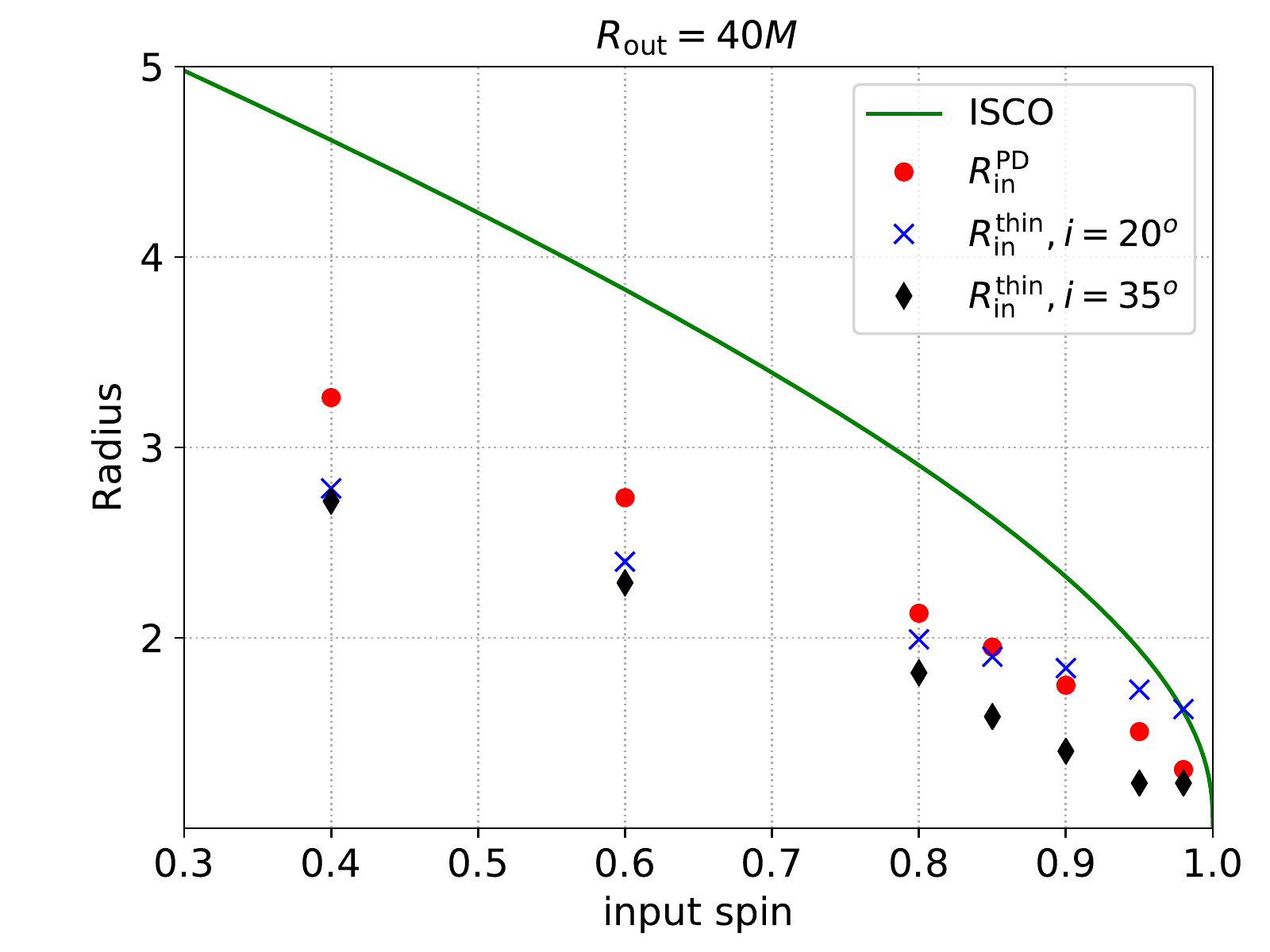}
\end{center}
\vspace{-0.4cm}
\caption{Simulations~A-D (power-law intensity profile) -- Input spin parameter of the simulations vs radial coordinate of the corresponding ISCO radius (green solid line), of the inner edge of the Polish donut disk (red dots), and of the inner edge obtained by {\sc relxill} (blue crosses for the simulations with $i = 20^\circ$ and black diamonds for the simulations with $i = 35^\circ$). Radius is in units of $M$. \label{f-rin}}
\end{figure*}

\begin{figure*}[t]
\begin{center}
\includegraphics[type=pdf,ext=.pdf,read=.pdf,width=8cm]{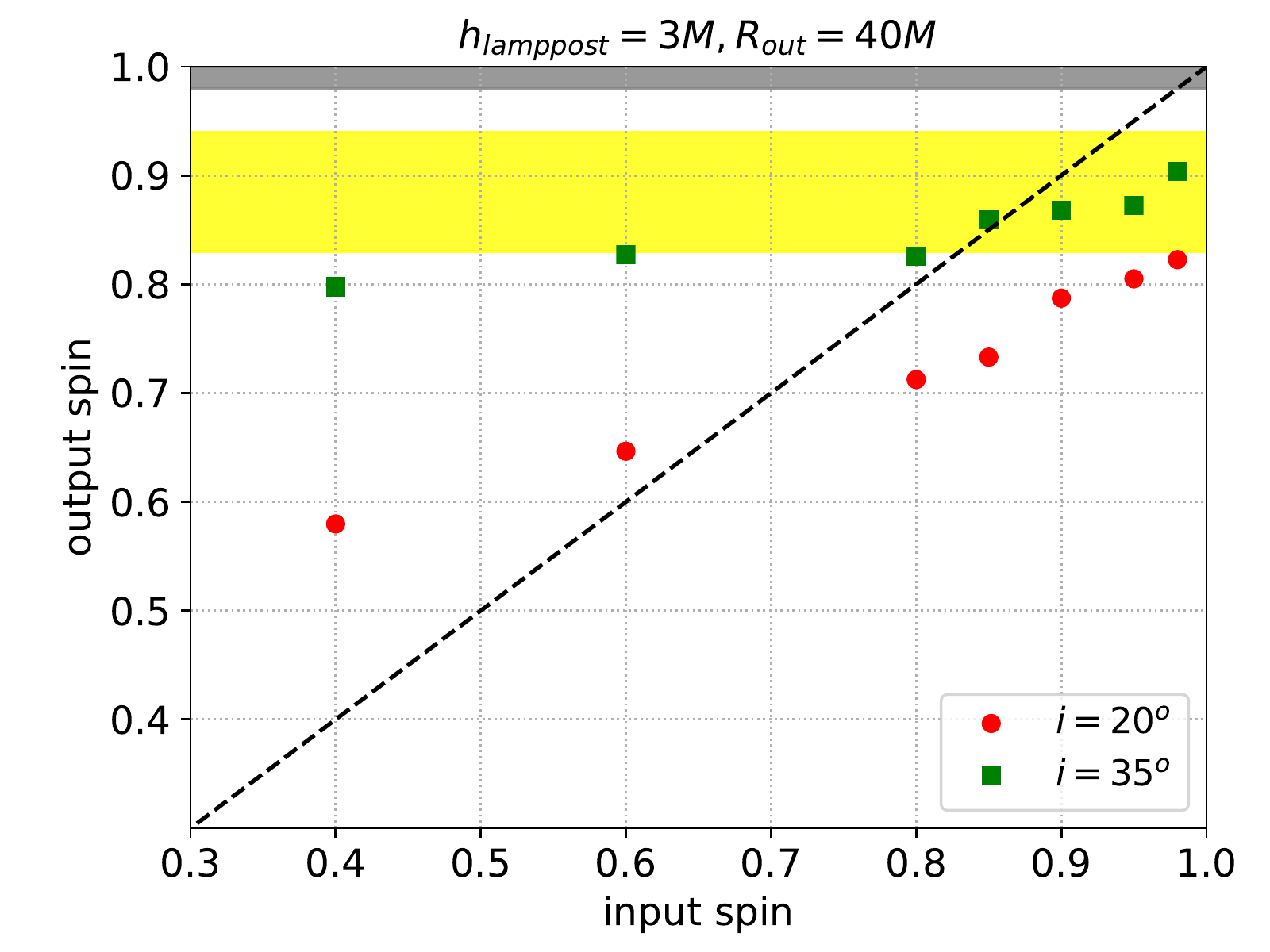}
\hspace{0.5cm}
\includegraphics[type=pdf,ext=.pdf,read=.pdf,width=8cm]{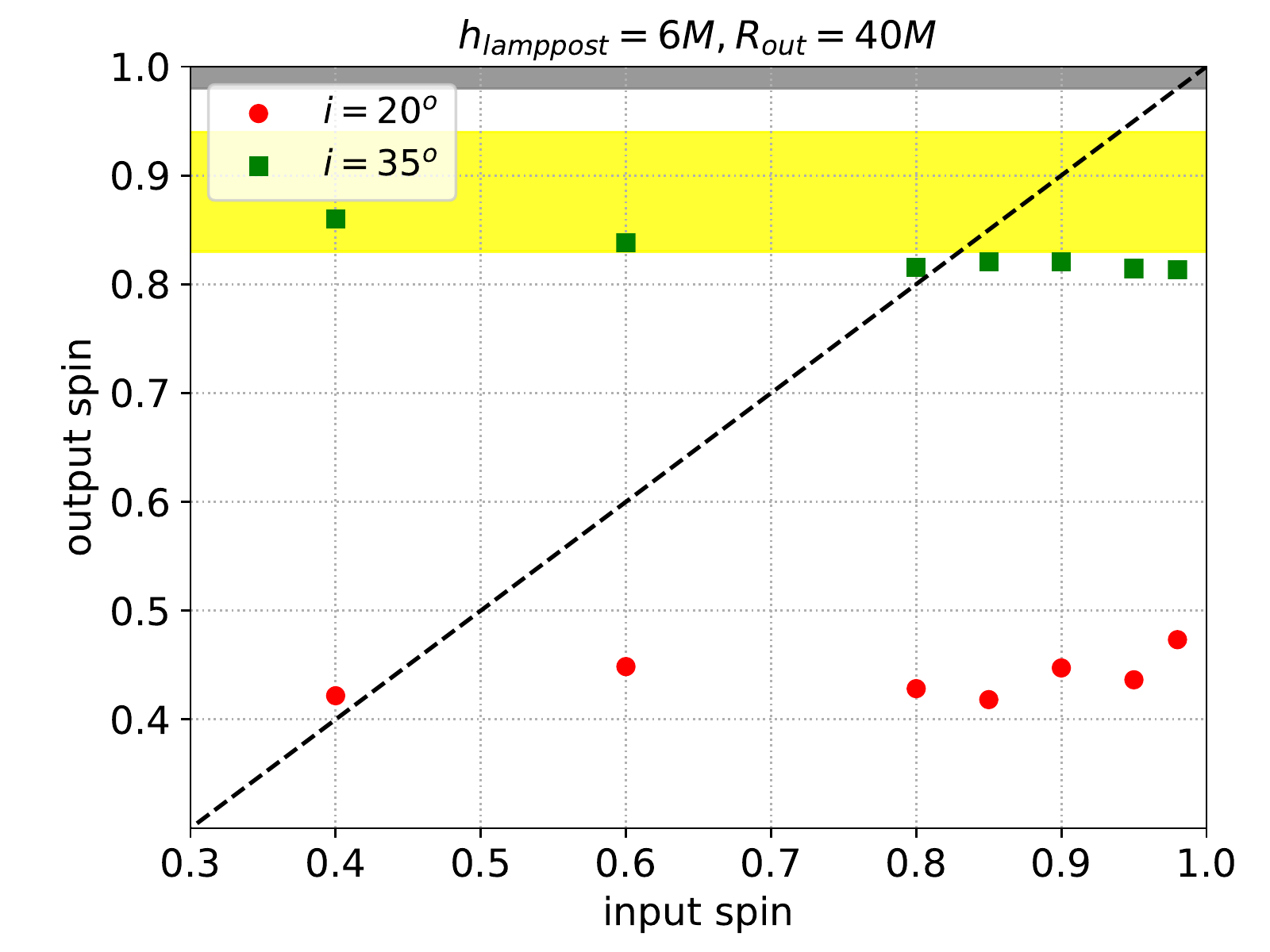}
\end{center}
\vspace{-0.4cm}
\caption{Simulations~E-H (lamppost corona intensity profile) -- Input spin parameter of the simulations vs best-fit spin parameter obtained from {\sc relxilllp}. Statistical uncertainties in the spin parameter are too small at this scale. The yellow horizontal region marks the spin measurement of Ton~S180 ($a_* = 0.92_{-0.09}^{+0.02}$ from \citet{s4}) and the gray horizontal region marks the spin measurement of 1H0707--495 [$a_* > 0.98$ from \citet{s4,zog}]. \label{f-spin-lp}}
\vspace{0.5cm}
\begin{center}
\includegraphics[type=pdf,ext=.pdf,read=.pdf,width=8cm]{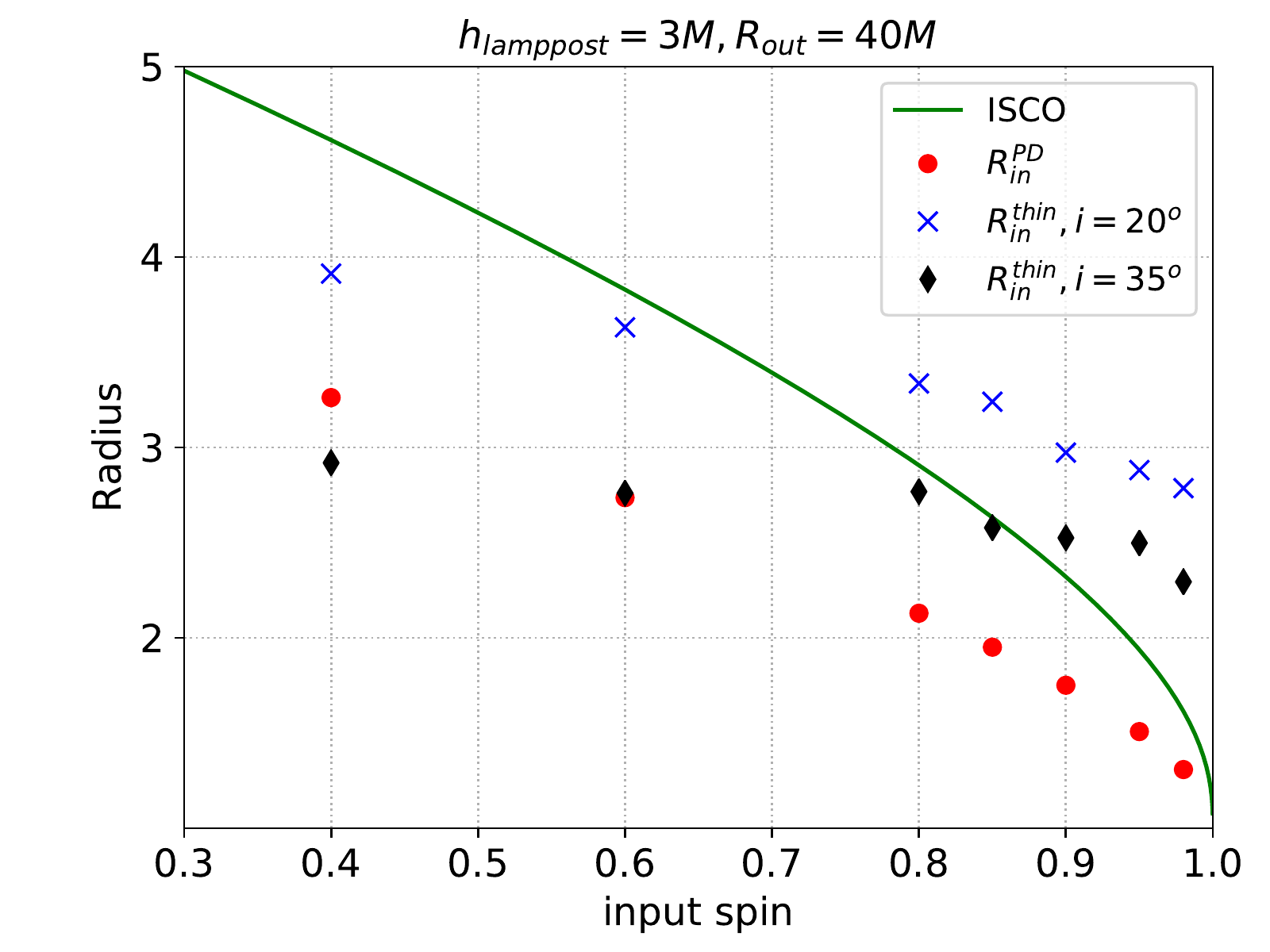}
\hspace{0.5cm}
\includegraphics[type=pdf,ext=.pdf,read=.pdf,width=8cm]{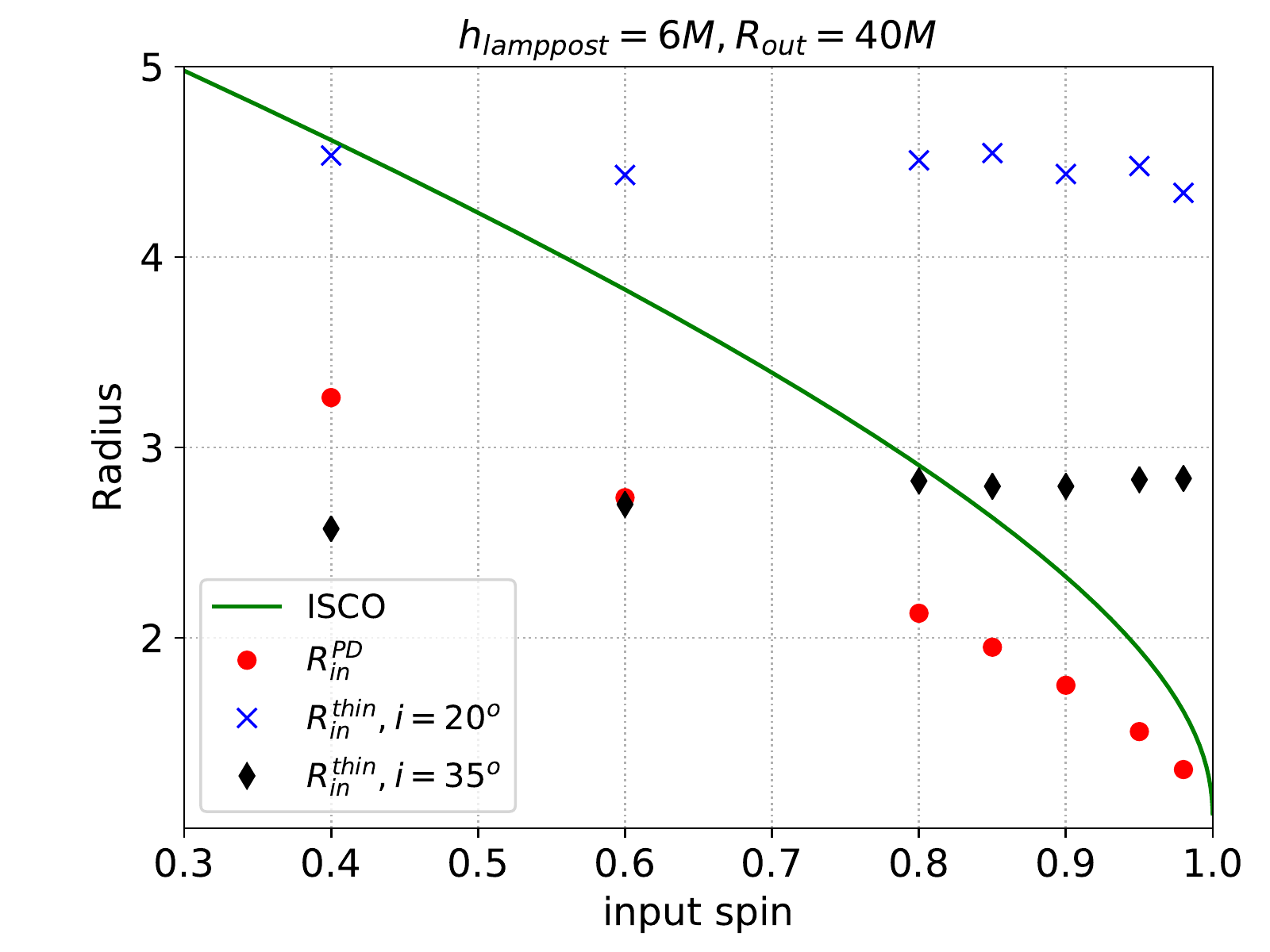}
\end{center}
\vspace{-0.4cm}
\caption{Simulations~E-H (lamppost corona intensity profile) -- Input spin parameter of the simulations vs radial coordinate of the corresponding ISCO radius (green solid line), of the inner edge of the Polish donut disk (red dots), and of the inner edge obtained by {\sc relxilllp} (blue crosses for the simulations with $i = 20^\circ$ and black diamonds for the simulations with $i = 35^\circ$). Radius is in units of $M$. \label{f-rin-lp}}
\end{figure*}

Our main focus is on the systematic effects on the spin measurement and the main result of our work is summarized in Fig.~\ref{f-spin} and Fig.~\ref{f-spin-lp}, which show the relation between input and output spin parameters, respectively for simulations A-D and simulations E-H. The fit errors on the output spin parameters are not shown in the figures because the uncertainties are too small at this scale (see Appendix for more details). We immediately see that the black hole spin parameter is always overestimated in the model with a power-law emissivity profile, while for the model with a lamppost corona we can overestimate the spin of slow-rotating black holes and underestimate the spin of fast-rotating black holes. In the lamppost models, the height of the corona and the viewing angle of the disk play an important role in the modeling bias.

As an illustration of the possible effect of a high mass accretion rate on current spin measurements, Figs.~\ref{f-spin} and \ref{f-spin-lp} also show two spin measurements of supermassive black holes that are thought to accrete near or above their Eddington limit. In the case of 1H0707--495, the spin measurement is $a_* > 0.98$ in \citet{s4,zog} (horizontal gray region in the figures). For Ton~S180, the spin measurement is $a_* = 0.92_{-0.09}^{+0.02}$ in \citet{s4} (horizontal yellow region in the figures). For example, the actual spin parameter value of 1H0707--495 may be as low as $a_* \approx 0.85$ in the case of a Polish donut disk with $R_{\rm out} = 40$~$M$, $q=9$, and $i = 35^\circ$. Ton~S180 may have $a_* \approx 0.4$ in the case of a Polish donut disk with $R_{\rm out} = 40$~$M$, $q=9$, and $i = 35^\circ$, as well as for a Polish donut disk observed at $i = 35^\circ$ and with $h_{\rm lamppost} = 6$~$M$.

%%%%%%%%%%%%%%%%%%%%lamppost spins and radius plots %%%%%%%

%%%%%%%%%%%%%%%%%%%%%%%%%%%%%%%%%%%%%%%%%%%%%

For simulations A-D, we chose quite a high value of the emissivity index in order to limit the effects related to the size of the disk. Regardless, high emissivity indices are often preferred in the fits of reflection dominated spectra. Besides, we have also run a few simulations with $q = 6$, finding the same qualitative results, so our conclusions do not change for emissivity indices somewhat lower than 9. For even lower emissivity indices, like $q = 3$, the effects of the size of the Polish donut disk become important and the fits are not good. We have also run a few simulations with the response files of \textsl{XMM-Newton} to see possible effects related to the choice of the X-ray mission, but we obtain similar results, confirming that our conclusions are robust.

While we can debate whether a Polish donut disk with a power-law intensity profile can describe a real accretion disk of a black hole accreting around the Eddington limit, the interpretation of Fig.~\ref{f-spin} is quite easy. Fig.~\ref{f-rin} shows the radial coordinates of the inner edges of the Polish donut disks employed in the simulations together with the ISCO radii corresponding to the measured black hole spins. Most of the radiation is emitted from the very inner part of the disk, so the fit measures something like the inner edge of the disk, while the actual disk structure is not very important. This also explains the high quality of the fits, which cannot probe well the disk geometry far from the inner edge. Such a consideration is not very sensitive to the exact viewing angle.

Fig.~\ref{f-rin-lp} is the counterpart of Fig.~\ref{f-rin} for the simulations with lamppost corona. In this case, the location of the inner edge of the disk does not play any clear role in the spin estimates. Even the actual spin parameter seems to play a minor rule: for $h_{\rm lamppost} = 3$~$M$, there is quite a weak dependence of the input spin parameter on the estimate of $a_*$, while the input spin parameter seems to be completely irrelevant for $h_{\rm lamppost} = 6$~$M$. The two key quantities are the height of the lamppost corona (which determines the exact intensity profile for a given accretion disk) and the viewing angle $i$ (which determines the effect of the Doppler boosting on the spectrum and the level of obscuration of the disk).

Modeling bias in black hole spin measurements due to the disk structure has been investigated in \citet{RF2008} and \citet{taylor}. In both works, the authors consider thin accretion disks, so with an Eddington-scaled accretion luminosity below 0.3, but some of their results are qualitatively close to ours. \citet{RF2008} simulate geometrically thin accretion disks in a pseudo-Newtonian potential. While they find that the density and vertical column density of the accretion flow drop significantly around the ISCO radius, the emission inside the ISCO affects black hole spin measurements, leading to an overestimate of $a_*$. The systematic error is larger for slow-rotating black holes and decreases as the spin parameter increases. Their Fig.~5 is qualitatively similar to our Fig.~\ref{f-spin}, but their systematic bias is smaller. For example, a black hole with $a_* = 0.4$ may be interpreted as a black hole with $a_* \approx 0.5$ for the model in \citet{RF2008} while here we find $a_* \approx 0.8$ for a Polish donut disk with power-law emissivity index. \citet{taylor} calculate synthetic relativistic reflection spectra of geometrically thin disks with finite thickness illuminated by lamppost coronae. Assuming a black hole spin $a_* = 0.9$, a viewing angle $i = 15^\circ$, and a lamppost height $h = 3$~$M$, they show that the black hole spin and the corona height are underestimated, and the inferred black hole spin and corona height decrease as the thickness of the disk increases. Here we find the same qualitative trend for our lamppost models. Lastly, \citet{2019ApJ...884L..21T} simulate super-Eddington accretion disks and find that the corresponding iron line profiles are systematically more blueshifted and symmetric in shape than the iron line calculated for geometrically thin disks. They do not consider the impact of the disk structure on the estimate of black hole spin, so it is not possible to easily compare their results with ours.

\section{Concluding remarks}\label{s-con}

All the current relativistic reflection models assume a geometrically thin accretion disk, but several sources accrete at a rate that makes their disks geometrically thick. This issue is particularly relevant for supermassive black holes, because X-ray reflection spectroscopy is currently the only technique that can measure the spin of these objects. Spin measurements obtained in this way are inevitably affected by systematic uncertainties, but surprisingly the problem has not garnered enough attention, probably because of the lack of alternative relativistic models. In this work, we have shown that employing a thin disk reflection model to fit the data of a source accreting with a thick disk can lead to significantly biased black hole spin measurements. The Polish donut model is surely a toy model, but we found that even high quality data of the reflection spectrum of a thick disk may be fit well with a thin disk model. Note that reliable spin measurements of supermassive black holes are important, for instance, for studying the growth and merger history of galaxies: very fast-rotating black holes are only possible if the black hole accretes for a long time from its accretion disk (low galaxy merger rate), while frequent black hole mergers lead to black holes with a moderate value of the spin parameter (high galaxy merger rate)~\citep{berti}.

%%%%%%%%%%%%%%%%%%%%%%%%%%%%%%%

\vspace{0.5cm}

{\bf Acknowledgments --}
This work was supported by the Innovation Program of the Shanghai Municipal Education Commission, Grant No.~2019-01-07-00-07-E00035, the National Natural Science Foundation of China (NSFC), Grant No.~11973019, and Fudan University, Grant No.~IDH1512060. S.N. thanks the Alexander von Humboldt Foundation.

%%%%%%%%%%%%%%%%%%%%%%%%%%%%%%%

\newpage

\appendix

\section{Data analysis}\label{app-1}

\begin{figure*}[b]
\begin{center}
\includegraphics[type=pdf,ext=.pdf,read=.pdf,width=10cm]{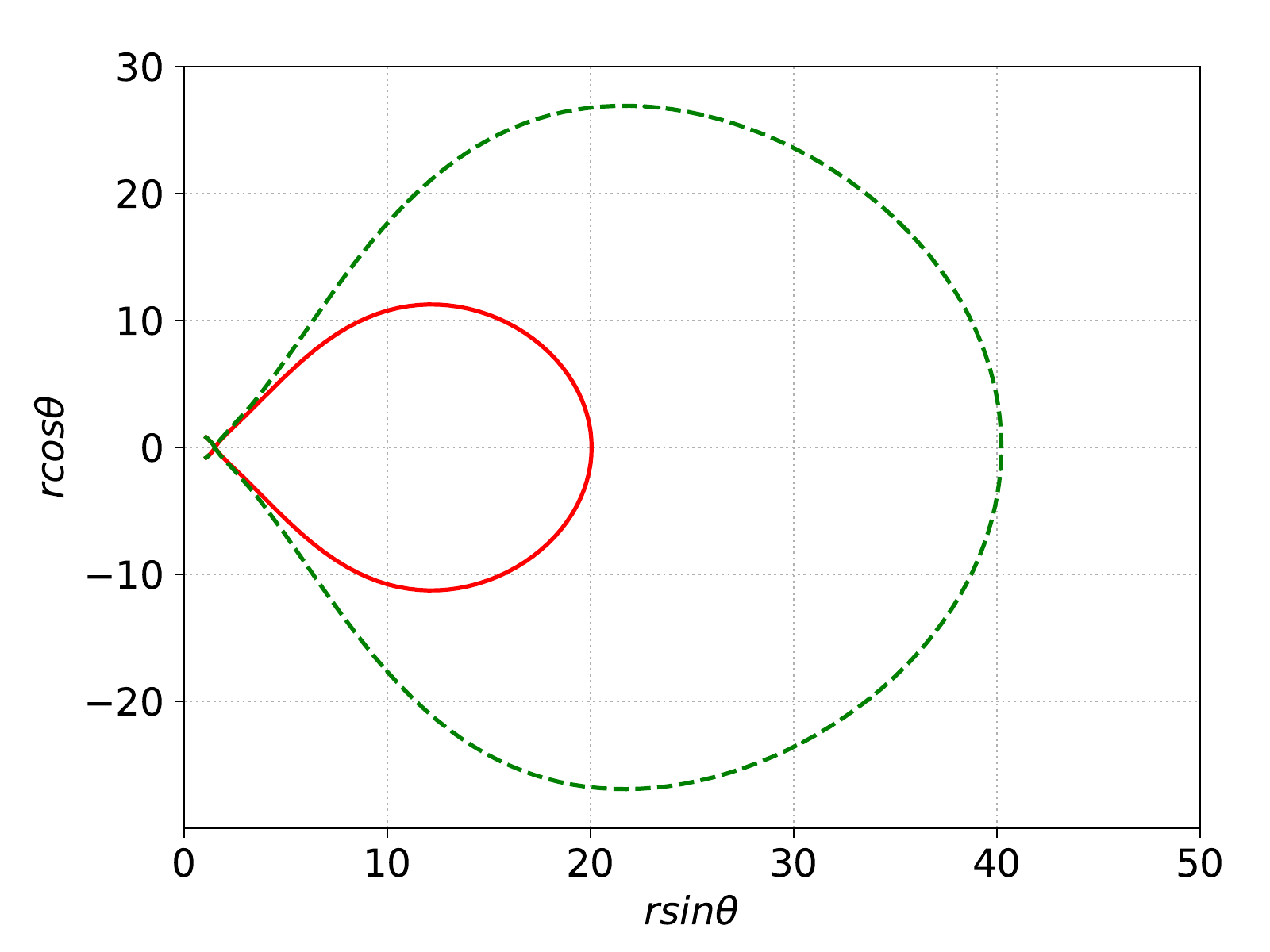}
\end{center}
\vspace{-0.4cm}
\caption{Shape of the Polish donut disks for $R_{\rm out} = 20$~$M$ (red solid curve) and 40~$M$ (green dashed curve) when the spin parameter is $a_* = 0.95$. \label{f-donut}}
\end{figure*}

The available relativistic reflection models employ the Novikov-Thorne accretion disk model, which completely ignores the pressure in the accretion flow and thus describes geometrically thin and optically thick disks~\citep{nt1,nt2}. Moreover, the disk is approximated as infinitesimally thin, so the particles of the gas follow nearly geodesic circular orbits in the equatorial plane. However, in reality, as the mass accretion rate increases, pressure becomes more and more important, and the thickness of the disk increases. The Polish donut model takes the pressure in the accretion flow into account and can describe geometrically and optically thick accretion disks, which form when the mass accretion rate is near or above the Eddington limit of the source~\citep{polish,zilong}. Fig.~\ref{f-donut} shows the shape of the Polish donut disks employed in the simulations for $a_* = 0.95$, but different spins have quite similar disks.

For the study presented in the paper, we simulate the following sets of observations (here $q$ is the emissivity index for the power-law intensity profile, $h_{\rm lamppost}$ is the height of the lamppost corona, $i$ is the viewing angle of the observer, and $R_{\rm out}$ is the outer edge of the Polish donut disk):
\vspace{-0.15cm}
\begin{enumerate}
\item Simulations~A: $q = 9$, $i = 20^\circ$, and $R_{\rm out} = 20$~$M$.
\vspace{-0.15cm}
\item Simulations~B: $q = 9$, $i = 20^\circ$, and $R_{\rm out} = 40$~$M$.
\vspace{-0.15cm}
\item Simulations~C: $q = 9$, $i = 35^\circ$, and $R_{\rm out} = 20$~$M$.
\vspace{-0.15cm}
\item Simulations~D: $q = 9$, $i = 35^\circ$, and $R_{\rm out} = 40$~$M$.
\vspace{-0.15cm}
\item Simulations~E: $h_{\rm lamppost} = 3$~$M$, $i = 20^\circ$, and $R_{\rm out} = 40$~$M$.
\vspace{-0.15cm}
\item Simulations~F: $h_{\rm lamppost} = 6$~$M$, $i = 20^\circ$, and $R_{\rm out} = 40$~$M$.
\vspace{-0.15cm}
\item Simulations~G: $h_{\rm lamppost} = 3$~$M$, $i = 35^\circ$, and $R_{\rm out} = 40$~$M$.
\vspace{-0.15cm}
\item Simulations~H: $h_{\rm lamppost} = 6$~$M$, $i = 35^\circ$, and $R_{\rm out} = 40$~$M$.
\end{enumerate}
\vspace{-0.15cm}
For every simulation set, we consider seven input spin parameters: $a_* = 0.4$, 0.6, 0.8, 0.85, 0.9, 0.95 and 0.98. For all simulations, the ionization parameter is set to $\log\xi = 3.1$ ($\xi$ in units erg~cm~s$^{-1}$), the iron abundance is assumed to have the Solar value ($A_{\rm Fe} = 1$), the coronal spectrum is described by a power-law component with photon index $\Gamma = 2$ and cutoff energy $E_{\rm cut} = 300$~keV. These are all typical values for AGNs.

For simulations~A-D, the data are fit with the XSPEC model {\sc tbabs$\times$relxill}. For simulations~E-H, we use the XSPEC model {\sc tbabs$\times$relxilllp}. In the fits, the column density $N_{\rm H}$ in {\sc tbabs} and the cutoff energy $E_{\rm cut}$ in {\sc relxill} and {\sc relxilllp} are frozen to their input value. Indeed, $N_{\rm H}$ can be inferred from other observations, while $E_{\rm cut}$ cannot be constrained and does not play a significant role for data up to 12~keV. The reflection fraction, $R_{\rm f}$, is always left free in the fit, so {\sc relxill} and {\sc relxilllp} are used to describe both the power-law component from the corona and the reflection spectrum from the disk.

\begin{figure*}[t]
\begin{center}
\includegraphics[type=pdf,ext=.pdf,read=.pdf,width=8.5cm]{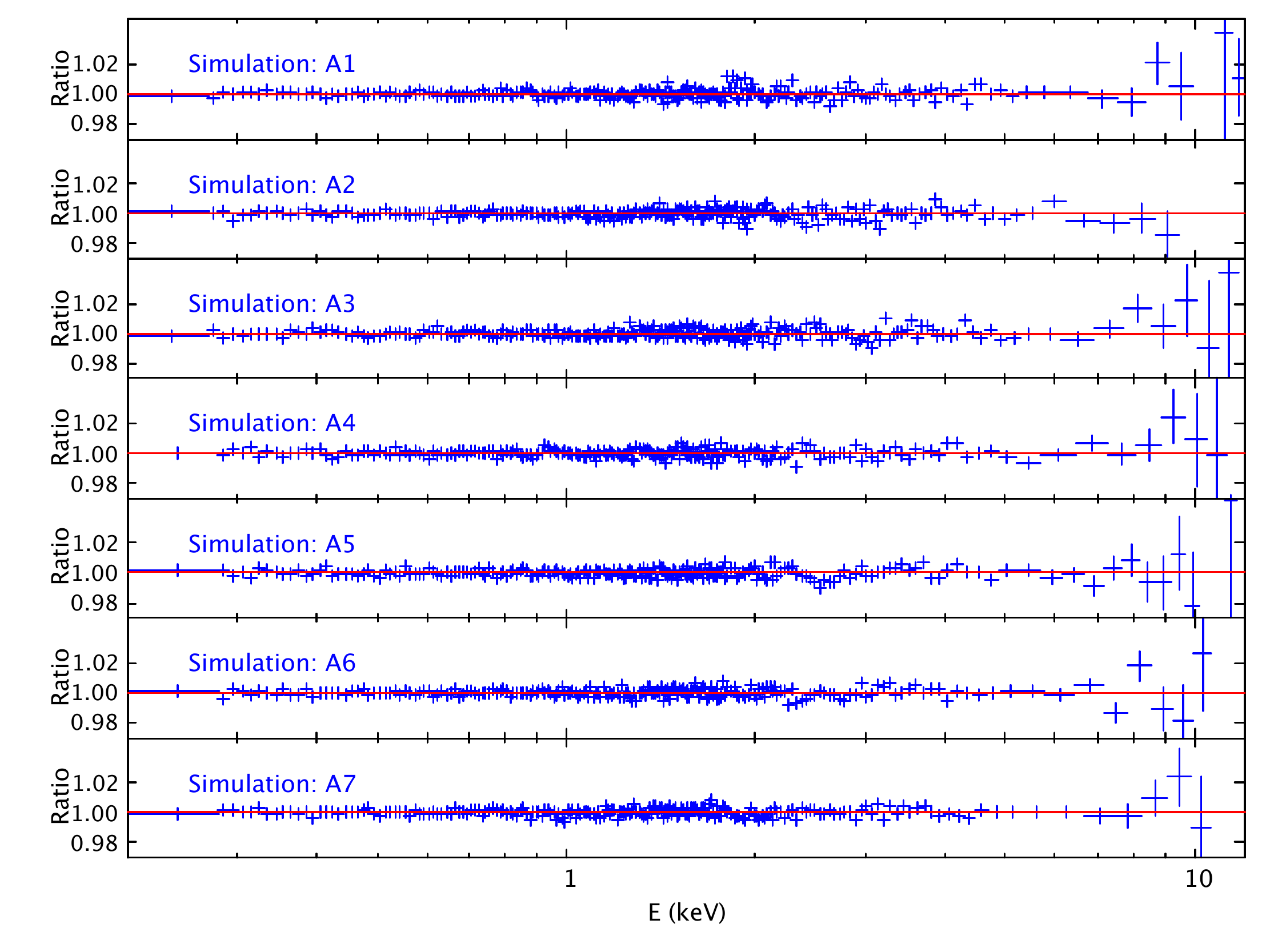}
\hspace{0.3cm}
\includegraphics[type=pdf,ext=.pdf,read=.pdf,width=8.5cm]{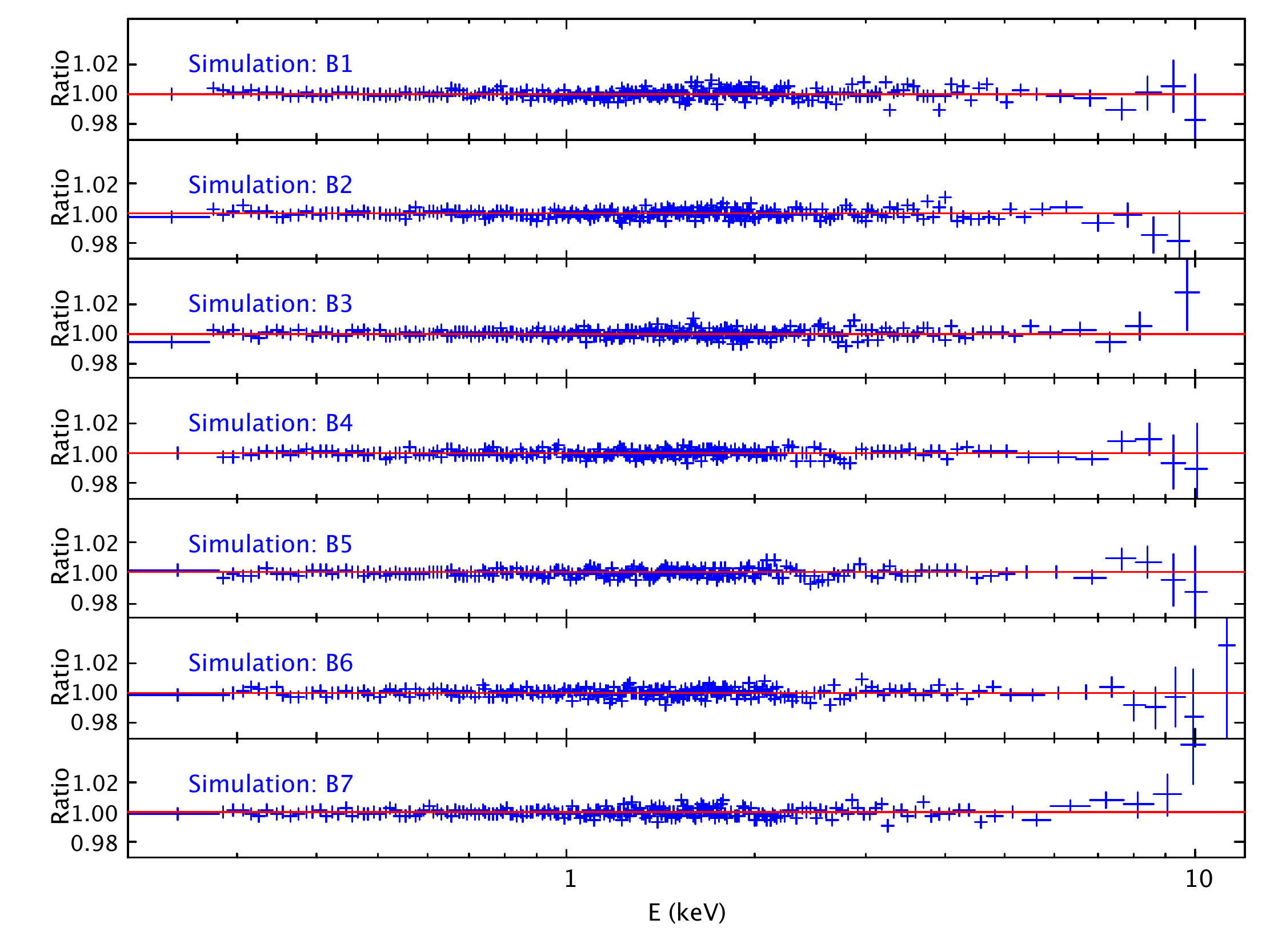}\\
\vspace{0.5cm}
\includegraphics[type=pdf,ext=.pdf,read=.pdf,width=8.5cm]{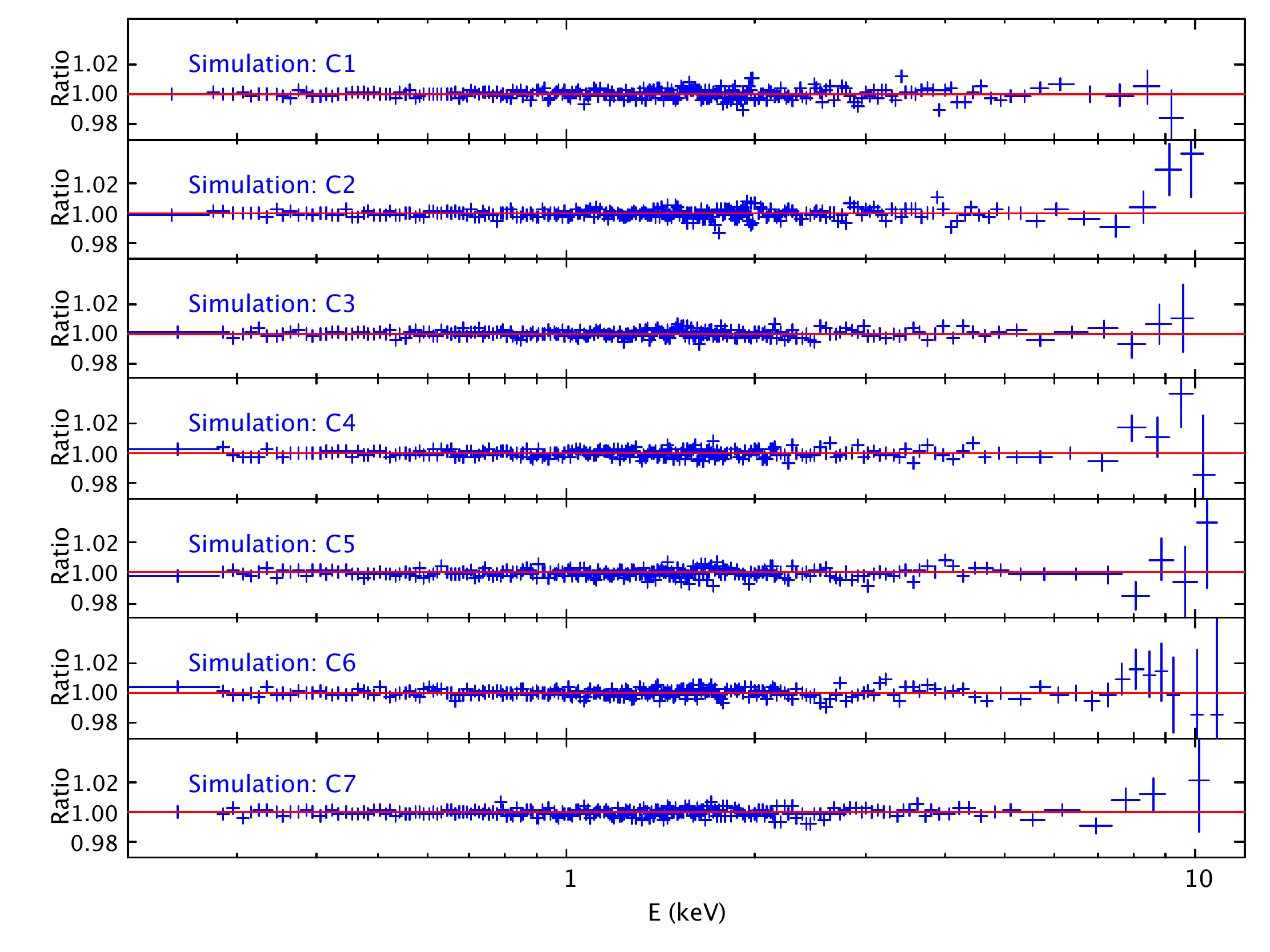}
\hspace{0.3cm}
\includegraphics[type=pdf,ext=.pdf,read=.pdf,width=8.5cm]{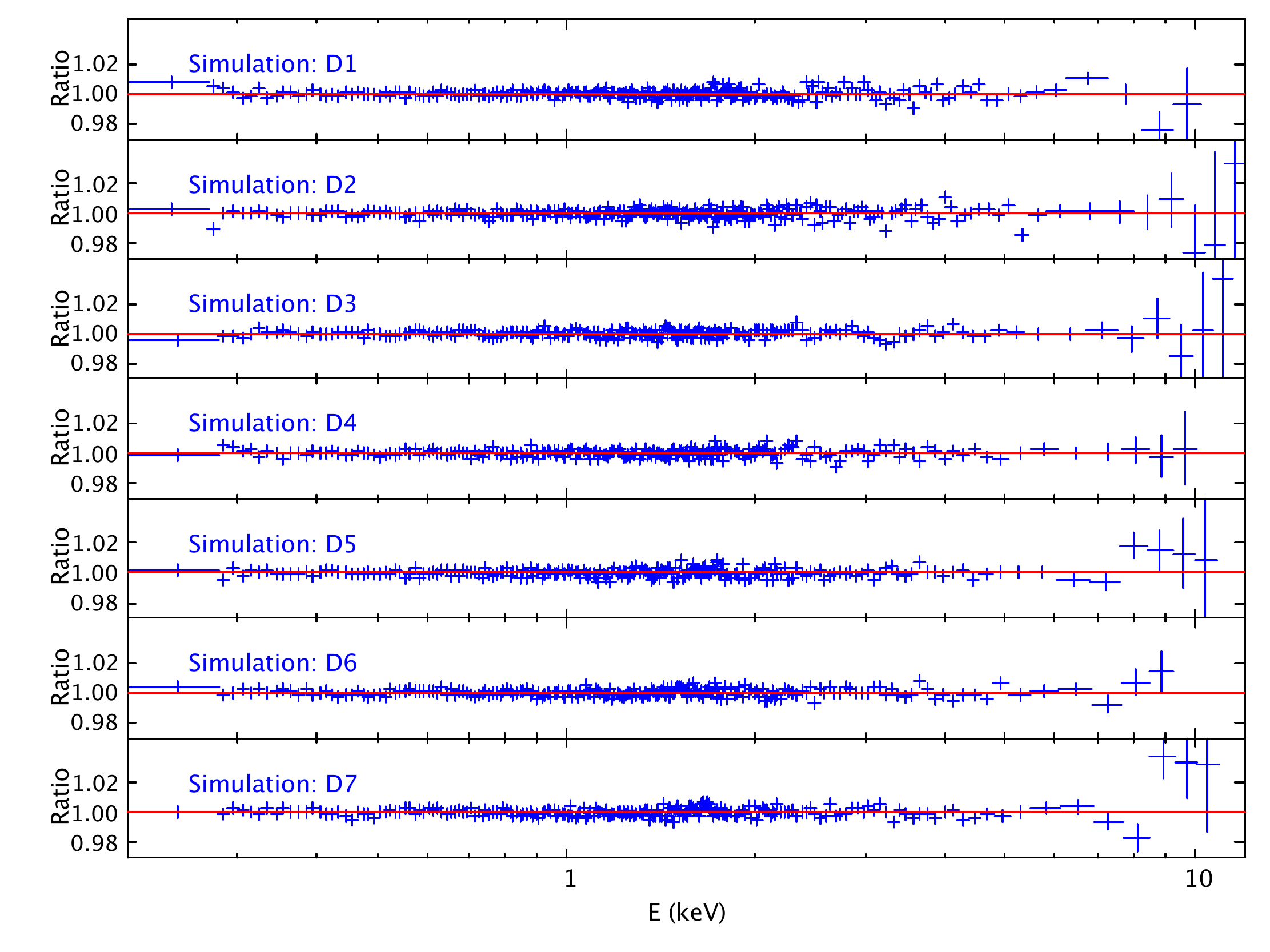}
\end{center}
\vspace{-0.4cm}
\caption{Data to best-fit model ratios for simulations~A1-A7 (top left panel), B1-B7 (top right panel), C1-C7 (bottom left panel), and D1-D7 (bottom right panel). \label{f-ratio}}
\end{figure*}

\begin{figure*}[t]
\begin{center}
\includegraphics[type=pdf,ext=.pdf,read=.pdf,width=8.5cm]{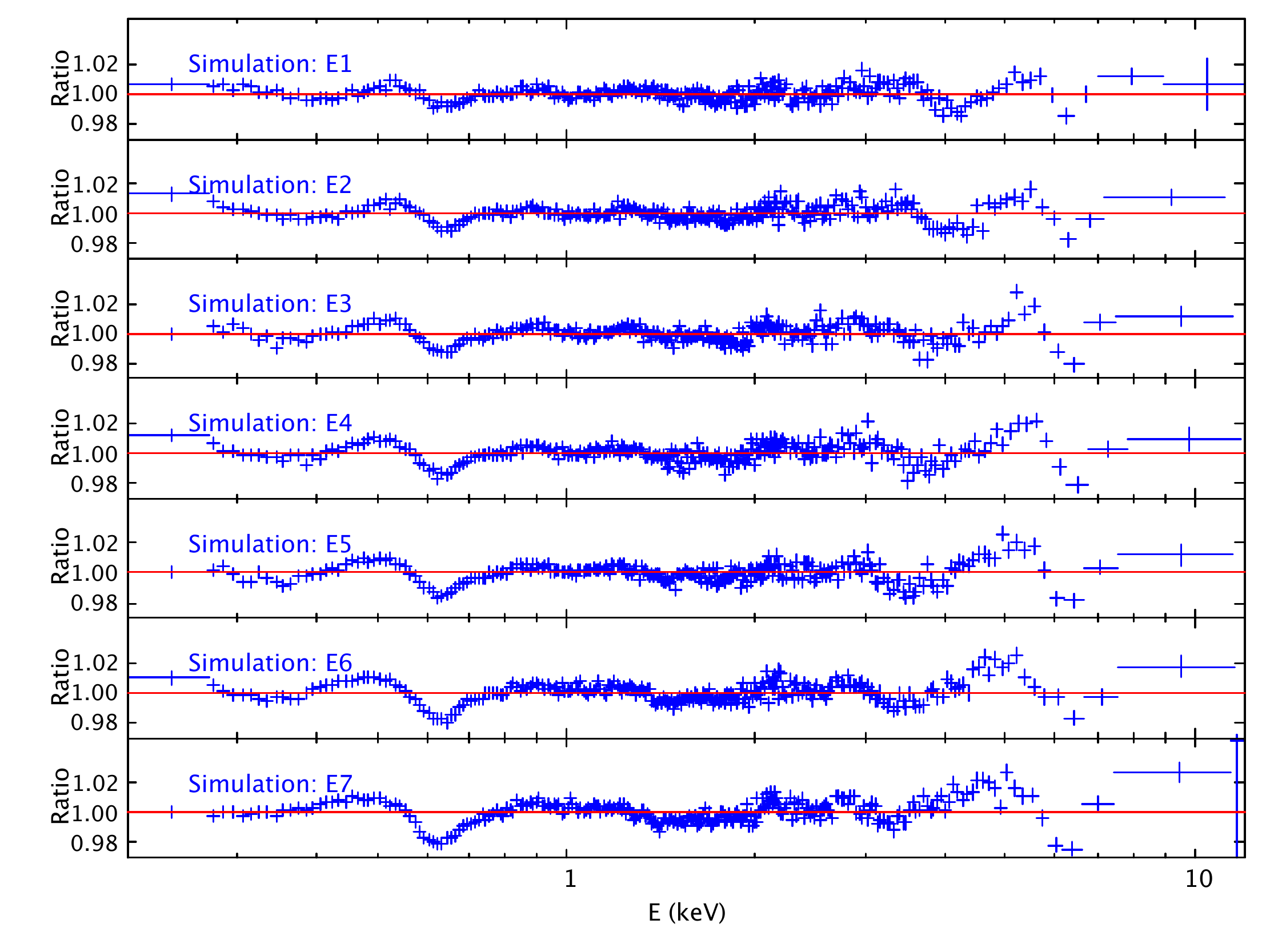}
\hspace{0.3cm}
\includegraphics[type=pdf,ext=.pdf,read=.pdf,width=8.5cm]{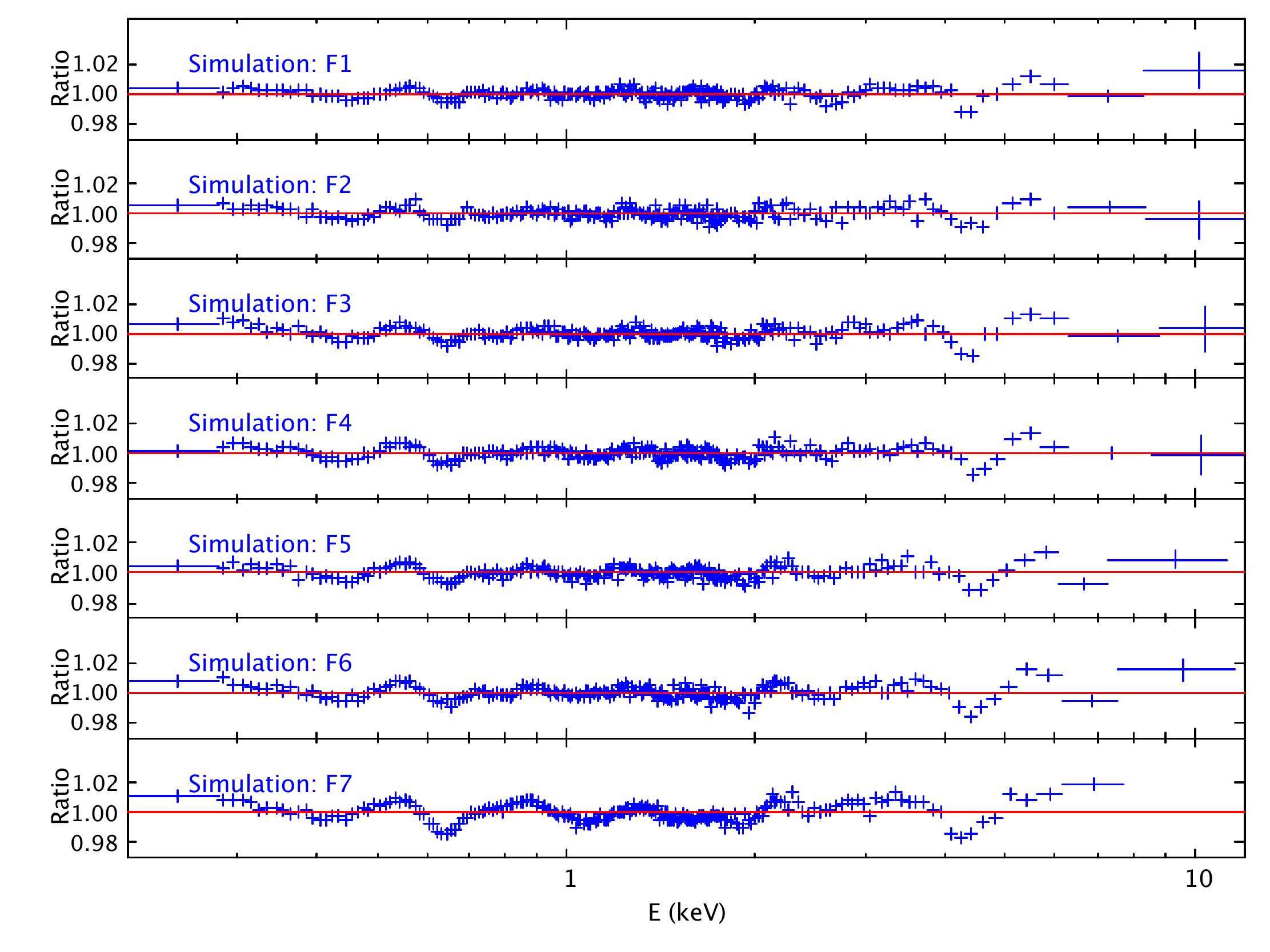}\\
\vspace{0.5cm}
\includegraphics[type=pdf,ext=.pdf,read=.pdf,width=8.5cm]{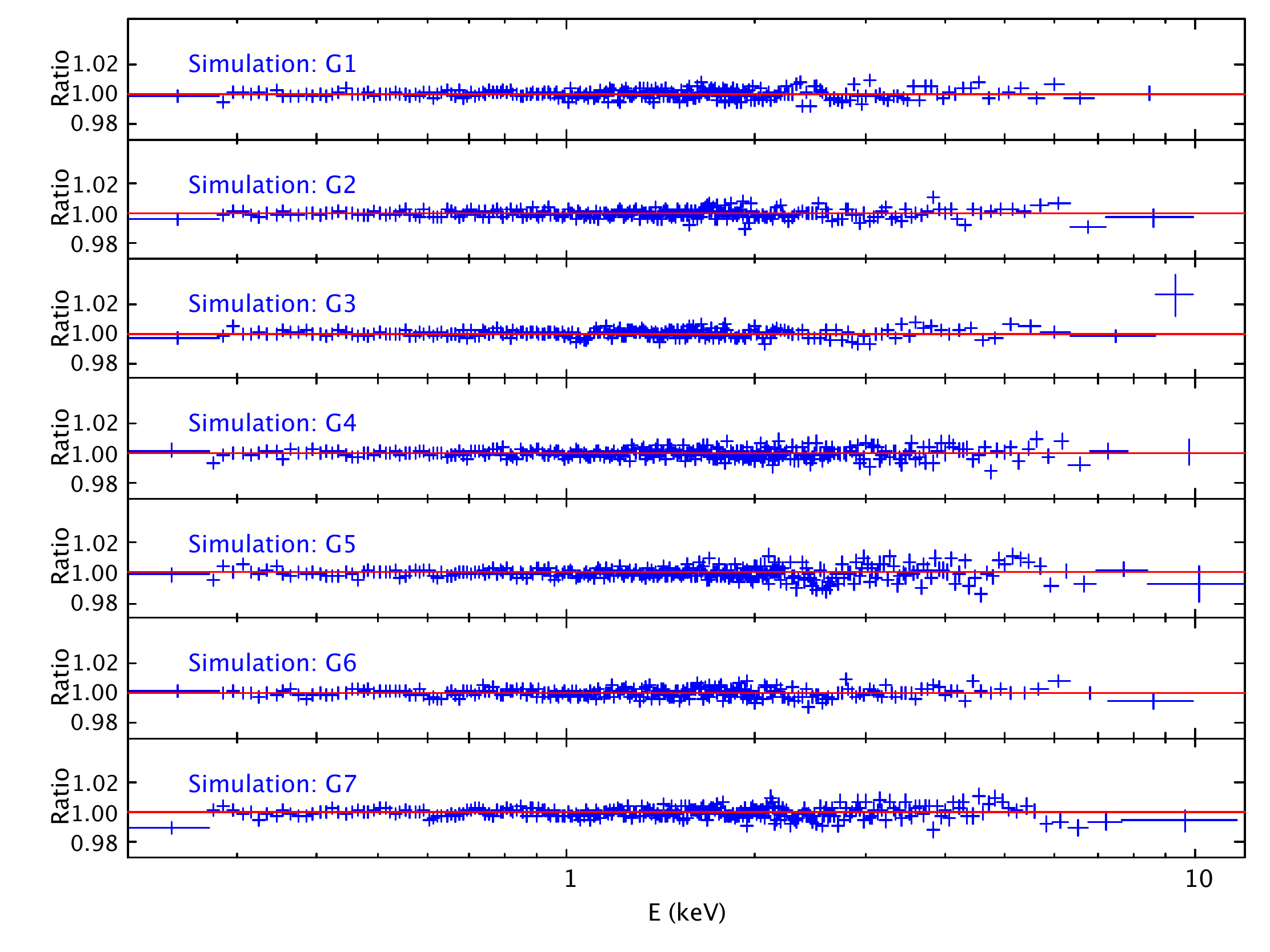}
\hspace{0.3cm}
\includegraphics[type=pdf,ext=.pdf,read=.pdf,width=8.5cm]{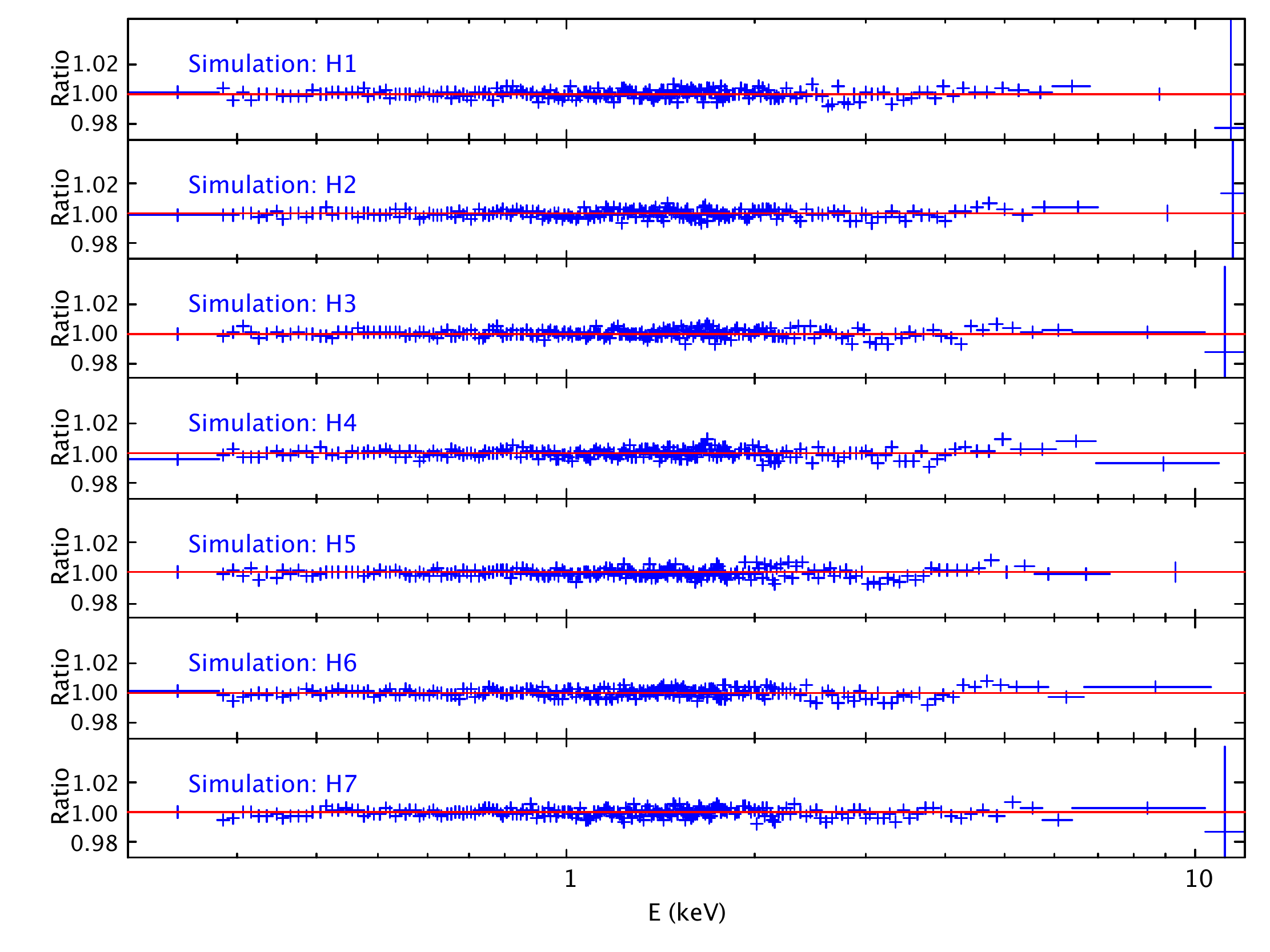}
\end{center}
\vspace{-0.4cm}
\caption{Data to best-fit model ratios for simulations~E1-E7 (top left panel), F1-F7 (top right panel), G1-G7 (bottom left panel), and H1-H7 (bottom right panel). \label{f-ratio-lp}}
\end{figure*}

The best-fit values of our fits are shown in Tab.~\ref{t-fit-a} (simulations~A1-A7), Tab.~\ref{t-fit-b} (simulations~B1-B7), Tab.~\ref{t-fit-c} (simulations~C1-C7), Tab.~\ref{t-fit-d} (simulations~D1-D7) Tab.~\ref{t-fit-e} (simulations~E1-E7), Tab.~\ref{t-fit-f} (simulations~F1-F7), Tab.~\ref{t-fit-g} (simulations~G1-G7), and Tab.~\ref{t-fit-h} (simulations~H1-H7). For simulations A-D, the ratio plots between simulated data and best-fit models are shown in Fig.~\ref{f-ratio}, and we see that there are no unresolved features, namely the theoretical model for thin disks can provide a good fit. For simulations E-H, the ratio plots are in Fig.~\ref{f-ratio-lp} and we clearly see that the fits are good for simulations G and H with input viewing angle $i = 35^\circ$ (their reduced $\chi^2$ is also close to 1) while there are unresolved features for simulations E and F with input viewing angle $i = 20^\circ$ (their reduced $\chi^2$ is also significantly larger than 1).

\begin{figure*}[t]
\begin{center}
\includegraphics[type=pdf,ext=.pdf,read=.pdf,width=8cm]{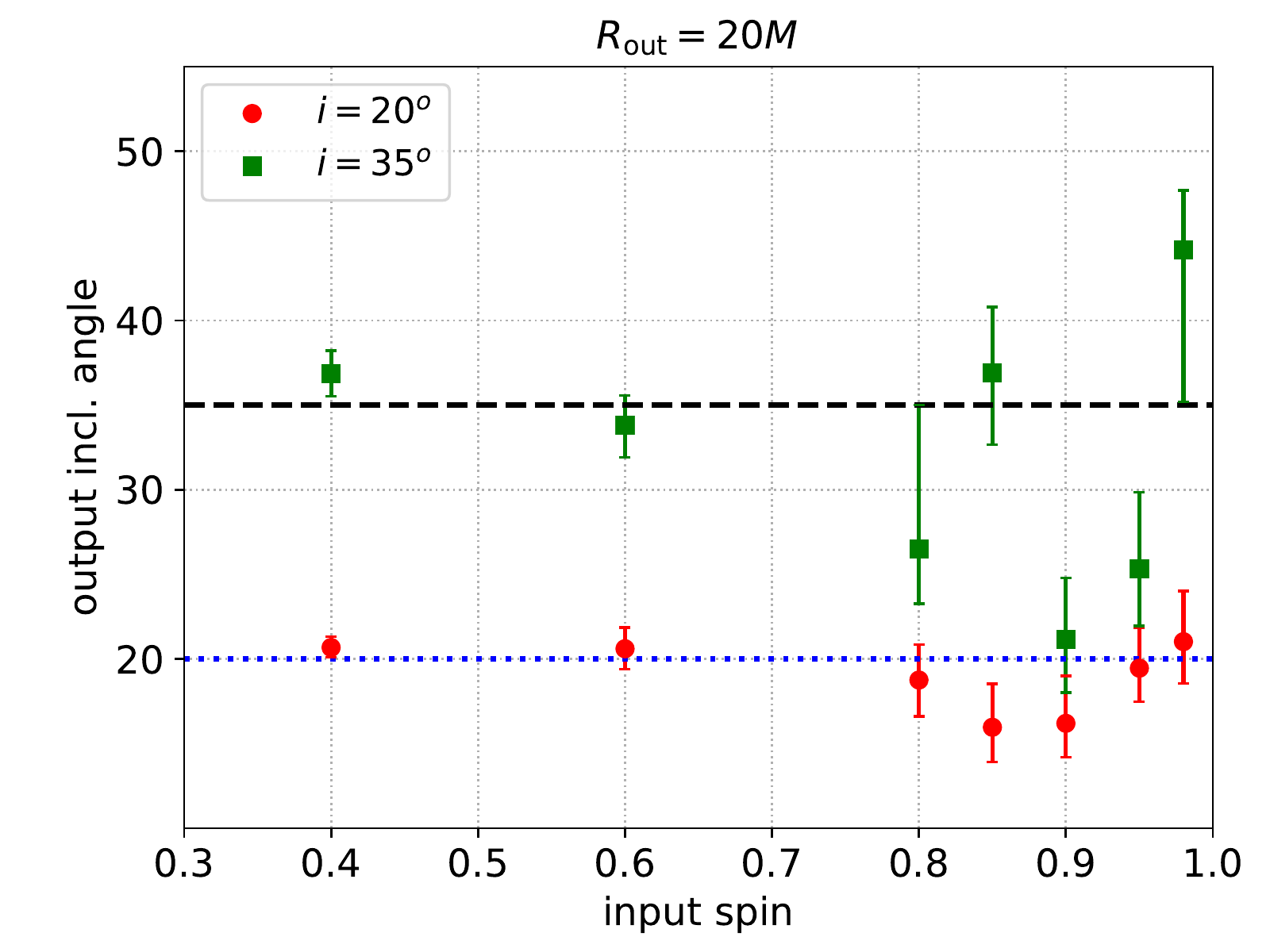}
\hspace{0.5cm}
\includegraphics[type=pdf,ext=.pdf,read=.pdf,width=8cm]{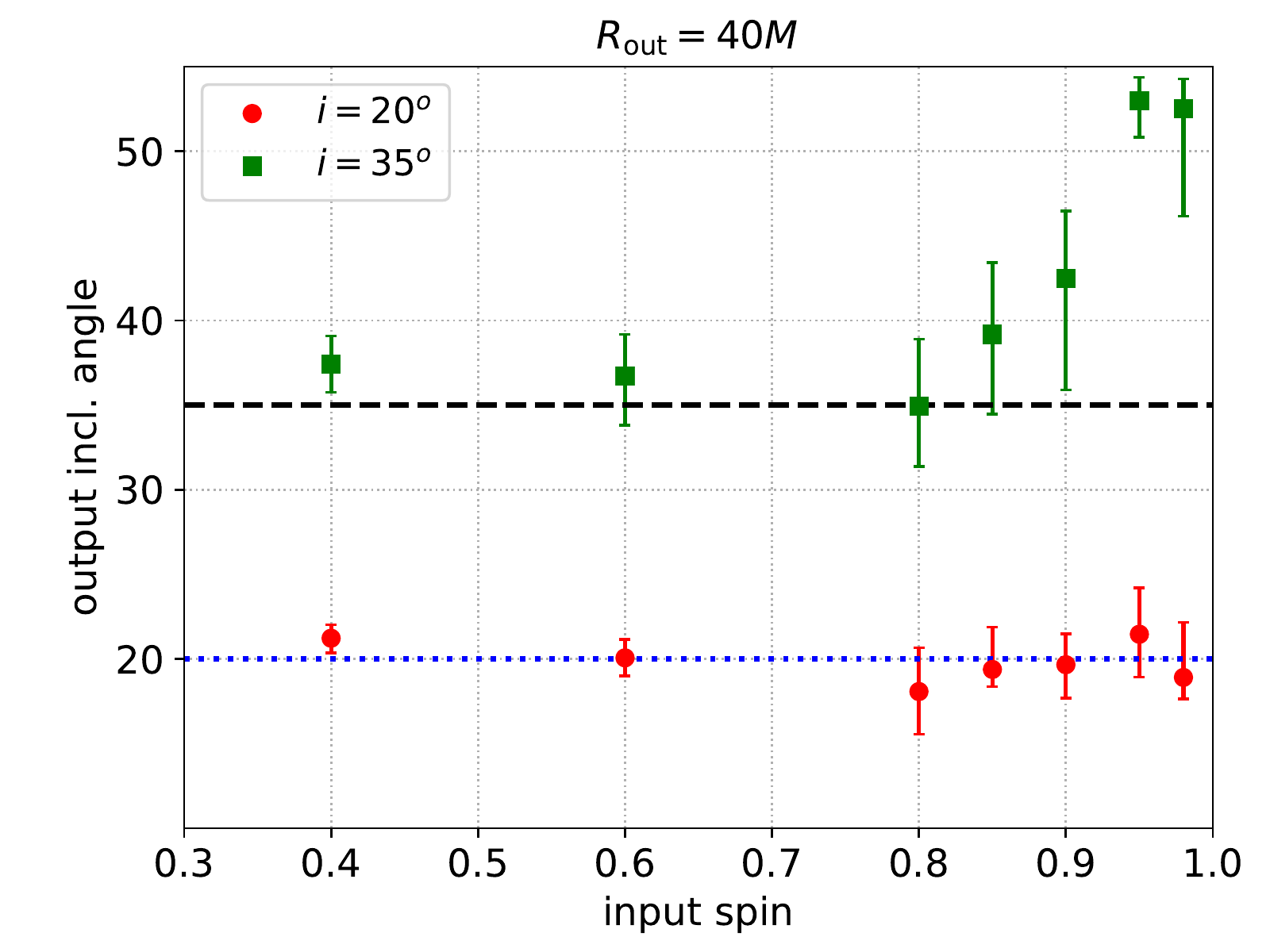}
\end{center}
\vspace{-0.4cm}
\caption{Simulations~A-D (power-law intensity profile) -- Input spin parameter of the simulations vs best-fit inclination angle obtained from {\sc relxill}. The error bars show the fit uncertainties. The horizontal black dashed and blue dotted lines mark the input inclination angles of the simulations, respectively $i = 35^\circ$ and $i = 20^\circ$. \label{f-incl}}
\vspace{0.5cm}
\begin{center}
\includegraphics[type=pdf,ext=.pdf,read=.pdf,width=8cm]{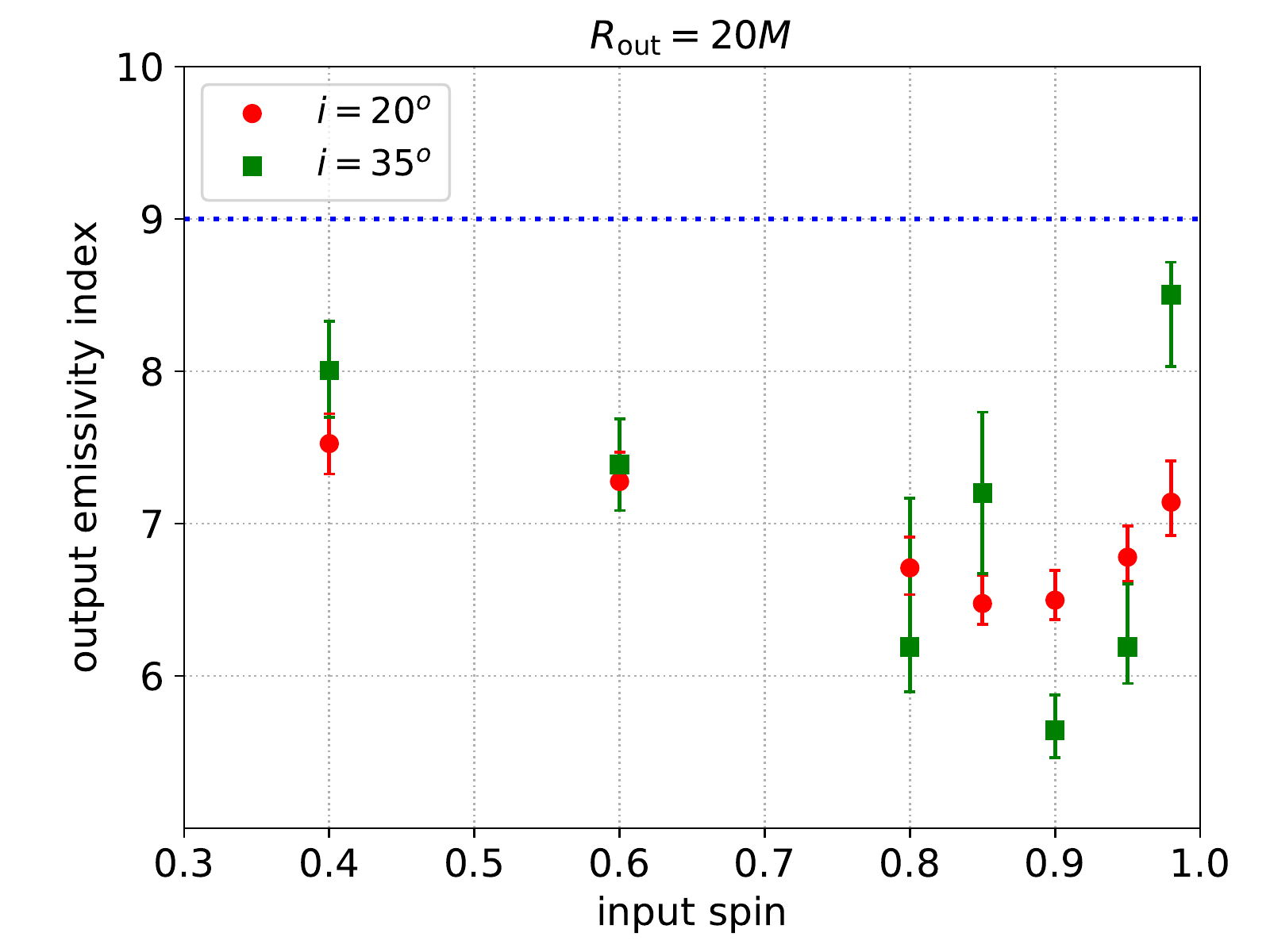}
\hspace{0.5cm}
\includegraphics[type=pdf,ext=.pdf,read=.pdf,width=8cm]{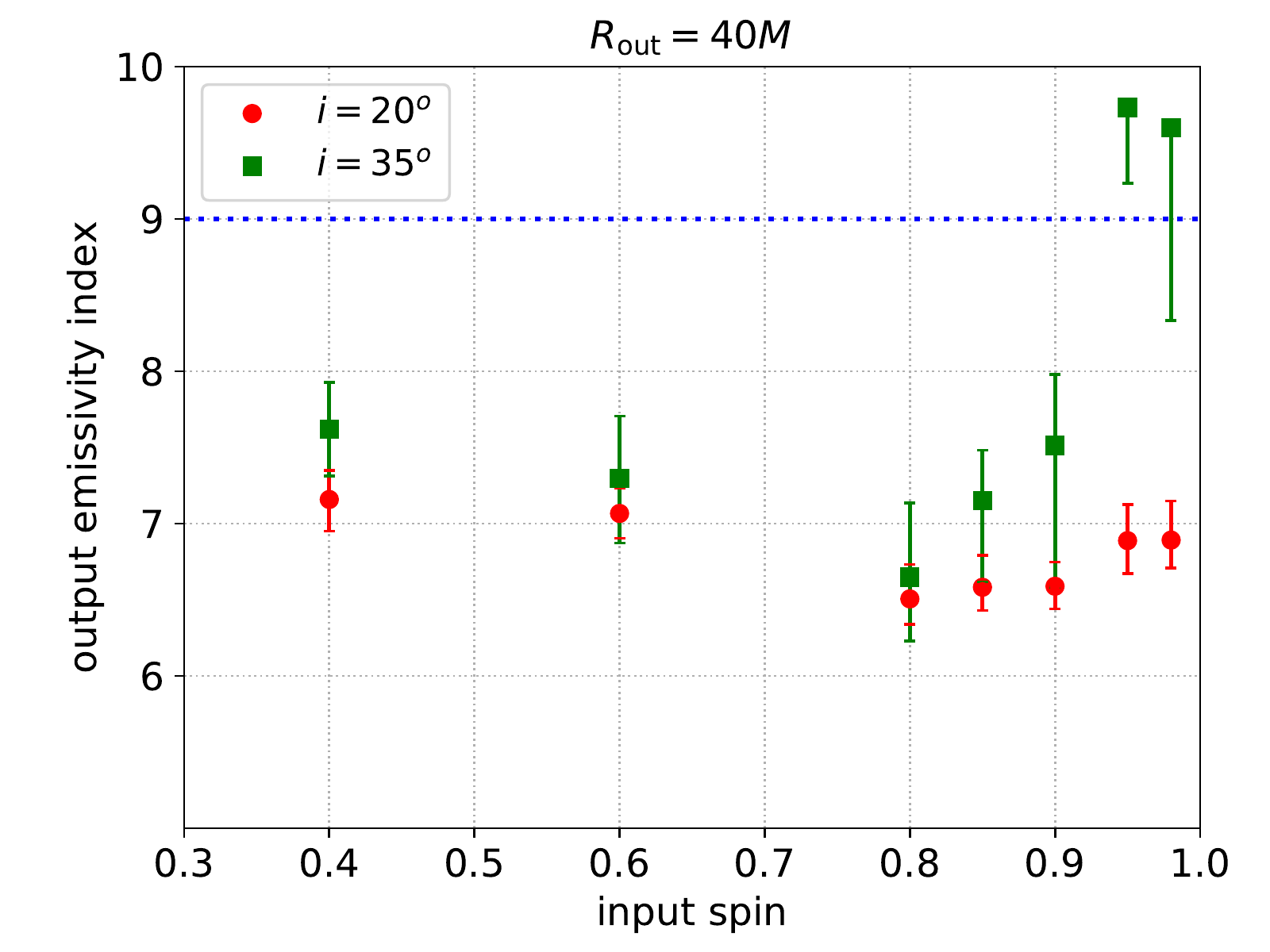}
\end{center}
\vspace{-0.4cm}
\caption{Simulations~A-D (power-law intensity profile) -- Input spin parameter of the simulations vs best-fit emissivity index obtained from {\sc relxill}. The error bars show the fit uncertainties. The horizontal blue dotted line marks the input emissivity index of the simulations. \label{f-index}}
\end{figure*}

\begin{figure*}[t]
\begin{center}
\includegraphics[type=pdf,ext=.pdf,read=.pdf,width=8.5cm]{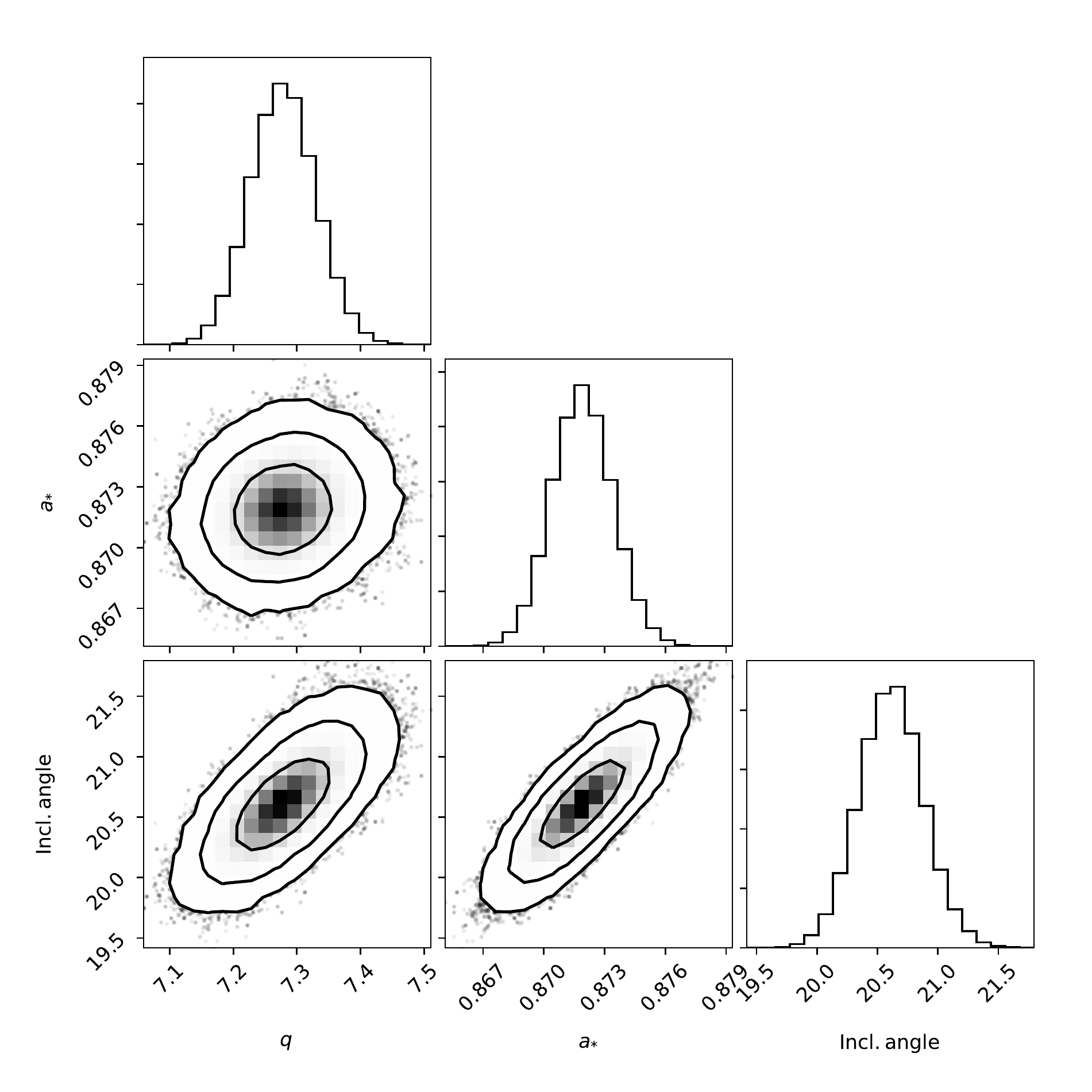}
%\hspace{0.5cm}
\includegraphics[type=pdf,ext=.pdf,read=.pdf,width=8.5cm]{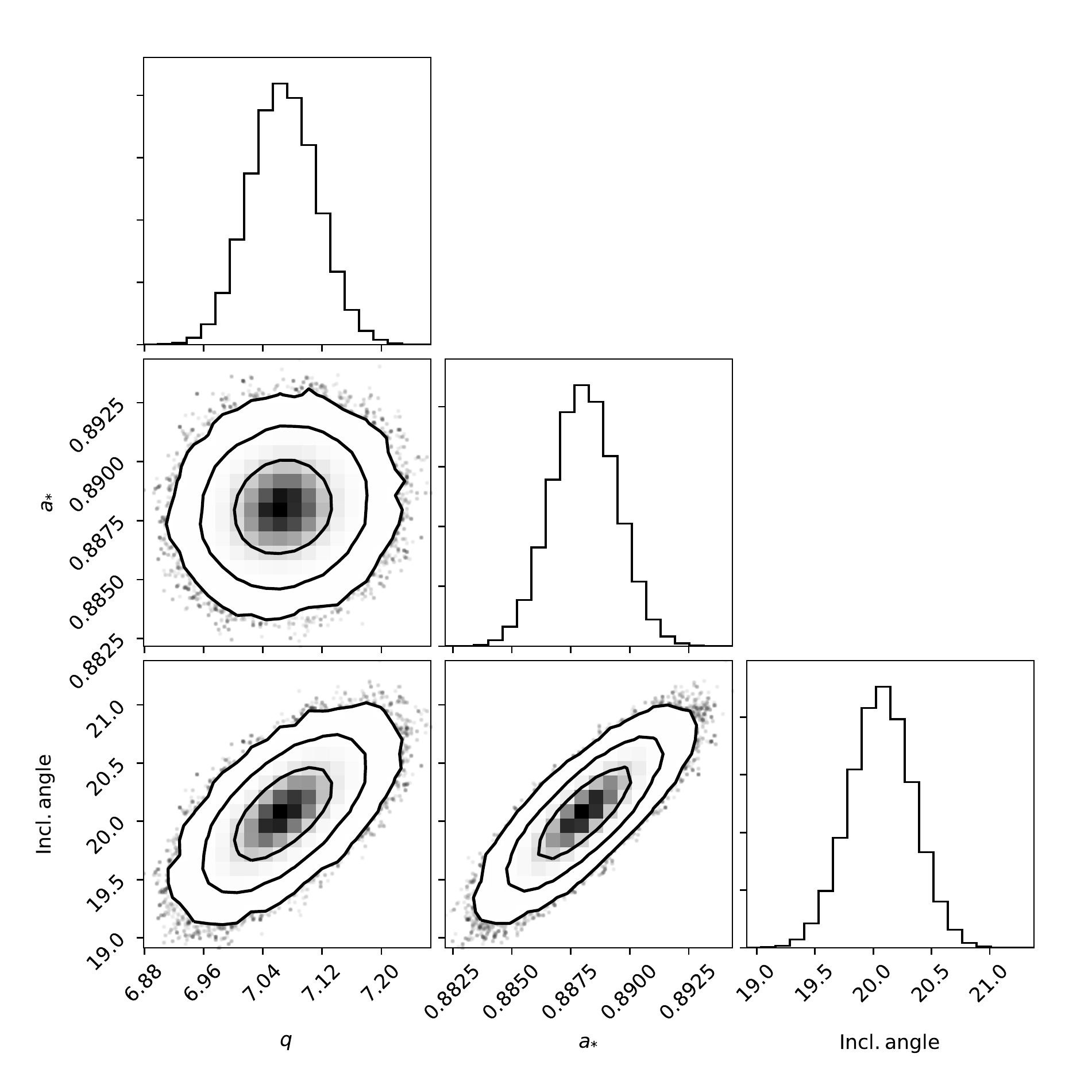}
\includegraphics[type=pdf,ext=.pdf,read=.pdf,width=8.5cm]{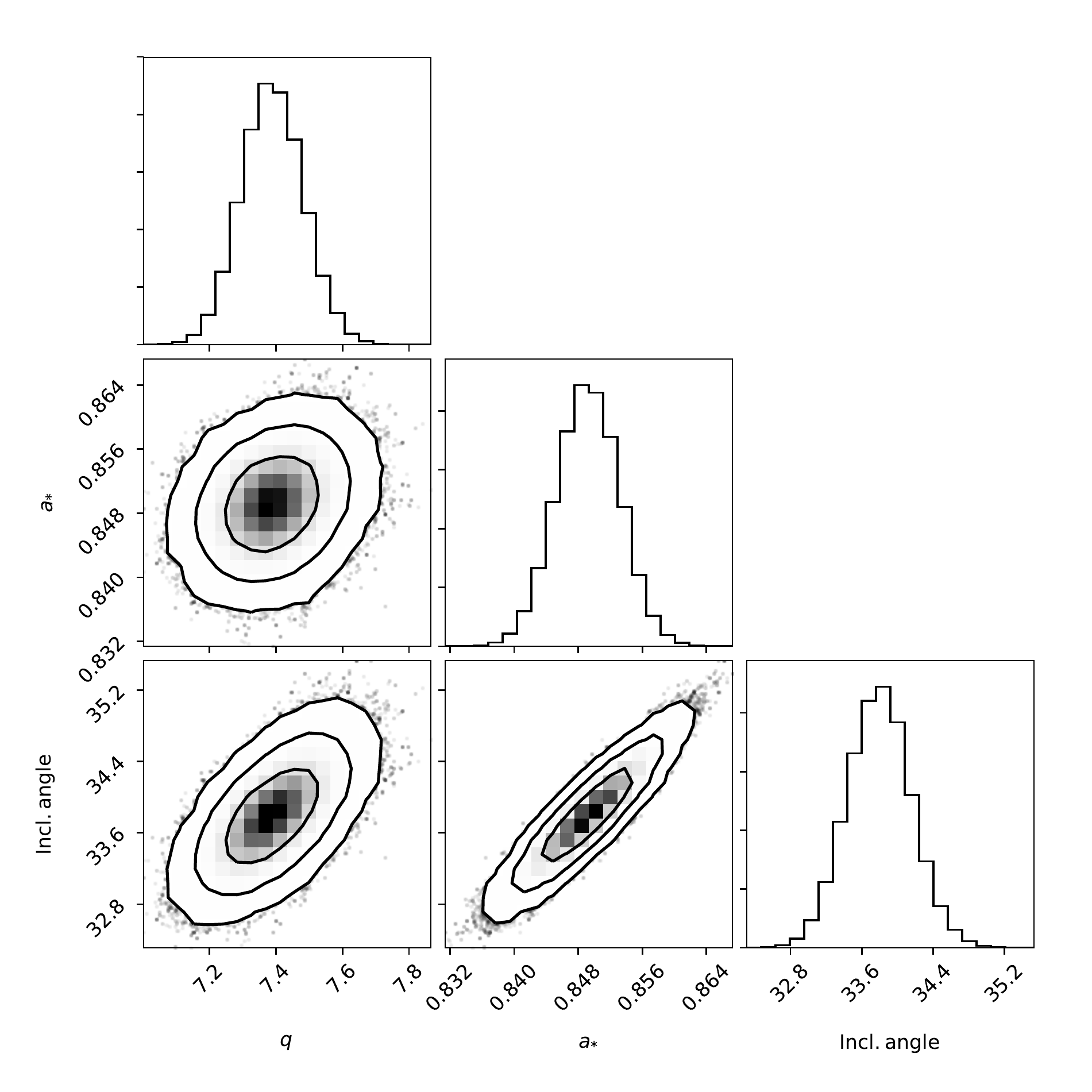}
%\hspace{0.5cm}
\includegraphics[type=pdf,ext=.pdf,read=.pdf,width=8.5cm]{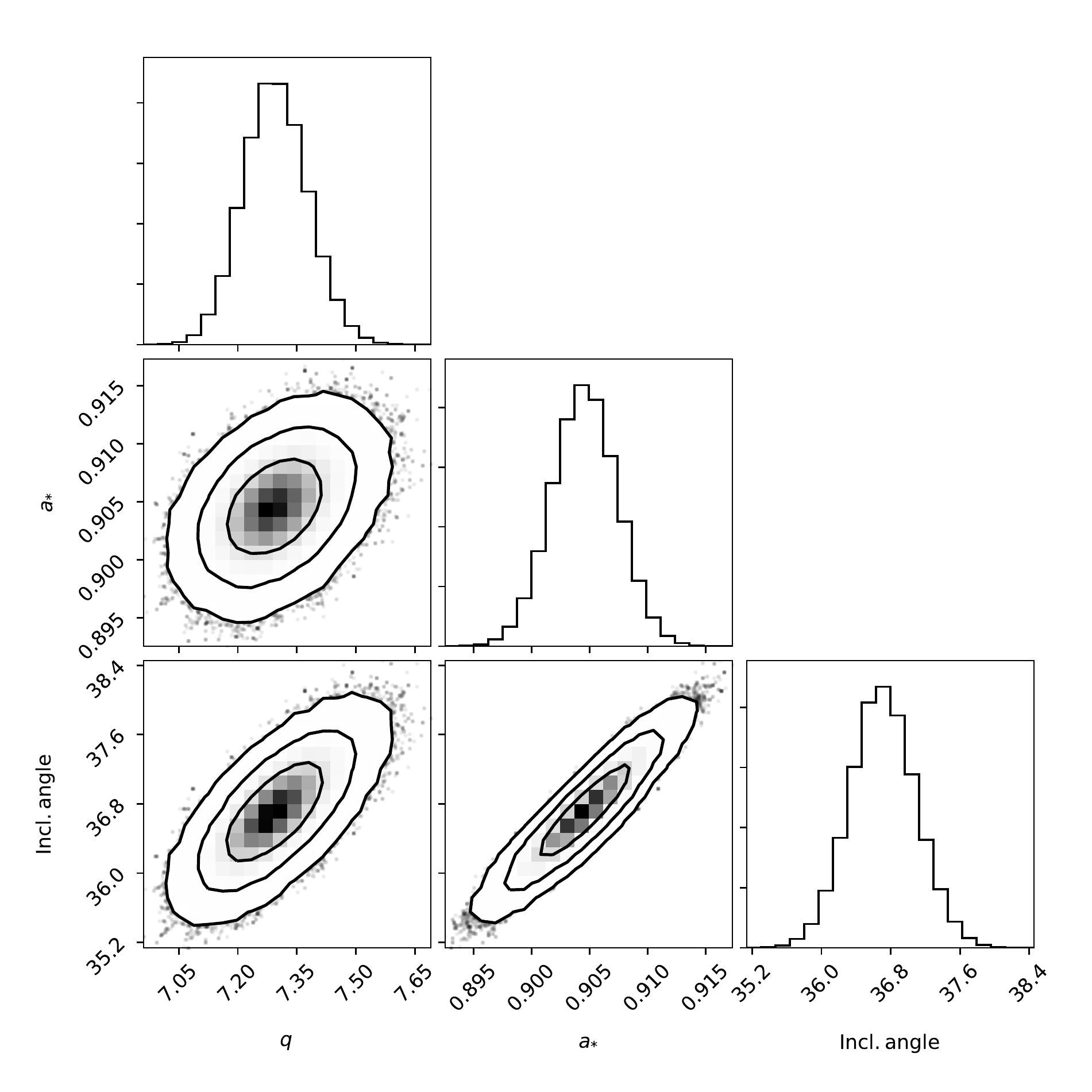}
\end{center}
\vspace{-0.4cm}
\caption{Corner plots for the emissivity index $q$, the spin parameter $a_*$, and the inclination angle after the MCMC run for simulations A2, B2, C2, and D2. The input spin parameter is $a_* = 0.60$. The input inclination angle is $i = 20^\circ$ in the top panels and $i = 35^\circ$ in the bottom panels. The outer radius of the Polish donut disk is 20~gravitational radii in the left panels and 40~gravitational radii in the right panels. \label{f-mcmc60}}
\end{figure*}

\begin{figure*}[t]
\begin{center}
\includegraphics[type=pdf,ext=.pdf,read=.pdf,width=8.5cm]{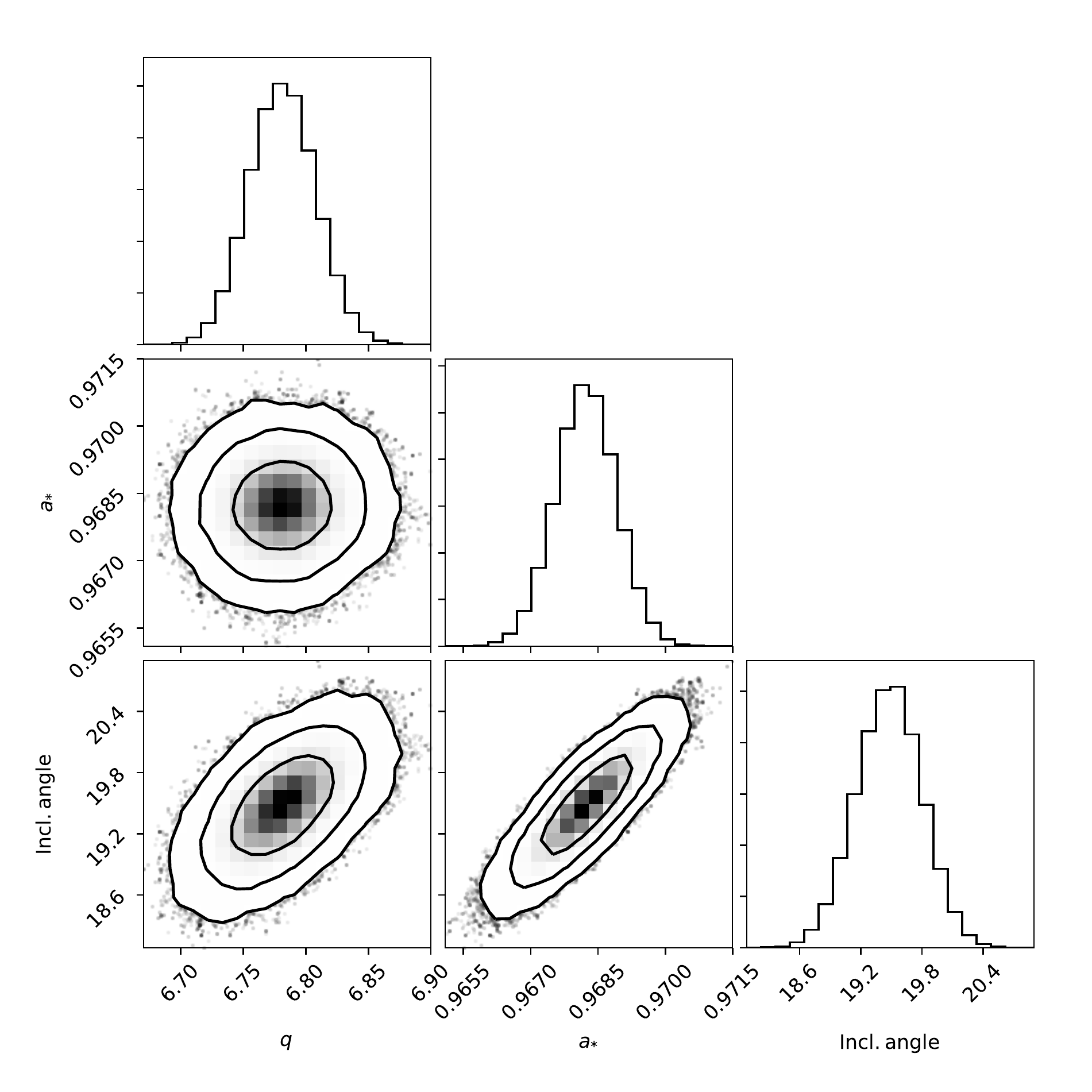}
%\hspace{0.5cm}
\includegraphics[type=pdf,ext=.pdf,read=.pdf,width=8.5cm]{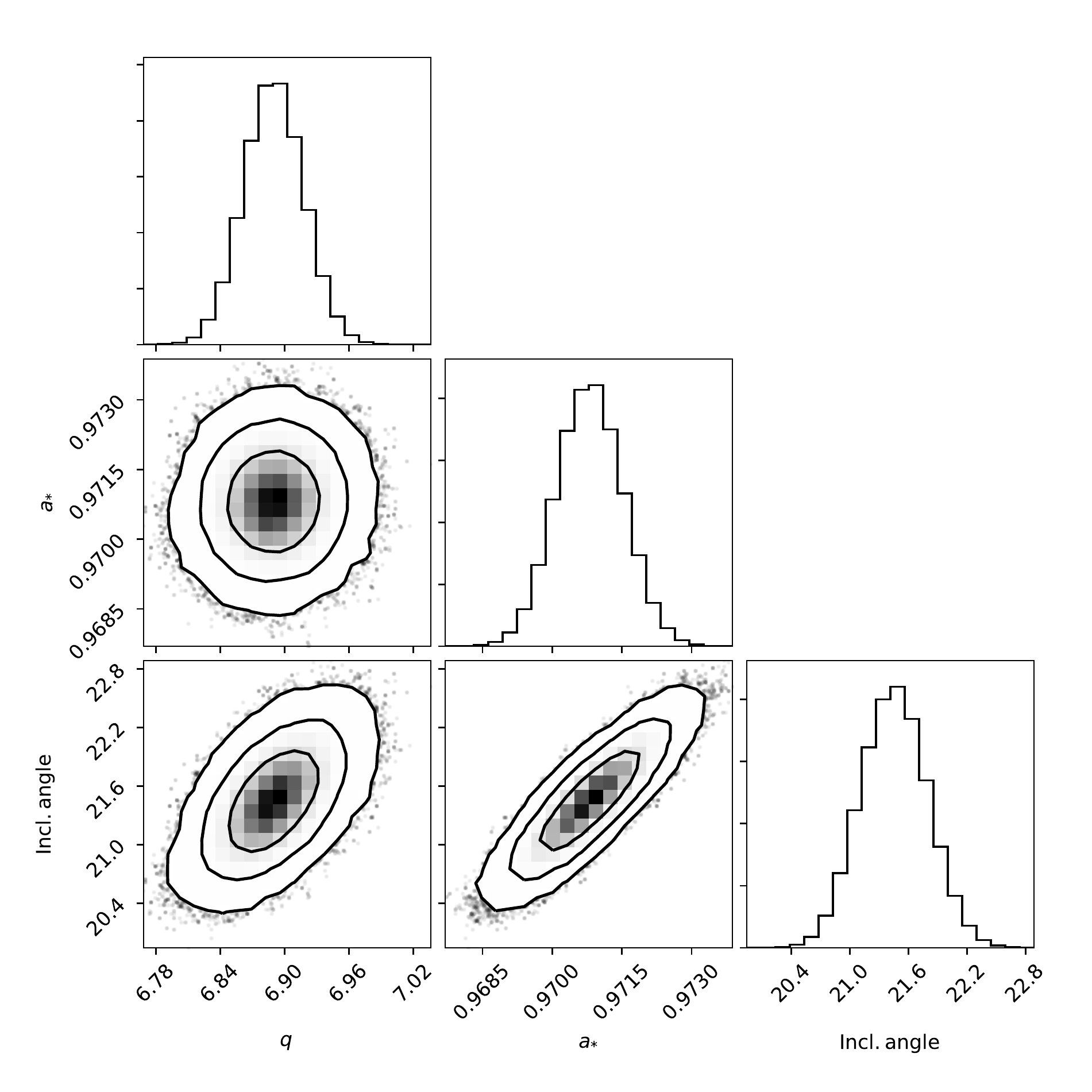}
\includegraphics[type=pdf,ext=.pdf,read=.pdf,width=8.5cm]{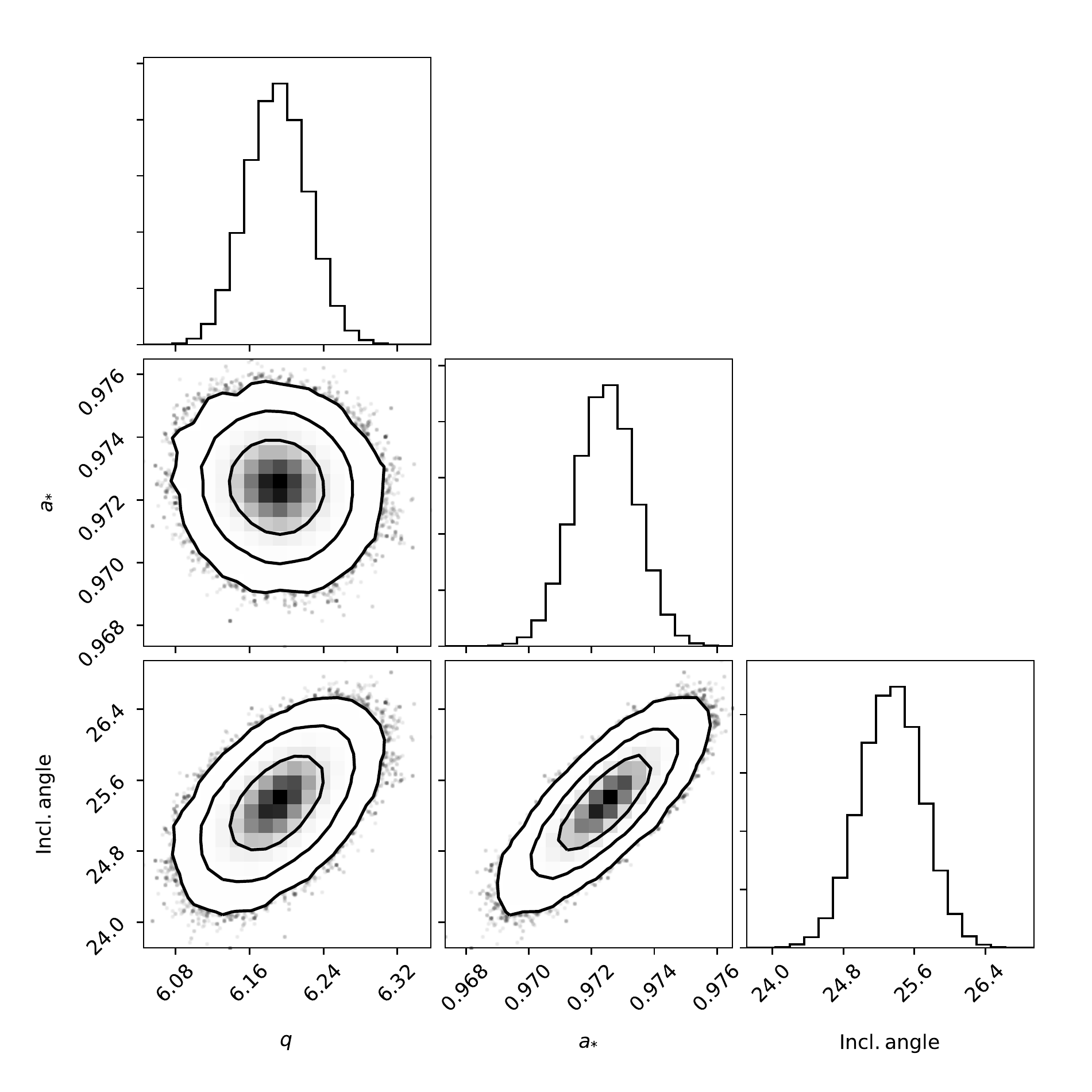}
%\hspace{0.5cm}
\includegraphics[type=pdf,ext=.pdf,read=.pdf,width=8.5cm]{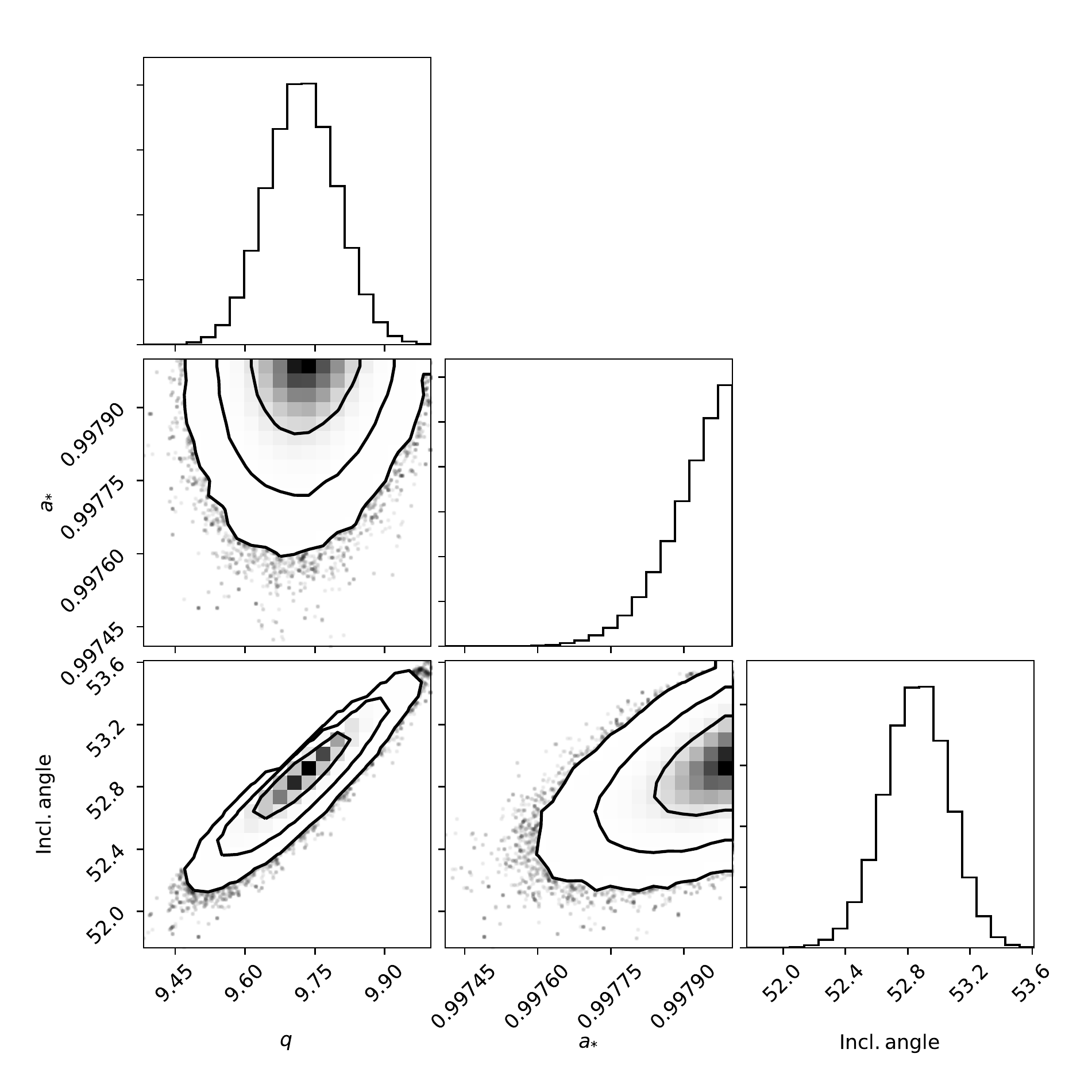}
\end{center}
\vspace{-0.4cm}
\caption{Corner plots for the emissivity index $q$, the spin parameter $a_*$, and the inclination angle after the MCMC run for simulations A6, B6, C6, and D6. The input spin parameter is $a_* = 0.95$. The input inclination angle is $i = 20^\circ$ in the top panels and $i = 35^\circ$ in the bottom panels. The outer radius of the Polish donut disk is 20~gravitational radii in the left panels and 40~gravitational radii in the right panels. \label{f-mcmc95}}
\end{figure*}

For simulations A-D (power-law intensity profile), Fig.~\ref{f-incl} shows the capability of our model to recover the input inclination angle and Fig.~\ref{f-index} shows the capability of recovering the emissivity index of the intensity profile of the reflection component. The input value of most model parameters are recovered pretty well, but there are a few exceptions. The emissivity index $q$ is always underestimated. The spin parameter can be significantly overestimated. The fits are all very good, with the reduced $\chi^2$ close to 1. In part, this is because we employ a high emissivity index, so most of the radiation is emitted from the very inner part of the accretion disk and the disk structure does not play a major role; only the location of the inner edge is important. In these cases, the quality of the fits cannot tell us that our disk model is incorrect. We have also run Markov chain Monte Carlo (MCMC) simulations for two different spin parameters: $a_* = 0.6$ (simulations~A2, B2, C2, and D2) and $a_* = 0.95$ (simulations A6, B6, C6, and D6). The resulting corner plots are reported, respectively, in Fig.~\ref{f-mcmc60} and Fig.~\ref{f-mcmc95}, where we can see that there is a correlation between the estimates of the inclination angle and of the emissivity index as well as between the estimates of the inclination angle and the spin parameter, while there is no correlation between the estimate of the spin parameter and the emissivity index.

\begin{figure*}[t]
\begin{center}
\includegraphics[type=pdf,ext=.pdf,read=.pdf,width=8cm]{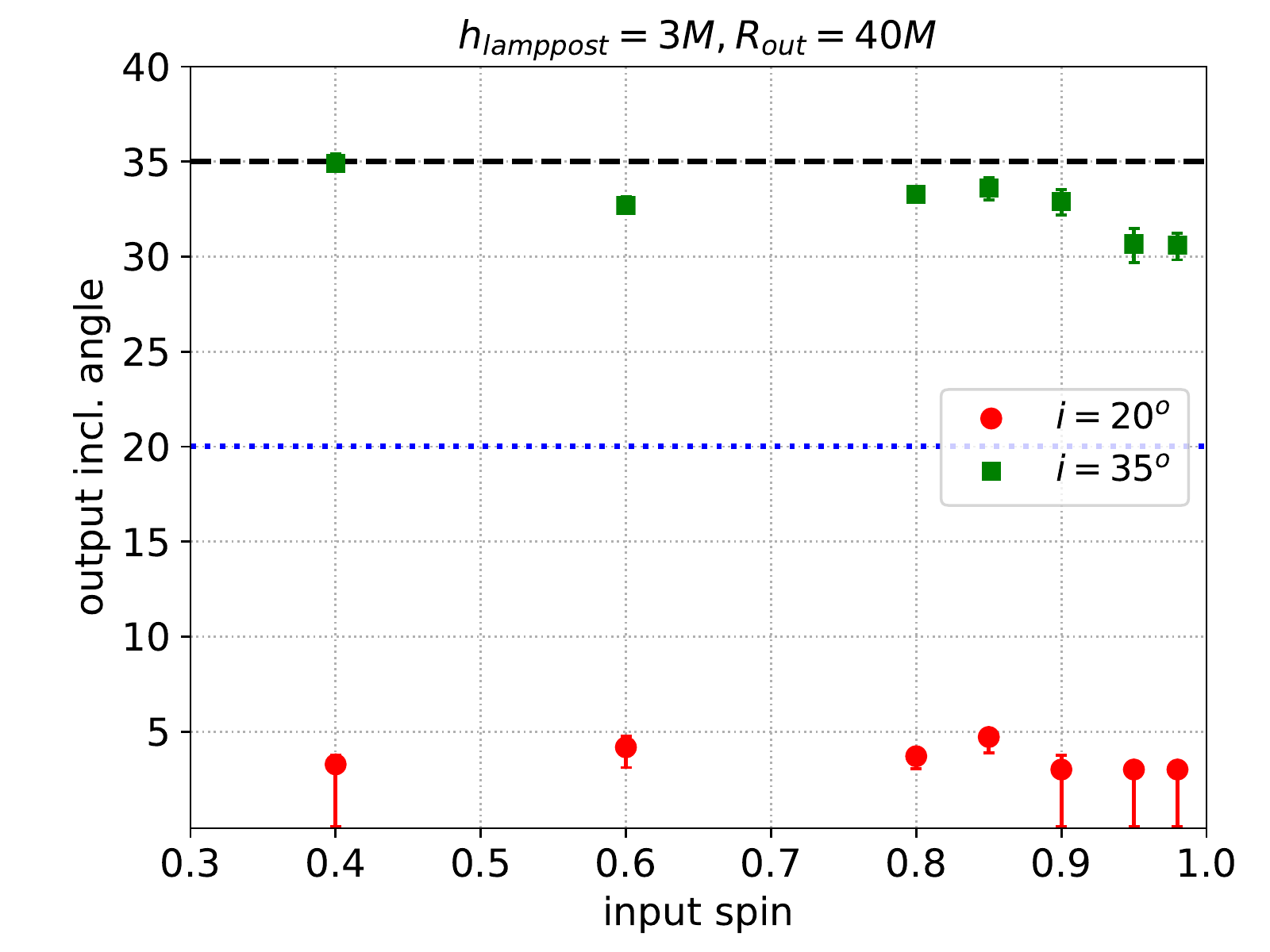}
\hspace{0.5cm}
\includegraphics[type=pdf,ext=.pdf,read=.pdf,width=8cm]{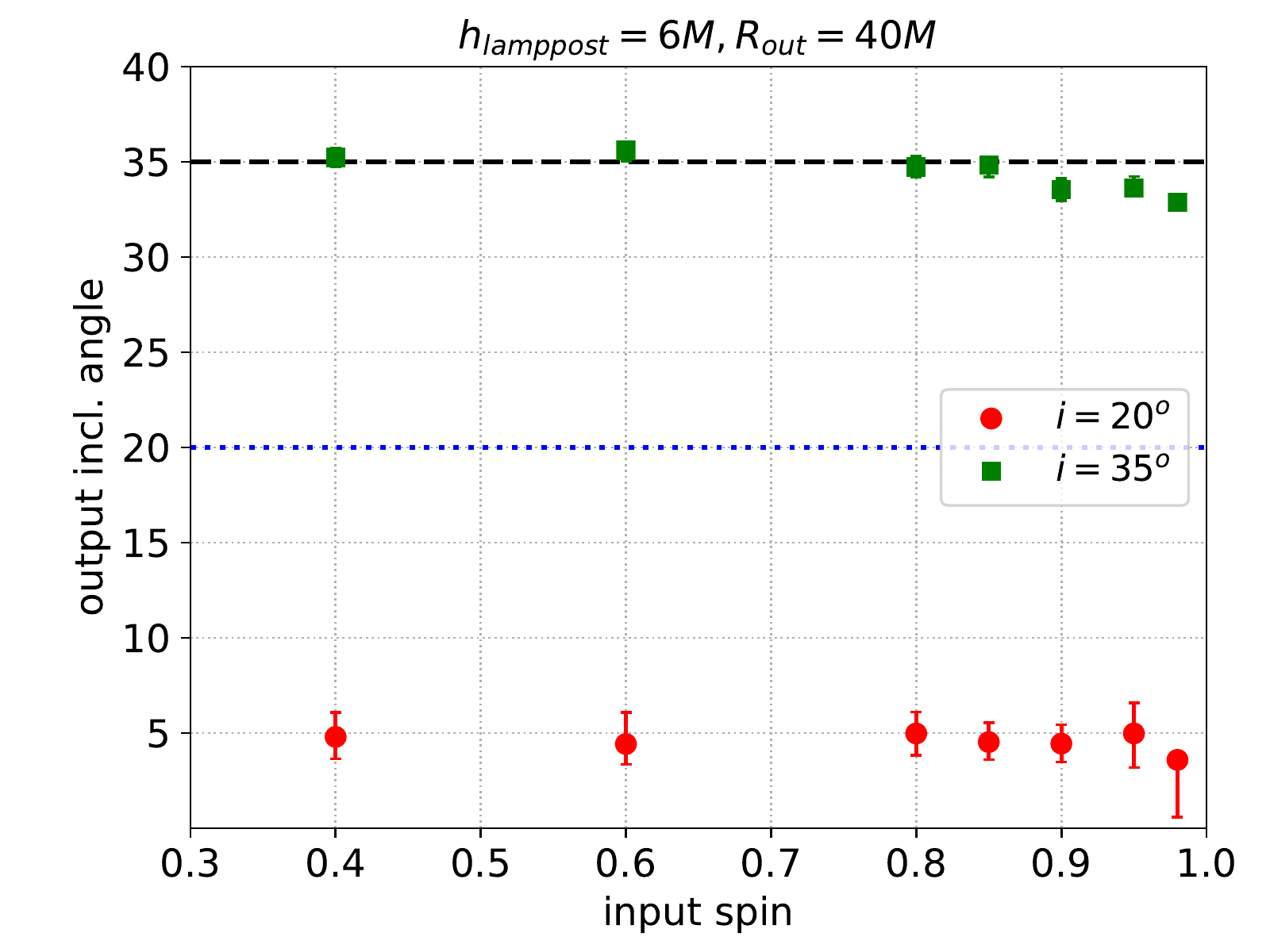}
\end{center}
\vspace{-0.4cm}
\caption{Simulations~E-H (lamppost corona intensity profile) -- Input spin parameter of the simulations vs best-fit inclination angle obtained from {\sc relxilllp}. The error bars show the fit uncertainties. The horizontal black dashed and blue dotted lines mark the input inclination angles of the simulations, respectively $i = 35^\circ$ and $i = 20^\circ$.   \label{f-spin-incl-lp}}
\vspace{0.5cm}
\begin{center}
\includegraphics[type=pdf,ext=.pdf,read=.pdf,width=8cm]{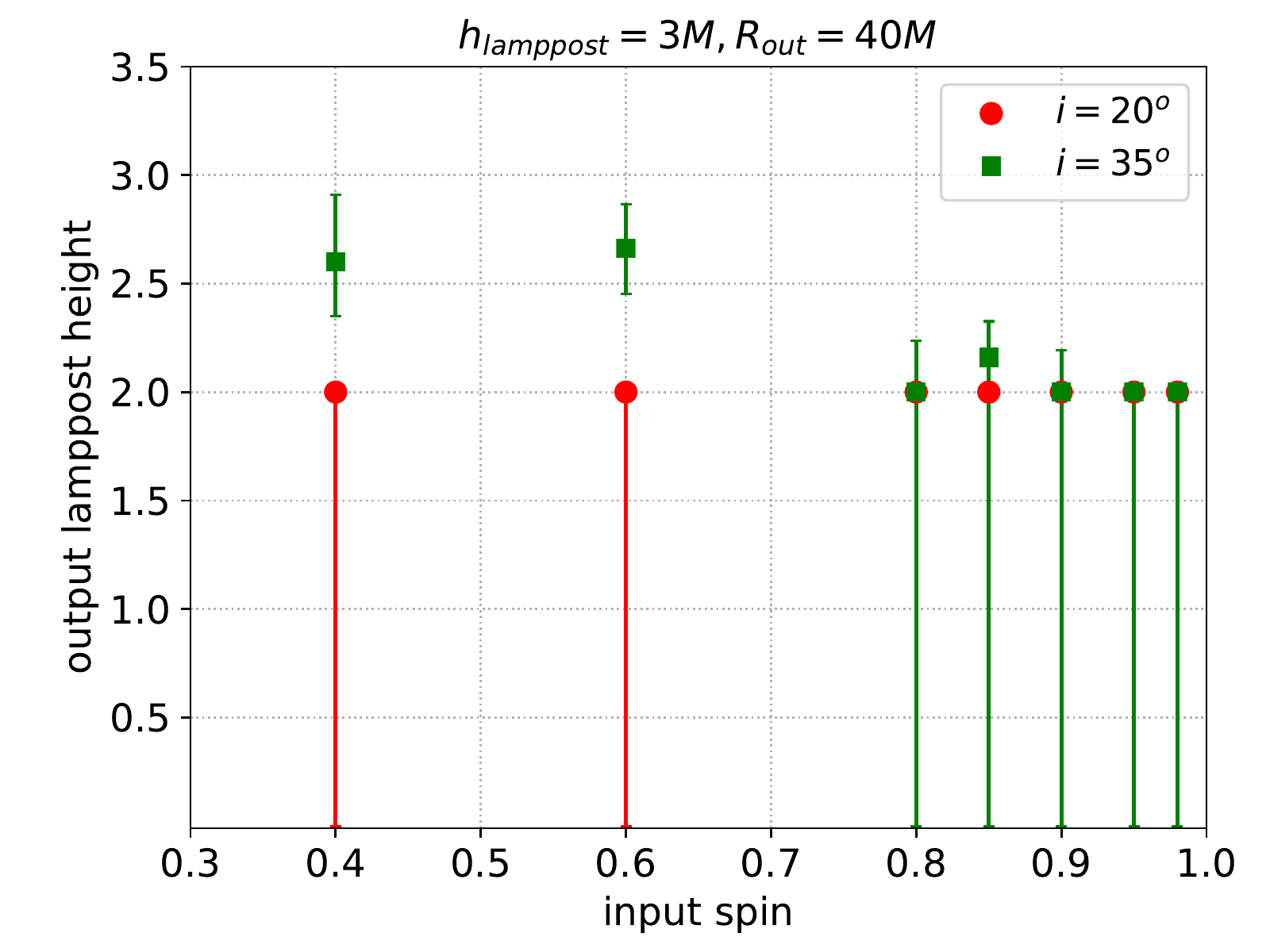}
\hspace{0.5cm}
\includegraphics[type=pdf,ext=.pdf,read=.pdf,width=8cm]{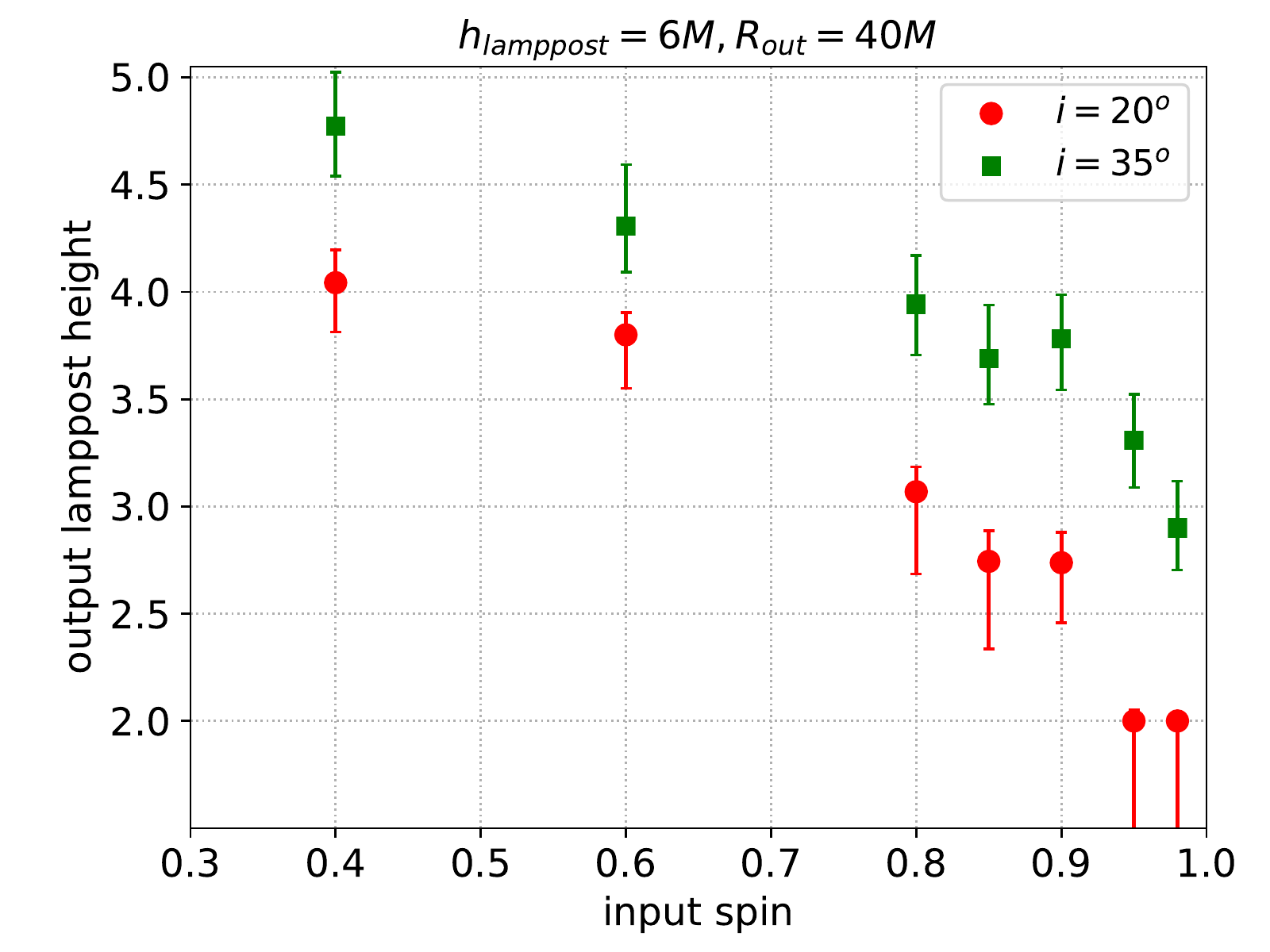}
\end{center}
\vspace{-0.4cm}
\caption{Simulations~E-H (lamppost corona intensity profile) --
Input spin parameter of the simulations vs best-fit lamppost height obtained from {\sc relxilllp}. The error bars show the fit uncertainties. The input values of the lamppost height are 3~$M$ (left panel) and 6~$M$ (right panel). \label{f-spin-h-lp}}
\end{figure*}

For simulations E-H (lamppost corona intensity profile), Fig.~\ref{f-spin-incl-lp} shows the capability of our model to recover the input inclination angle and Fig.~\ref{f-spin-h-lp} shows the capability of recovering the input corona height. Even for the lamppost set-up, we can recover the input value of most model parameters. The corona height $h_{\rm lamppost}$ is always underestimated. The spin parameter is significantly biased, with a remarkable dependence on the spin value, corona height, and inclination angle. Generally speaking, for slow-rotating black holes we overestimate the spin parameter and for fast-rotating black holes we underestimate it. The transition between overestimating and underestimating depends on the values of the viewing angle and corona height. For simulations G-H, the quality of the fits is good, so the same considerations as simulations A-D are valid. For simulations E-F, the quality of the fits is not good, but we should remember that we have analyzed overoptimistic observations with 50~million counts, which is not achievable for most AGNs.

%%%%%%%% Simulations set A table i = 20, Rout = 20 %%%%%%%%%%%%%%%%%%%

\begin{table*}
\centering
\caption{ \label{t-fit-a}}
{\renewcommand{\arraystretch}{1.3}
\begin{tabular}{lccccccccccc}
\hline\hline
 & \multicolumn{2}{c}{Simulation~A1} && \multicolumn{2}{c}{Simulation~A2} && \multicolumn{2}{c}{Simulation~A3} && \multicolumn{2}{c}{Simulation~A4}\\
 & Input & Fit && Input & Fit && Input & Fit && Input & Fit \\
\hline
{\sc tbabs} &&&&&&&&&& \\
$N_{\rm H} / 10^{20}$ cm$^{-2}$ & $6.74$ & $6.74^\star$ && $6.74$ & $6.74^\star$ && $6.74$ & $6.74^\star$ && $6.74$ & $6.74^\star$ \\
\hline
{\sc relxill} &&&&&&&& \\
$q$ & $9$ & $7.53^{+0.20}_{-0.20}$ && $9$ & $7.3^{+0.2}_{-0.2}$ && $9$ & $6.71^{+0.20}_{-0.17}$ && $9$ & $6.48^{+0.18}_{-0.14}$ \\
$i$ [deg] & $20$ & $20.7^{+0.6}_{-0.6}$ && $20$ & $20.6^{+1.3}_{-1.2}$ && $20$ & $18.8^{+2.1}_{-2.1}$ && $20$ & $16.0^{+2.6}_{-2.1}$ \\
$a_*$ & $0.40$ & $0.764^{+0.007}_{-0.007}$ && $0.60$ & $0.872^{+0.007}_{-0.007}$ && $0.80$ & $0.940^{+0.005}_{-0.004}$ && $0.85$ & $0.945^{+0.004}_{-0.003}$ \\
$\log\xi$ & $3.1$ & $3.080^{+0.009}_{-0.008}$ && $3.1$ & $3.090^{+0.008}_{-0.009}$ && $3.1$ & $3.093^{+0.008}_{-0.008}$ && $3.1$ & $3.089^{+0.008}_{-0.008}$ \\
$A_{\rm Fe}$ & $1$ & $0.972^{+0.011}_{-0.011}$ && $1$ & $0.977^{+0.010}_{-0.011}$ && $1$ & $0.980^{+0.012}_{-0.012}$ && $1$ & $0.969^{+0.012}_{-0.012}$ \\
$\Gamma$ & $2$ & $1.997^{+0.002}_{-0.002}$ && $2$ & $1.995^{+0.002}_{-0.002}$ && $2$ & $1.995^{+0.002}_{-0.002}$ && $2$ & $1.996^{+0.002}_{-0.002}$ \\
$E_{\rm cut}$ [keV] & $300$ & $300^\star$ && $300$ & $300^\star$ && $300$ & $300^\star$ && $300$ & $300^\star$ \\
$R_{\rm f}$ & $-1.0$ & $0.668^{+0.019}_{-0.021}$ && $-1.0$ & $0.80^{+0.03}_{-0.03}$ && $-1.0$ & $0.91^{+0.04}_{-0.04}$ && $-1.0$ & $0.90^{+0.04}_{-0.03}$ \\
\hline
$\chi^2/\nu$ && $\quad 1244.57/1171 \quad$ &&& $\quad 1196.93/1171 \quad$ &&& $\quad 1219.05/1171 \quad$ &&& $\quad 1140.97/1171 \quad$ \\
&& =1.06283 &&& =1.02215 &&& =1.04103 &&& =0.974354 \\
\hline\hline
\vspace{0.4cm}
\end{tabular}}
{\renewcommand{\arraystretch}{1.3}
\begin{tabular}{lccccccccccc}
\hline\hline
 & \multicolumn{2}{c}{Simulation~A5} && \multicolumn{2}{c}{Simulation~A6} && \multicolumn{2}{c}{Simulation~A7} \\
 & Input & Fit && Input & Fit && Input & Fit\\
\hline
{\sc tbabs} &&&&&&&&&& \\
$N_{\rm H} / 10^{20}$ cm$^{-2}$ & $6.74$ & $6.74^\star$ && $6.74$ & $6.74^\star$ && $6.74$ & $6.74^\star$  \\
\hline
{\sc relxill} &&&&&&&& \\
$q$ & $9$ & $6.5^{+0.2}_{-0.1}$ && $9$ & $6.78^{+0.20}_{-0.16}$ && $9$ & $7.1^{+0.3}_{-0.2}$  \\
$i$ [deg] & $20$ & $16.2^{+2.8}_{-2.0}$ && $20$ & $19.5^{+2.4}_{-2.0}$ && $20$ & $21.0^{+3.0}_{-2.5}$ \\
$a_*$ & $0.90$ & $0.955^{+0.004}_{-0.002}$ && $0.95$ & $0.968^{+0.003}_{-0.003}$ && $0.98$ & $0.981^{+0.003}_{-0.003}$ \\
$\log\xi$ & $3.1$ & $3.090^{+0.009}_{-0.007}$ && $3.1$ & $3.098^{+0.008}_{-0.008}$ && $3.1$ & $3.085^{+0.008}_{-0.005}$ \\
$A_{\rm Fe}$ & $1$ & $0.971^{+0.013}_{-0.012}$ && $1$ & $0.977^{+0.013}_{-0.013}$ && $1$ & $0.969^{+0.013}_{-0.015}$  \\
$\Gamma$ & $2$ & $1.995^{+0.001}_{-0.003}$ && $2$ & $1.993^{+0.003}_{-0.003}$ && $2$ & $1.996^{+0.002}_{-0.003}$\\
$E_{\rm cut}$ [keV] & $300$ & $300^\star$ && $300$ & $300^\star$ && $300$ & $300^\star$ \\
$R_{\rm f}$ & $-1.0$ & $0.95^{+0.05}_{-0.03}$ && $-1.0$ & $1.06^{+0.05}_{-0.05}$ && $-1.0$ & $1.11^{+0.06}_{-0.06}$\\
\hline
$\chi^2/\nu$ && $\quad 1176.69/1171 \quad$ &&& $\quad 1188.97/1171 \quad$ &&& $\quad 1179.94/1171 \quad$ \\
&& =1.00486 &&& =1.01534 &&& =1.00764  \\
\hline\hline
\end{tabular}
\vspace{0.2cm}
\tablenotetext{0}{Best-fit values for simulations~A1-A7. In all these simulations, the disk inclination angle is $i = 20^{\circ}$ and the outer radius of the Polish donut disk is $R_{\rm out}=20$~$M$. The reported uncertainties correspond to 90\% confidence level for one relevant parameter. $^\star$ indicates that the parameter is frozen in the fit.}
}
\end{table*}

%%%%%%%% Simulations set B table i = 20, Rout = 40 %%%%%%%%%%%%%%%%%%%

\begin{table*}
\centering
\caption{ \label{t-fit-b}}
{\renewcommand{\arraystretch}{1.3}
\begin{tabular}{lccccccccccc}
\hline\hline
 & \multicolumn{2}{c}{Simulation~B1} && \multicolumn{2}{c}{Simulation~B2} && \multicolumn{2}{c}{Simulation~B3} && \multicolumn{2}{c}{Simulation~B4} \\
 & Input & Fit && Input & Fit && Input & Fit && Input & Fit \\
\hline
{\sc tbabs} &&&&&&&&&& \\
$N_{\rm H} / 10^{20}$ cm$^{-2}$ & $6.74$ & $6.74^\star$ && $6.74$ & $6.74^\star$ && $6.74$ & $6.74^\star$ && $6.74$ & $6.74^\star$ \\
\hline
{\sc relxill} &&&&&&&& \\
$q$ & $9$ & $7.16^{+0.19}_{-0.21}$ && $9$ & $7.07^{+0.17}_{-0.16}$ && $9$ & $6.5^{+0.2}_{-0.2}$ && $9$ & $6.58^{+0.21}_{-0.15}$ \\
$i$ [deg] & $20$ & $21.2^{+0.8}_{-0.9}$ && $20$ & $20^{+1}_{-1}$ && $20$ & $18.1^{+2.6}_{-2.5}$ && $20$ & $19.4^{+2.5}_{-1.0}$ \\
$a_*$ & $0.40$ & $0.823^{+0.008}_{-0.007}$ && $0.60$ & $0.888^{+0.006}_{-0.005}$ && $0.80$ & $0.944^{+0.005}_{-0.004}$ && $0.85$ & $0.954^{+0.004}_{-0.003}$ \\
$\log\xi$ & $3.1$ & $3.082^{+0.010}_{-0.008}$ && $3.1$ & $3.073^{+0.009}_{-0.008}$ && $3.1$ & $3.084^{+0.008}_{-0.008}$ && $3.1$ & $3.088^{+0.008}_{-0.007}$ \\
$A_{\rm Fe}$ & $1$ & $0.972^{+0.010}_{-0.013}$ && $1$ & $0.963^{+0.012}_{-0.010}$ && $1$ & $0.976^{+0.012}_{-0.012}$ && $1$ & $0.970^{+0.013}_{-0.012}$ \\
$\Gamma$ & $2$ & $1.998^{+0.002}_{-0.002}$ && $2$ & $2.000^{+0.002}_{-0.002}$ && $2$ & $1.998^{+0.003}_{-0.002}$ && $2$ & $1.996^{+0.002}_{-0.003}$ \\
$E_{\rm cut}$ [keV] & $300$ & $300^\star$ && $300$ & $300^\star$ && $300$ & $300^\star$ && $300$ & $300^\star$ \\
$R_{\rm f}$ & $-1.0$ & $0.692^{+0.027}_{-0.020}$ && $-1.0$ & $0.75^{+0.03}_{-0.03}$ && $-1.0$ & $0.86^{+0.04}_{-0.04}$ && $-1.0$ & $0.93^{+0.05}_{-0.03}$ \\
\hline
$\chi^2/\nu$ && $\quad 1213.41/1171 \quad$ &&& $\quad 1231.81/1171 \quad$ &&& $\quad 1197.38/1171 \quad$ &&& $\quad 1090.89/1171 \quad$ \\
&& =1.03622 &&& =1.05193 &&& =1.02253 &&& =0.931588 \\
\hline\hline
\vspace{0.4cm}
\end{tabular}}
{\renewcommand{\arraystretch}{1.3}
\begin{tabular}{lccccccccccc}
\hline\hline
 & \multicolumn{2}{c}{Simulation~B5} && \multicolumn{2}{c}{Simulation~B6} && \multicolumn{2}{c}{Simulation~B7} \\
 & Input & Fit && Input & Fit && Input & Fit\\
\hline
{\sc tbabs} &&&&&&&&&& \\
$N_{\rm H} / 10^{20}$ cm$^{-2}$ & $6.74$ & $6.74^\star$ && $6.74$ & $6.74^\star$ && $6.74$ & $6.74^\star$  \\
\hline
{\sc relxill} &&&&&&&& \\
$q$ & $9$ & $6.6^{+0.2}_{-0.1}$ && $9$ & $6.9^{+0.2}_{-0.2}$ && $9$ & $6.9^{+0.3}_{-0.2}$  \\
$i$ [deg] & $20$ & $19.7^{+1.8}_{-2.0}$ && $20$ & $21.5^{+2.7}_{-2.5}$ && $20$ & $18.9^{+3.3}_{-1.3}$ \\
$a_*$ & $0.90$ & $0.960^{+0.003}_{-0.003}$ && $0.95$ & $0.971^{+0.004}_{-0.003}$ && $0.98$ & $0.979^{+0.004}_{-0.002}$ \\
$\log\xi$ & $3.1$ & $3.098^{+0.006}_{-0.006}$ && $3.1$ & $3.089^{+0.008}_{-0.008}$ && $3.1$ & $3.081^{+0.008}_{-0.008}$ \\
$A_{\rm Fe}$ & $1$ & $0.983^{+0.010}_{-0.010}$ && $1$ & $0.973^{+0.014}_{-0.014}$ && $1$ & $0.968^{+0.017}_{-0.014}$  \\
$\Gamma$ & $2$ & $1.994^{+0.003}_{-0.003}$ && $2$ & $1.996^{+0.003}_{-0.003}$ && $2$ & $1.996^{+0.002}_{-0.003}$\\
$E_{\rm cut}$ [keV] & $300$ & $300^\star$ && $300$ & $300^\star$ && $300$ & $300^\star$ \\
$R_{\rm f}$ & $-1.0$ & $2.01^{+0.14}_{-0.14}$ && $-1.0$ & $1.02^{+0.05}_{-0.05}$ && $-1.0$ & $1.03^{+0.06}_{-0.05}$\\
\hline
$\chi^2/\nu$ && $\quad 1275.99/1171 \quad$ &&& $\quad 1205.22/1171 \quad$ &&& $\quad 1189.23/1171 \quad$ \\
&& =1.08966 &&& =1.02922 &&& =1.01556  \\
\hline\hline
\end{tabular}
\vspace{0.2cm}
\tablenotetext{0}{Best-fit values for simulations~B1-B7. In all these simulations, the disk inclination angle is $i = 20^{\circ}$ and the outer radius of the Polish donut disk is $R_{\rm out}=40$~$M$. The reported uncertainties correspond to 90\% confidence level for one relevant parameter. $^\star$ indicates that the parameter is frozen in the fit.}
}
\end{table*}

%\newpage

%%%%%%%% Simulations set C table i = 35, Rout = 20 %%%%%%%%%%%%%%%%%%%

\begin{table*}
\centering
\caption{ \label{t-fit-c}}
{\renewcommand{\arraystretch}{1.3}
\begin{tabular}{lccccccccccc}
\hline\hline
 & \multicolumn{2}{c}{Simulation~C1} && \multicolumn{2}{c}{Simulation~C2} && \multicolumn{2}{c}{Simulation~C3} && \multicolumn{2}{c}{Simulation~C4} \\
 & Input & Fit && Input & Fit && Input & Fit && Input & Fit \\
\hline
{\sc tbabs} &&&&&&&&&& \\
$N_{\rm H} / 10^{20}$ cm$^{-2}$ & $6.74$ & $6.74^\star$ && $6.74$ & $6.74^\star$ && $6.74$ & $6.74^\star$ && $6.74$ & $6.74^\star$ \\
\hline
{\sc relxill} &&&&&&&& \\
$q$ & $9$ & $8.0^{+0.3}_{-0.3}$ && $9$ & $7.4^{+0.3}_{-0.3}$ && $9$ & $6.2^{+1.0}_{-0.3}$ && $9$ & $7.2^{+0.5}_{-0.5}$ \\
$i$ [deg] & $35$ & $36.9^{+1.3}_{-1.3}$ && $35$ & $33.8^{+1.8}_{-1.9}$ && $35$ & $26^{+8}_{-3}$ && $35$ & $37^{+4}_{-4}$ \\
$a_*$ & $0.40$ & $0.792^{+0.021}_{-0.023}$ && $0.60$ & $0.849^{+0.019}_{-0.022}$ && $0.80$ & $0.913^{+0.033}_{-0.015}$ && $0.85$ & $0.967^{+0.007}_{-0.011}$ \\
$\log\xi$ & $3.1$ & $3.081^{+0.010}_{-0.010}$ && $3.1$ & $3.075^{+0.013}_{-0.012}$ && $3.1$ & $3.070^{+0.018}_{-0.013}$ && $3.1$ & $3.087^{+0.012}_{-0.012}$ \\
$A_{\rm Fe}$ & $1$ & $0.977^{+0.016}_{-0.016}$ && $1$ & $0.974^{+0.019}_{-0.018}$ && $1$ & $0.968^{+0.025}_{-0.016}$ && $1$ & $0.990^{+0.034}_{-0.017}$ \\
$\Gamma$ & $2$ & $1.996^{+0.002}_{-0.002}$ && $2$ & $1.998^{+0.002}_{-0.002}$ && $2$ & $2.001^{+0.003}_{-0.004}$ && $2$ & $1.996^{+0.002}_{-0.003}$ \\
$E_{\rm cut}$ [keV] & $300$ & $300^\star$ && $300$ & $300^\star$ && $300$ & $300^\star$ && $300$ & $300^\star$ \\
$R_{\rm f}$ & $-1.0$ & $0.436^{+0.013}_{-0.013}$ && $-1.0$ & $0.456^{+0.020}_{-0.018}$ && $-1.0$ & $0.49^{+0.06}_{-0.03}$ && $-1.0$ & $0.60^{+0.03}_{-0.03}$ \\
\hline
$\chi^2/\nu$ && $\quad 1185.89/1171 \quad$ &&& $\quad 1127.37/1171 \quad$ &&& $\quad 1129.73/1171 \quad$ &&& $\quad 1128.35/1171 \quad$ \\
&& =1.01271 &&& =0.962738 &&& =0.964759 &&& =0.963577 \\
\hline\hline
\vspace{0.4cm}
\end{tabular}}
{\renewcommand{\arraystretch}{1.3}
\begin{tabular}{lccccccccccc}
\hline\hline
 & \multicolumn{2}{c}{Simulation~C5} && \multicolumn{2}{c}{Simulation~C6} && \multicolumn{2}{c}{Simulation~C7} \\
 & Input & Fit && Input & Fit && Input & Fit\\
\hline
{\sc tbabs} &&&&&&&&&& \\
$N_{\rm H} / 10^{20}$ cm$^{-2}$ & $6.74$ & $6.74^\star$ && $6.74$ & $6.74^\star$ && $6.74$ & $6.74^\star$  \\
\hline
{\sc relxill} &&&&&&&& \\
$q$ & $9$ & $5.64^{+0.23}_{-0.18}$ && $9$ & $6.2^{+0.4}_{-0.2}$ && $9$ & $8.5^{+0.2}_{-0.5}$ \\
$i$ [deg] & $35$ & $21^{+4}_{-3}$ && $35$ & $25^{+4}_{-3}$ && $35$ & $44^{+3}_{-9}$ \\
$a_*$ & $0.90$ & $0.956^{+0.007}_{-0.005}$ && $0.95$ & $0.972^{+0.007}_{-0.005}$ && $0.98$ & $0.995^{+0.002}_{-0.005}$ \\
$\log\xi$ & $3.1$ & $3.079^{+0.015}_{-0.012}$ && $3.1$ & $3.069^{+0.011}_{-0.012}$ && $3.1$ & $3.094^{+0.005}_{-0.010}$ \\
$A_{\rm Fe}$ & $1$ & $0.993^{+0.063}_{-0.016}$ && $1$ & $0.979^{+0.017}_{-0.017}$ && $1$ & $0.970^{+0.015}_{-0.016}$  \\
$\Gamma$ & $2$ & $2.000^{+0.003}_{-0.003}$ && $2$ & $2.003^{+0.003}_{-0.003}$ && $2$ & $1.992^{+0.003}_{-0.002}$\\
$E_{\rm cut}$ [keV] & $300$ & $300^\star$ && $300$ & $300^\star$ && $300$ & $300^\star$ \\
$R_{\rm f}$ & $-1.0$ & $0.56^{+0.03}_{-0.03}$ && $-1.0$ & $0.58^{+0.05}_{-0.03}$ && $-1.0$ & $0.800^{+0.026}_{-0.085}$\\
\hline
$\chi^2/\nu$ && $\quad 1216.36/1171 \quad$ &&& $\quad 1178.76/1171 \quad$ &&& $\quad 1111.73/1171 \quad$ \\
&& =1.03874 &&& =1.00663 &&& =0.949386  \\
\hline\hline
\end{tabular}}
\vspace{0.2cm}
\tablenotetext{0}{Best-fit values for simulations~C1-C7. In all these simulations, the disk inclination angle is $i = 35^{\circ}$ and the outer radius of the Polish donut disk is $R_{\rm out}=20$~$M$. The reported uncertainties correspond to 90\% confidence level for one relevant parameter. $^\star$ indicates that the parameter is frozen in the fit.}
\end{table*}

%%%%%%%% Simulations set D table i = 35, Rout = 40 %%%%%%%%%%%%%%%%%%%

\begin{table*}
\centering
\caption{ \label{t-fit-d}}
{\renewcommand{\arraystretch}{1.3}
\begin{tabular}{lccccccccccc}
\hline\hline
 & \multicolumn{2}{c}{Simulation~D1} && \multicolumn{2}{c}{Simulation~D2} && \multicolumn{2}{c}{Simulation~D3} && \multicolumn{2}{c}{Simulation~D4} \\
 & Input & Fit && Input & Fit && Input & Fit && Input & Fit \\
\hline
{\sc tbabs} &&&&&&&&&& \\
$N_{\rm H} / 10^{20}$ cm$^{-2}$ & $6.74$ & $6.74^\star$ && $6.74$ & $6.74^\star$ && $6.74$ & $6.74^\star$ && $6.74$ & $6.74^\star$ \\
\hline
{\sc relxill} &&&&&&&& \\
$q$ & $9$ & $7.6^{+0.3}_{-0.3}$ && $9$ & $7.3^{+0.4}_{-0.4}$ && $9$ & $6.6^{+0.5}_{-0.4}$ && $9$ & $7.2^{+0.3}_{-0.5}$ \\
$i$ [deg] & $35$ & $37.4^{+1.7}_{-1.7}$ && $35$ & $36.7^{+2.5}_{-2.9}$ && $35$ & $35^{+4}_{-4}$ && $35$ & $39^{+4}_{-5}$ \\
$a_*$ & $0.40$ & $0.835^{+0.020}_{-0.023}$ && $0.60$ & $0.905^{+0.016}_{-0.022}$ && $0.80$ & $0.963^{+0.009}_{-0.009}$ && $0.85$ & $0.982^{+0.005}_{-0.007}$ \\
$\log\xi$ & $3.1$ & $3.092^{+0.016}_{-0.011}$ && $3.1$ & $3.086^{+0.016}_{-0.012}$ && $3.1$ & $3.081^{+0.014}_{-0.012}$ && $3.1$ & $3.085^{+0.013}_{-0.008}$ \\
$A_{\rm Fe}$ & $1$ & $0.991^{+0.054}_{-0.017}$ && $1$ & $1.001^{+0.071}_{-0.019}$ && $1$ & $0.991^{+0.043}_{-0.017}$ && $1$ & $0.995^{+0.027}_{-0.017}$ \\
$\Gamma$ & $2$ & $1.995^{+0.002}_{-0.003}$ && $2$ & $1.996^{+0.002}_{-0.003}$ && $2$ & $1.998^{+0.003}_{-0.003}$ && $2$ & $1.997^{+0.002}_{-0.003}$ \\
$E_{\rm cut}$ [keV] & $300$ & $300^\star$ && $300$ & $300^\star$ && $300$ & $300^\star$ && $300$ & $300^\star$ \\
$R_{\rm f}$ & $-1.0$ & $0.459^{+0.022}_{-0.016}$ && $-1.0$ & $0.496^{+0.025}_{-0.023}$ && $-1.0$ & $0.56^{+0.03}_{-0.03}$ && $-1.0$ & $0.62^{+0.03}_{-0.02}$ \\
\hline
$\chi^2/\nu$ && $\quad 1199.27/1171 \quad$ &&& $\quad 1204.24/1171 \quad$ &&& $\quad  1156.41/1171 \quad$ &&& $\quad 1221.69/1171 \quad$ \\
&& =1.02414 &&& =1.02839 &&& =0.987538 &&& =1.04329 \\
\hline\hline
\vspace{0.4cm}
\end{tabular}}
{\renewcommand{\arraystretch}{1.3}
\begin{tabular}{lccccccccccc}
\hline\hline
 & \multicolumn{2}{c}{Simulation~D5} && \multicolumn{2}{c}{Simulation~D6} && \multicolumn{2}{c}{Simulation~D7} \\
 & Input & Fit && Input & Fit && Input & Fit\\
\hline
{\sc tbabs} &&&&&&&&&& \\
$N_{\rm H} / 10^{20}$ cm$^{-2}$ & $6.74$ & $6.74^\star$ && $6.74$ & $6.74^\star$ && $6.74$ & $6.74^\star$  \\
\hline
{\sc relxill} &&&&&&&& \\
$q$ & $9$ & $7.5^{+0.5}_{-0.9}$ && $9$ & $9.7^{}_{-0.5}$ && $9$ & $9.6^{}_{-1.3}$ \\
$i$ [deg] & $35$ & $42^{+4}_{-7}$ && $35$ & $53.0^{+1.4}_{-2.1}$ && $35$ & $52.5^{+1.8}_{-6.4}$ \\
$a_*$ & $0.90$ & $0.992^{+0.003}_{-0.005}$ && $0.95$ & $0.9980^{}_{-0.0003}$ && $0.98$ & $0.9980^{}_{-0.0003}$ \\
$\log\xi$ & $3.1$ & $3.076^{+0.012}_{-0.013}$ && $3.1$ & $3.091^{+0.013}_{-0.012}$ && $3.1$ & $3.072^{+0.014}_{-0.013}$ \\
$A_{\rm Fe}$ & $1$ & $0.980^{+0.031}_{-0.020}$ && $1$ & $0.971^{+0.020}_{-0.024}$ && $1$ & $0.980^{+0.125}_{-0.026}$  \\
$\Gamma$ & $2$ & $2.000^{+0.005}_{-0.003}$ && $2$ & $1.995^{+0.003}_{-0.004}$ && $2$ & $2.001^{+0.007}_{-0.004}$\\
$E_{\rm cut}$ [keV] & $300$ & $300^\star$ && $300$ & $300^\star$ && $300$ & $300^\star$ \\
$R_{\rm f}$ & $-1.0$ & $0.66^{+0.04}_{-0.07}$ && $-1.0$ & $0.767^{+0.042}_{-0.022}$ && $-1.0$ & $0.70^{+0.04}_{-0.05}$\\
\hline
$\chi^2/\nu$ && $\quad 1169.03/1171 \quad$ &&& $\quad 1145.59/1171 \quad$ &&& $\quad 1075.16/1171 \quad$ \\
&& =0.998319 &&& =0.978301 &&& =0.918156  \\
\hline\hline
\end{tabular}}
\vspace{0.2cm}
\tablenotetext{0}{Best-fit values for simulations~D1-D7. In all these simulations, the disk inclination angle is $i = 35^{\circ}$ and the outer radius of the Polish donut disk is $R_{\rm out}=40$~$M$. The reported uncertainties correspond to 90\% confidence level for one relevant parameter. $^\star$ indicates that the parameter is frozen in the fit.}
\end{table*}

%%%%%%%% lamppost simulations tables %%%%%%%%%%%%%%%%%%%%

%% %%%%%%%%%%%% E: h = 3, i = 20 %%%%%%%%%%%%%%

{\begin{table*}
\centering
\caption{ \label{t-fit-e}}
{\renewcommand{\arraystretch}{1.3}
\begin{tabular}{lccccccccccc}
\hline\hline
 & \multicolumn{2}{c}{Simulation~E1} && \multicolumn{2}{c}{Simulation~E2} && \multicolumn{2}{c}{Simulation~E3} && \multicolumn{2}{c}{Simulation~E4} \\
 & Input & Fit && Input & Fit && Input & Fit && Input & Fit \\
\hline
{\sc tbabs} &&&&&&&&&& \\
$N_{\rm H} / 10^{20}$ cm$^{-2}$ & $6.74$ & $6.74^\star$ && $6.74$ & $6.74^\star$ && $6.74$ & $6.74^\star$ && $6.74$ & $6.74^\star$ \\
\hline
{\sc relxilllp} &&&&&&&& \\
$h$ [$M$] & $3$ & $2.000^{+0.010}$ && $3$ & $2.000^{+0.011}$ && $3$ & $2.000^{+0.007}$  && $3$ & $2.000^{+0.004} $ \\
$i$ [deg] & $20$ & $3.3^{+0.5}_{\rm -(P)}$ && $20$ & $4.2^{+0.6}_{-1.1}$ && $20$ & $3.70^{+0.15}_{-0.66}$ && $20$ & $4.71^{+0.36}_{-0.84}$ \\
$a_*$ & $0.40$ & $0.580^{+0.008}_{-0.005}$ && $0.60$ & $0.646^{+0.004}_{-0.006}$ && $0.80$ & $0.712^{+0.004}_{-0.005}$ && $0.85$ & $0.733^{+0.008}_{-0.006}$ \\
$\log\xi$ & $3.1$ & $3.1138^{+0.0022}_{-0.0017}$ && $3.1$ & $3.107^{+0.013}_{-0.003}$ && $3.1$ & $3.127^{+0.004}_{-0.003}$ && $3.1$ & $3.139^{+0.004}_{-0.003}$ \\
$A_{\rm Fe}$ & $1$ & $1.00^{+0.009}_{-0.005}$ && $1$ & $0.988^{+0.012}_{-0.004}$ && $1$ & $1.003^{+0.015}_{-0.005}$ && $1$ & $1.000^{+0.007}_{-0.005}$ \\
$\Gamma$ & $2$ & $1.98748^{+0.00004}_{-0.00018}$ && $2$ & $1.9879^{+0.0011}_{-0.0012}$ && $2$ & $1.98384^{+0.00019}_{-0.00045}$ && $2$ & $1.9838^{+0.0003}_{-0.0011}$ \\
$E_{\rm cut}$ [keV] & $300$ & $300^\star$ && $300$ & $300^\star$ && $300$ & $300^\star$ && $300$ & $300^\star$ \\
$R_{\rm f}$ & $-1.0$ & $1.807^{+0.025}_{-0.009}$ && $-1.0$ & $1.868^{+0.049}_{-0.022}$ && $-1.0$ & $2.185^{+0.014}_{-0.052}$ && $-1.0$ & $2.29^{+0.03}_{-0.04}$ \\
\hline
$\chi^2/\nu$ && $\quad 1692.89/1171 \quad$ &&& $\quad 1805.42/1171\quad$ &&& $\quad  2246.72/1171 \quad$ &&& $\quad 2642.44/1171 \quad$ \\
&& =1.44568 &&& =1.54178 &&& =1.91864 &&& = 2.25657\\
\hline\hline
\vspace{0.4cm}
\end{tabular}}
{\renewcommand{\arraystretch}{1.3}
\begin{tabular}{lcccccccccc}
\hline\hline
 & \multicolumn{2}{c}{Simulation~E5} && \multicolumn{2}{c}{Simulation~E6}&& \multicolumn{2}{c}{Simulation~E7}\\
 & Input & Fit && Input & Fit && Input & Fit\\
\hline
{\sc tbabs} &&&&&&&&&& \\
$N_{\rm H} / 10^{20}$ cm$^{-2}$ & $6.74$ & $6.74^\star$ && $6.74$ & $6.74^\star$ && $6.74$ & $6.74^\star$ \\
\hline
{\sc relxilllp} &&&&&&&& \\
$h$ [$M$]& $3$ & $2.000^{+0.006}$ && $3$ & $2.0000^{+0.0023}$ && $3$ & $2.0000^{+0.0017}$   \\
$i$ [deg] & $20$ & $3.0^{+0.8}$ && $20$ & $3.00^{+0.16}$  && $20$ & $3.00^{+0.16}$\\
$a_*$ & $0.90$ & $0.787^{+0.004}_{-0.007}$ && $0.95$ & $0.805^{+0.008}_{-0.007}$ && $0.98$ & $0.823^{+0.005}_{-0.002}$ \\
$\log\xi$ & $3.1$ & $3.148^{+0.004}_{-0.007}$ && $3.1$ & $3.1462^{+0.0024}_{-0.0044}$ && $3.1$ & $3.150^{+0.004}_{-0.005}$ \\
$A_{\rm Fe}$ & $1$ & $1.008^{+0.018}_{-0.011}$ && $1$ & $1.000^{+0.010}_{-0.005}$ && $1$ & $1.00^{+0.01}_{-0.01}$  \\
$\Gamma$ & $2$ & $1.9801^{+0.0011}_{-0.0003}$ && $2$ & $1.9803^{+0.0009}_{-0.0009}$ && $2$ & $1.9811^{+0.0011}_{-0.0004}$\\
$E_{\rm cut}$ [keV] & $300$ & $300^\star$ && $300$ & $300^\star$ && $300$ & $300^\star$  \\
$R_{\rm f}$ & $-1.0$ & $2.73^{+0.09}_{-0.04}$ && $-1.0$ & $2.78^{+0.03}_{-0.06}$ && $-1.0$ & $3.00^{+0.02}_{-0.02}$\\
\hline
$\chi^2/\nu$ && $\quad 2581.80/1171 \quad$ &&& $\quad 3022.72/1171 \quad$ &&& $\quad 3504.85/1171 \quad$ \\
&& =2.20478 &&& =2.58131 &&& =2.99304\\
\hline\hline
\end{tabular}}
\vspace{0.2cm}
\tablenotetext{0}{Best-fit values for simulations~E1-E7. In all these simulations, the disk inclination angle is $i = 20^{\circ}$, the height of the corona is $h = 3$~$M$, and the outer radius of the Polish donut disk is $R_{\rm out}=40$~$M$. The reported uncertainties correspond to 90\% confidence level for one relevant parameter. $^\star$ indicates that the parameter is frozen in the fit.}
\end{table*}}

%% %%%%%%%%%%%% F: h = 6, i = 20 %%%%%%%%%%%%%%

{\begin{table*}
\centering
\caption{ \label{t-fit-f}}
{\renewcommand{\arraystretch}{1.3}
\begin{tabular}{lccccccccccc}
\hline\hline
 & \multicolumn{2}{c}{Simulation~F1} && \multicolumn{2}{c}{Simulation~F2} && \multicolumn{2}{c}{Simulation~F3} && \multicolumn{2}{c}{Simulation~F4} \\
 & Input & Fit && Input & Fit && Input & Fit && Input & Fit \\
\hline
{\sc tbabs} &&&&&&&&&& \\
$N_{\rm H} / 10^{20}$ cm$^{-2}$ & $6.74$ & $6.74^\star$ && $6.74$ & $6.74^\star$ && $6.74$ & $6.74^\star$ && $6.74$ & $6.74^\star$ \\
\hline
{\sc relxilllp} &&&&&&&& \\
$h$ [$M$] & $6$ & $4.04^{+0.15}_{-0.23}$ && $6$ & $3.80^{+0.10}_{-0.25}$ && $6$ & $3.07^{+0.12}_{-0.38}$  && $6$ & $2.74^{+0.14}_{-0.41} $ \\
$i$ [deg] & $20$ & $4.8^{+1.3}_{-1.4}$ && $20$ & $4.4^{+1.7}_{-1.1}$ && $20$ & $5.0^{+1.1}_{-1.1}$ && $20$ & $4.5^{+1.0}_{-0.9}$ \\
$a_*$ & $0.40$ & $0.422^{+0.019}_{-0.017}$ && $0.60$ & $0.449^{+0.012}_{-0.015}$ && $0.80$ & $0.428^{+0.013}_{-0.014}$ && $0.85$ & $0.418^{+0.009}_{-0.010}$ \\
$\log\xi$ & $3.1$ & $3.096^{+0.007}_{-0.007}$ && $3.1$ & $3.102^{+0.005}_{-0.006}$ && $3.1$ & $3.098^{+0.004}_{-0.006}$ && $3.1$ & $3.099^{+0.005}_{-0.005}$ \\
$A_{\rm Fe}$ & $1$ & $0.991^{+0.006}_{-0.005}$ && $1$ & $1.00^{+0.03}_{-0.01}$ && $1$ & $0.995^{+0.028}_{-0.009}$ && $1$ & $0.995^{+0.013}_{-0.006}$ \\
$\Gamma$ & $2$ & $1.9935^{+0.0011}_{-0.0013}$ && $2$ & $1.9930^{+0.0010}_{-0.0013}$ && $2$ & $1.9928^{+0.0012}_{-0.0015}$ && $2$ & $1.9917^{+0.0009}_{-0.0013}$ \\
$E_{\rm cut}$ [keV] & $300$ & $300^\star$ && $300$ & $300^\star$ && $300$ & $300^\star$ && $300$ & $300^\star$ \\
$R_{\rm f}$ & $-1.0$ & $1.370^{+0.024}_{-0.022}$ && $-1.0$ & $1.418^{+0.034}_{-0.021}$ && $-1.0$ & $1.435^{+0.036}_{-0.021}$ && $-1.0$ & $1.470^{+0.038}_{-0.018}$ \\
\hline
$\chi^2/\nu$ && $\quad 1388.15/1171 \quad$ &&& $\quad 1513.61/1171\quad$ &&& $\quad  1545.76/1171 \quad$ &&& $\quad 1599.63/1171 \quad$ \\
&& =1.18544 &&& =1.29258 &&& =1.32004 &&& =1.36604\\
\hline\hline
\vspace{0.4cm}
\end{tabular}}
{\renewcommand{\arraystretch}{1.3}
\begin{tabular}{lcccccccccc}
\hline\hline
 & \multicolumn{2}{c}{Simulation~F5} && \multicolumn{2}{c}{Simulation~F6}&& \multicolumn{2}{c}{Simulation~F7}\\
 & Input & Fit && Input & Fit && Input & Fit\\
\hline
{\sc tbabs} &&&&&&&&&& \\
$N_{\rm H} / 10^{20}$ cm$^{-2}$ & $6.74$ & $6.74^\star$ && $6.74$ & $6.74^\star$ && $6.74$ & $6.74^\star$ \\
\hline
{\sc relxilllp} &&&&&&&& \\
$h$ [$M$] & $6$ & $2.74^{+0.14}_{-0.28}$ && $6$ & $2.00^{+0.05}$ && $6$ & $2.000^{+0.009}$   \\
$i$ [deg] & $20$ & $4.4^{+1.0}_{-1.0}$ && $20$ & $5.0^{+1.6}_{-1.8}$  && $20$ & $3.6^{+0}_{-3}$\\
$a_*$ & $0.90$ & $0.447^{+0.009}_{-0.011}$ && $0.95$ & $0.437^{+0.018}_{-0.009}$ && $0.98$ & $0.468^{+0.004}_{-0.003}$ \\
$\log\xi$ & $3.1$ & $3.104^{+0.008}_{-0.005}$ && $3.1$ & $3.110^{+0.015}_{-0.009}$ && $3.1$ & $3.3280^{+0.0021}_{-0.0174}$ \\
$A_{\rm Fe}$ & $1$ & $0.998^{+0.022}_{-0.006}$ && $1$ & $1.007^{+0.086}_{-0.021}$ && $1$ & $1.468^{+0.016}_{-0.016}$  \\
$\Gamma$ & $2$ & $1.9902^{+0.0010}_{-0.0007}$ && $2$ & $1.9904^{+0.0022}_{-0.0024}$ && $2$ & $1.96488^{+0.00196}_{-0.00018}$\\
$E_{\rm cut}$ [keV] & $300$ & $300^\star$ && $300$ & $300^\star$ && $300$ & $300^\star$  \\
$R_{\rm f}$ & $-1.0$ & $1.523^{+0.035}_{-0.010}$ && $-1.0$ & $1.54^{+0.06}_{-0.05}$ && $-1.0$ & $3.73^{+0.05}_{-0.01}$\\
\hline
$\chi^2/\nu$ && $\quad 1604.58/1171 \quad$ &&& $\quad 1730.59/1171 \quad$ &&& $\quad 2731.15/1171 \quad$ \\
&& =1.37027 &&& =1.47788 &&& =2.33232\\
\hline\hline
\end{tabular}}
\vspace{0.2cm}
\tablenotetext{0}{Best-fit values for simulations~F1-F7. In all these simulations, the disk inclination angle is $i = 20^{\circ}$, the height of the corona is $h = 6$~$M$, and the outer radius of the Polish donut disk is $R_{\rm out}=40$~$M$. The reported uncertainties correspond to 90\% confidence level for one relevant parameter. $^\star$ indicates that the parameter is frozen in the fit.}
\end{table*}}

%% %%%%%%%%%%%% G: h = 3, i = 35 %%%%%%%%%%%%%%

{\begin{table*}
\centering
\caption{ \label{t-fit-g}}
{\renewcommand{\arraystretch}{1.3}
\begin{tabular}{lccccccccccc}
\hline\hline
 & \multicolumn{2}{c}{Simulation~G1} && \multicolumn{2}{c}{Simulation~G2} && \multicolumn{2}{c}{Simulation~G3} && \multicolumn{2}{c}{Simulation~G4} \\
 & Input & Fit && Input & Fit && Input & Fit && Input & Fit \\
\hline
{\sc tbabs} &&&&&&&&&& \\
$N_{\rm H} / 10^{20}$ cm$^{-2}$ & $6.74$ & $6.74^\star$ && $6.74$ & $6.74^\star$ && $6.74$ & $6.74^\star$ && $6.74$ & $6.74^\star$ \\
\hline
{\sc relxilllp} &&&&&&&& \\
$h$ [$M$] & $3$ & $2.6^{+0.3}_{-0.3}$ && $3$ & $2.66^{+0.20}_{-0.21}$ && $3$ & $2.00^{+0.24}$ && $3$ & $2.17^{+0.17}_{\rm -(P)} $ \\
$i$ [deg] & $35$ & $34.9^{+0.5}_{-0.4}$ && $35$ & $32.7^{+0.5}_{-0.3}$ && $35$ & $33.28^{+0.43}_{-0.23}$ && $35$ & $33.62^{+0.54}_{-0.64}$ \\
$a_*$ & $0.40$ & $0.798^{+0.014}_{-0.015}$ && $0.60$ & $0.827^{+0.013}_{-0.015}$ && $0.80$ & $0.826^{+0.011}_{-0.002}$ && $0.85$ & $0.859^{+0.009}_{-0.011}$ \\
$\log\xi$ & $3.1$ & $3.090^{+0.011}_{-0.010}$ && $3.1$ & $3.076^{+0.010}_{-0.010}$ && $3.1$ & $3.075^{+0.011}_{-0.003}$ && $3.1$ & $3.092^{+0.012}_{-0.011}$ \\
$A_{\rm Fe}$ & $1$ & $1.008^{+0.045}_{-0.017}$ && $1$ & $0.988^{+0.021}_{-0.013}$ && $1$ & $0.980^{+0.008}_{-0.007}$ && $1$ & $1.000^{+0.056}_{-0.016}$ \\
$\Gamma$ & $2$ & $1.9935^{+0.0009}_{-0.0009}$ && $2$ & $1.9958^{+0.0011}_{-0.0010}$ && $2$ & $1.9967^{+0.0004}_{-0.0016}$ && $2$ & $1.9937^{+0.0018}_{-0.0019}$ \\
$E_{\rm cut}$ [keV] & $300$ & $300^\star$ && $300$ & $300^\star$ && $300$ & $300^\star$ && $300$ & $300^\star$ \\
$R_{\rm f}$ & $-1.0$ & $1.26^{+0.06}_{-0.04}$ && $-1.0$ & $1.26^{+0.05}_{-0.04}$ && $-1.0$ & $1.32^{+0.05}_{-0.06}$ && $-1.0$ & $1.48^{+0.07}_{-0.06}$ \\
\hline
$\chi^2/\nu$ && $\quad 1216.72/1171 \quad$ &&& $\quad 1199.80/1171\quad$ &&& $\quad  1004.67/1171 \quad$ &&& $\quad 1170.98/1171 \quad$ \\
&& =1.03904 &&& =1.02459 &&& =1.03467 &&& =0.99998 \\
\hline\hline
\vspace{0.4cm}
\end{tabular}}
{\renewcommand{\arraystretch}{1.3}
\begin{tabular}{lcccccccccc}
\hline\hline
 & \multicolumn{2}{c}{Simulation~G5} && \multicolumn{2}{c}{Simulation~G6}&& \multicolumn{2}{c}{Simulation~G7}\\
 & Input & Fit && Input & Fit && Input & Fit\\
\hline
{\sc tbabs} &&&&&&&&&& \\
$N_{\rm H} / 10^{20}$ cm$^{-2}$ & $6.74$ & $6.74^\star$ && $6.74$ & $6.74^\star$ && $6.74$ & $6.74^\star$ \\
\hline
{\sc relxilllp} &&&&&&&& \\
$h$ [$M$] & $3$ & $2.00^{+0.19}$ && $3$ & $2.00^{+0.03}$ && $3$ & $2.000^{+0.019}$   \\
$i$ [deg] & $35$ & $32.9^{+0.6}_{-0.7}$ && $35$ & $30.7^{+0.8}_{-1.0}$ && $35$ & $30.6^{+0.6}_{-0.8}$\\
$a_*$ & $0.90$ & $0.868^{+0.009}_{-0.012}$ && $0.95$ & $0.872^{+0.010}_{-0.013}$ && $0.98$ & $0.904^{+0.007}_{-0.008}$ \\
$\log\xi$ & $3.1$ & $3.087^{+0.015}_{-0.013}$ && $3.1$ & $3.082^{+0.011}_{-0.012}$ && $3.1$ & $3.086^{+0.013}_{-0.009}$ \\
$A_{\rm Fe}$ & $1$ & $1.000^{+0.072}_{-0.019}$ && $1$ & $1.000^{+0.036}_{-0.019}$ && $1$ & $1.000^{+0.062}_{-0.015}$  \\
$\Gamma$ & $2$ & $1.9941^{+0.0021}_{-0.0022}$ && $2$ & $1.9965^{+0.0020}_{-0.0019}$ && $2$ & $1.9942^{+0.0017}_{-0.0020}$\\
$E_{\rm cut}$ [keV] & $300$ & $300^\star$ && $300$ & $300^\star$ && $300$ & $300^\star$  \\
$R_{\rm f}$ & $-1.0$ & $1.50^{+0.08}_{-0.07}$ && $-1.0$ & $1.51^{+0.06}_{-0.08}$ && $-1.0$ & $1.69^{+0.08}_{-0.06}$\\
\hline
$\chi^2/\nu$ && $\quad 1177.31/1171 \quad$ &&& $\quad 1260.59/1171 \quad$ &&& $\quad 1190.89/1171 \quad$ \\
&& =1.00539 &&& =1.07651 &&& =1.01699\\
\hline\hline
\end{tabular}}
\vspace{0.2cm}
\tablenotetext{0}{Best-fit values for simulations~G1-G7. In all these simulations, the disk inclination angle is $i = 35^{\circ}$, the height of the corona is $h = 3$~$M$, and the outer radius of the Polish donut disk is $R_{\rm out}=40$~$M$. The reported uncertainties correspond to 90\% confidence level for one relevant parameter. $^\star$ indicates that the parameter is frozen in the fit.}
\end{table*}}

%%%%%%%%%%%%%% H: h = 6, i = 35 %%%%%%%%%%%%%%%%

{\begin{table*}
\centering
\caption{ \label{t-fit-h}}
{\renewcommand{\arraystretch}{1.3}
\begin{tabular}{lccccccccccc}
\hline\hline
 & \multicolumn{2}{c}{Simulation~H1} && \multicolumn{2}{c}{Simulation~H2} && \multicolumn{2}{c}{Simulation~H3} && \multicolumn{2}{c}{Simulation~H4} \\
 & Input & Fit && Input & Fit && Input & Fit && Input & Fit \\
\hline
{\sc tbabs} &&&&&&&&&& \\
$N_{\rm H} / 10^{20}$ cm$^{-2}$ & $6.74$ & $6.74^\star$ && $6.74$ & $6.74^\star$ && $6.74$ & $6.74^\star$ && $6.74$ & $6.74^\star$ \\
\hline
{\sc relxilllp} &&&&&&&& \\
$h$ [$M$] & $6$ & $4.77^{+0.25}_{-0.23}$ && $6$ & $4.31^{+0.29}_{-0.22}$ && $6$ & $3.94^{+0.23}_{-0.24}$ && $6$ & $3.69^{+0.25}_{-0.21} $ \\
$i$ [deg] & $35$ & $35.2^{+0.5}_{-0.5}$ && $35$ & $35.6^{+0.5}_{-0.6}$ && $35$ & $34.7^{+0.6}_{-0.5}$ && $35$ & $34.8^{+0.3}_{-0.6}$ \\
$a_*$ & $0.40$ & $0.86^{+0.04}_{-0.04}$ && $0.60$ & $0.84^{+0.04}_{-0.03}$ && $0.80$ & $0.816^{+0.014}_{-0.028}$ && $0.85$ & $0.82^{+0.03}_{-0.03}$ \\
$\log\xi$ & $3.1$ & $3.078^{+0.008}_{-0.009}$ && $3.1$ & $3.088^{+0.009}_{-0.009}$ && $3.1$ & $3.092^{+0.012}_{-0.010}$ && $3.1$ & $3.094^{+0.012}_{-0.011}$ \\
$A_{\rm Fe}$ & $1$ & $0.980^{+0.014}_{-0.015}$ && $1$ & $0.999^{+0.041}_{-0.014}$ && $1$ & $0.993^{+0.040}_{-0.016}$ && $1$ & $1.02^{+0.05}_{-0.03}$ \\
$\Gamma$ & $2$ & $1.9962^{+0.0017}_{-0.0014}$ && $2$ & $1.9930^{+0.0016}_{-0.0014}$ && $2$ & $1.9947^{+0.0017}_{-0.0015}$ && $2$ & $1.9946^{+0.0018}_{-0.0017}$ \\
$E_{\rm cut}$ [keV] & $300$ & $300^\star$ && $300$ & $300^\star$ && $300$ & $300^\star$ && $300$ & $300^\star$ \\
$R_{\rm f}$ & $-1.0$ & $1.05^{+0.03}_{-0.03}$ && $-1.0$ & $1.12^{+0.03}_{-0.03}$ && $-1.0$ & $1.13^{+0.04}_{-0.04}$ && $-1.0$ & $1.16^{+0.04}_{-0.04}$ \\
\hline
$\chi^2/\nu$ && $\quad 1177.85/1171 \quad$ &&& $\quad 1120.05/1171\quad$ &&& $\quad  1151.56/1171 \quad$ &&& $\quad 1311.55/1171 \quad$ \\
&& =1.00585 &&& =0.956487 &&& =0.983401 &&& =1.12002 \\
\hline\hline
\vspace{0.4cm}
\end{tabular}}
{\renewcommand{\arraystretch}{1.3}
\begin{tabular}{lcccccccccc}
\hline\hline
 & \multicolumn{2}{c}{Simulation~H5} && \multicolumn{2}{c}{Simulation~H6}&& \multicolumn{2}{c}{Simulation~H7}\\
 & Input & Fit && Input & Fit && Input & Fit\\
\hline
{\sc tbabs} &&&&&&&&&& \\
$N_{\rm H} / 10^{20}$ cm$^{-2}$ & $6.74$ & $6.74^\star$ && $6.74$ & $6.74^\star$ && $6.74$ & $6.74^\star$ \\
\hline
{\sc relxilllp} &&&&&&&& \\
$h$ [$M$] & $6$ & $3.78^{+0.20}_{-0.24}$ && $6$ & $3.31^{+0.21}_{-0.22}$ && $6$ & $2.90^{+0.22}_{-0.20}$   \\
$i$ [deg] & $35$ & $33.5^{+0.6}_{-0.6}$ && $35$ & $33.6^{+0.6}_{-0.3}$ && $35$ & $32.9^{+0.4}_{-0.4}$\\
$a_*$ & $0.90$ & $0.82^{+0.03}_{-0.03}$ && $0.95$ & $0.814^{+0.022}_{-0.023}$ && $0.98$ & $0.813^{+0.015}_{-0.016}$ \\
$\log\xi$ & $3.1$ & $3.082^{+0.013}_{-0.009}$ && $3.1$ & $3.082^{+0.012}_{-0.005}$ && $3.1$ & $3.109^{+0.012}_{-0.013}$ \\
$A_{\rm Fe}$ & $1$ & $0.996^{+0.056}_{-0.015}$ && $1$ & $0.996^{+0.051}_{-0.015}$ && $1$ & $1.10^{+0.05}_{-0.06}$  \\
$\Gamma$ & $2$ & $1.9944^{+0.0016}_{-0.0018}$ && $2$ & $1.9941^{+0.0017}_{-0.0017}$ && $2$ & $1.9904^{+0.0017}_{-0.0018}$\\
$E_{\rm cut}$ [keV] & $300$ & $300^\star$ && $300$ & $300^\star$ && $300$ & $300^\star$  \\
$R_{\rm f}$ & $-1.0$ & $1.41^{+0.05}_{-0.03}$ && $-1.0$ & $1.19^{+0.05}_{-0.04}$ && $-1.0$ & $1.34^{+0.06}_{-0.05}$\\
\hline
$\chi^2/\nu$ && $\quad 1165.47/1171 \quad$ &&& $\quad 1166.52/1171 \quad$ &&& $\quad 1242.74/1171 \quad$ \\
&& =0.995275 &&& =0.996178 &&& =1.06126\\
\hline\hline
\end{tabular}}
\vspace{0.2cm}
\tablenotetext{0}{Best-fit values for simulations~H1-H7. In all these simulations, the disk inclination angle is $i = 35^{\circ}$, the height of the corona is $h = 6$~$M$, and the outer radius of the Polish donut disk is $R_{\rm out}=40$~$M$. The reported uncertainties correspond to 90\% confidence level for one relevant parameter. $^\star$ indicates that the parameter is frozen in the fit.}
\end{table*}}

%%%%%%%%%%%%%%%%%%%%%%%%%%%%%%%%%%%%%%%%%%%%

%%%%%%%%%%% lamppost plots %%%%%%%%%%%%%%%%%%

%%%%%%%% lamppost ratio plots%%%%%%%%%%%%%%%%%%%%

%%%%%%%%%%%%%%%%%%%%%%%%%%%%%%%%%%%%%%%%

%%%%%%%%%%%%%%%%%%%%%%%%%%%%%%%

%%%%%%%%%% Lamppost plots %%%%%%%%%%%%

\section{Lamppost corona} \label{LP_corona}

The lamppost set-up is quite a popular choice in literature when we want to study a specific coronal geometry. In its simplest form, the corona is an isotropically emitting, point-like source along the black hole spin axis and located at a certain height $h_{\rm lamppost}$, which is the only parameter of the coronal geometry. \citet{rel1} calculated the irradiation flux on infinitesimally thin accretion disks and constructed the {\sc relxilllp} model. \citet{taylor} calculated the intensity profile for thin disks with finite thickness, considering Eddington-scaled accretion luminosities up to 30\%; they showed that the the convex geometry of thin disks with finite thickness casts important features in the irradiation profile such as flattening of the flux at larger radius due to self shadowing. \citet{relxill_nk} calculated the irradiation flux for infinitesimally thin accretion disks in deformed Kerr spacetimes.

Here we follow the procedure described in \citet{rel1} and \citet{relxill_nk} and we compute the irradiation intensity profile for Polish donut disks. The photons hitting the surface of the disk are binned into annuli along the disk surface. The comoving area of each annulus is calculated as in \citet{taylor}  
\begin{equation}
A = 2\pi \gamma \left[ g_{rr} + g_{\theta \theta} \left(\frac{d\theta}{d\rho}\right)^2 \sin^2\theta \right ]^2 \left( \frac{g_{\phi \phi}}{\sin^2\theta} \right)^{1/2} d\rho \, .
\end{equation}
Here $\rho = r \sin\theta$ is the projection of the radial coordinate on the equatorial plane and  $\gamma$ is the Lorentz factor, which is computed as in \citet{relxill_nk} 
\begin{equation}
\gamma =  \left [ \frac{ \left(\Omega - \frac{g_{t\phi}}{g_{\phi\phi}} \right)^2 g_{\phi\phi}^2}{g_{tt}g_{t\phi} - g_{t\phi}^2} + 1    \right]^{-1/2} \, 
\end{equation}
where $\Omega$ is the disk's angular velocity
\begin{equation}
\Omega = -\left( \frac{l g_{tt} + g_{t\phi}}{ l g_{t\phi} + g_{\phi\phi}}    \right) \, .
\end{equation}

\begin{figure*}[t]
\begin{center}
\includegraphics[type=pdf,ext=.pdf,read=.pdf,width=8cm]{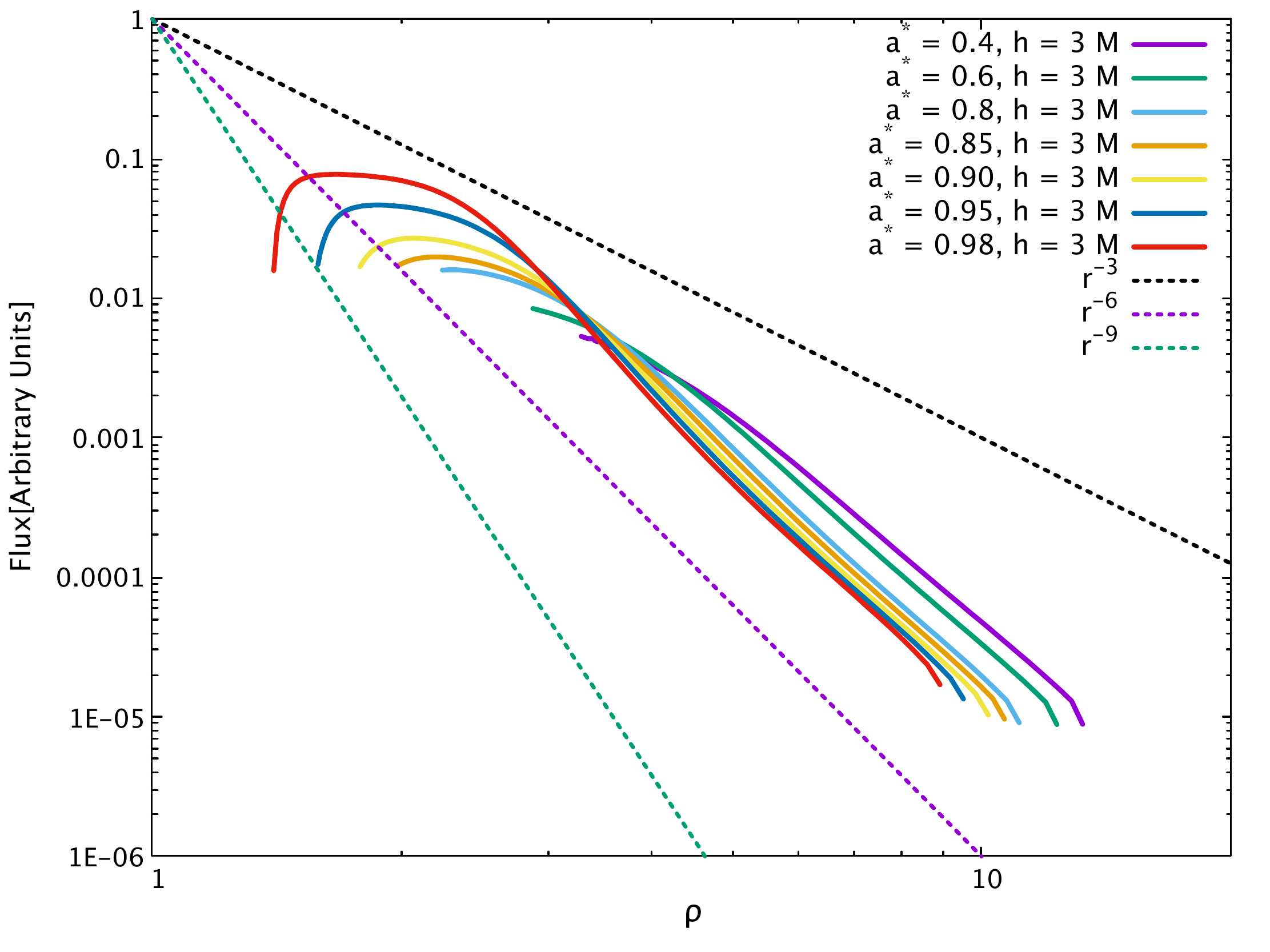}
\hspace{0.8cm}
\includegraphics[type=pdf,ext=.pdf,read=.pdf,width=8cm]{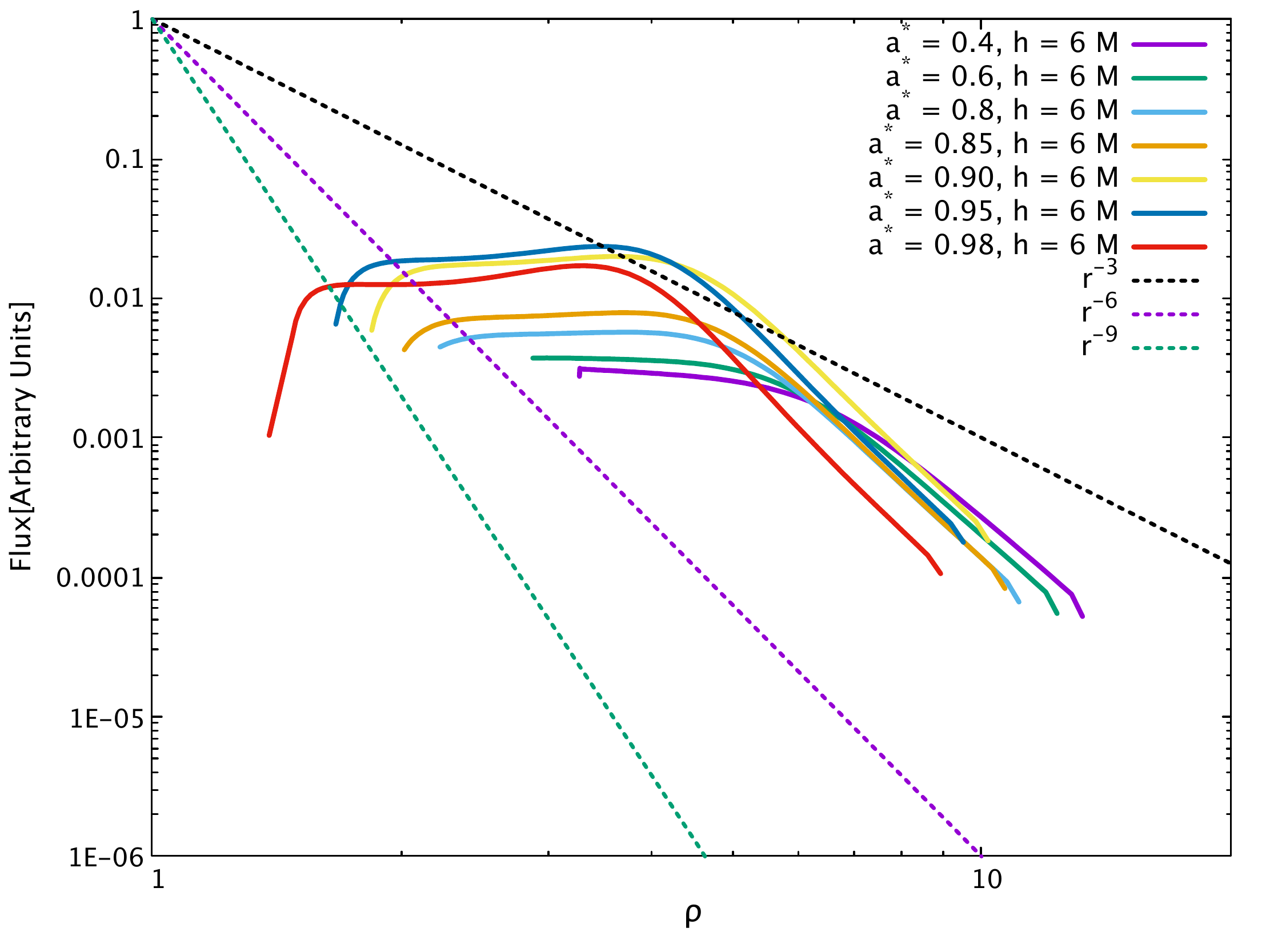}\\
%\vspace{0.4cm}
\caption{Emissivity profiles (arbitrary units) with respect to the projection of the radial coordinate on the equatorial plane, $\rho$. The input spin parameter is $a_* = 0.4$, 0.6, 0.8, 0.85, 0.90, 0.95, and 0.98. The corona height is 3~$M$ (left panel) and 6~$M$ (right panel). The dotted straight lines (black, violet, and green) are the power-law irradiation profile (with emissivity index 3, 6, and 9, respectively). The outer edge of the Polish donut is at $R_{\rm out} = 40$~$M$.}
\label{flux}
\end{center}
\end{figure*}

The resulting lamppost emissivity profiles for the Polish donut disks studied in this work are reported in Fig.~\ref{flux} and compared with the power-law intensity profiles with emissivity index 3, 6, and 9.

%%%%%%%% lamppost flux plots%%%%%%%%%%%%%%%%%%%%

\end{document}